\documentclass[onecolumn,11pt]{article}
\usepackage{amsmath,amssymb}
\usepackage[dvipdfmx]{graphicx}
\usepackage{setspace,atbegshi,mediabb,color}
\usepackage{cite}
\usepackage{ascmac}
\newdimen\tbaselineshift
\usepackage{framed}
\usepackage{fancybox}
\AtBeginShipoutFirst{}
\usepackage{authblk}

\textwidth=\paperwidth
\advance \textwidth by -1.5in
\textheight=\paperheight
\advance \textheight by -2.2in
\def\bra#1{\langle #1|}
\def\ket#1{|#1\rangle}

\newcommand{\rA}{{\mathrm{A}}}
\newcommand{\rB}{{\mathrm{B}}}
\newcommand{\rC}{{\mathrm{C}}}

\newcommand{\rF}{{\mathrm{F}}}

\newcommand{\rI}{{\mathrm{I}}}

\newcommand{\rR}{{\mathrm{R}}}
\newcommand{\rS}{{\mathrm{S}}}

\newcommand{\rX}{{\mathrm{X}}}
\newcommand{\rY}{{\mathrm{Y}}}
\newcommand{\rZ}{{\mathrm{Z}}}
\newcommand{\ra}{{\mathrm{a}}}
\newcommand{\rb}{{\mathrm{b}}}

\newcommand{\ri}{{\mathrm{i}}}
\newcommand{\rj}{{\mathrm{j}}}

\newcommand{\rt}{{\mathrm{t}}}

\newcommand{\rv}{{\mathrm{v}}}

\newcommand{\rx}{{\mathrm{x}}}
\newcommand{\ry}{{\mathrm{y}}}
\newcommand{\rz}{{\mathrm{z}}}

\newcommand{\can}{{\mathrm{can}}}
\newcommand{\tr}{{\mathrm{tr}}}

\newcommand{\dist}{{\mathrm{dist}}}
\newcommand{\supp}{{\mathrm{supp}}}

\newcommand{\trunc}{{\mathrm{T}}}

\newcommand{\defL}{:=}

\newcommand{\trtotal}{\mathrm{tr}_{\mathrm{S}\cup\mathrm{B}}}

\newcommand{\sbra}[1]{\left(#1\right)}

\newcommand{\bbra}[1]{\left[#1\right]}
\newcommand{\tbra}[1]{\left\langle#1\right\rangle}
\newcommand{\abra}[1]{\left|#1\right|}
\newcommand{\nbra}[1]{\left\|#1\right\|}

\DeclareMathOperator*{\argmin}{arg\,min}
\DeclareMathOperator*{\argmax}{arg\,max}

\newcounter{cnt:Lemma}
\newcommand{\cntLemma}{\arabic{cnt:Lemma}}
\newcommand{\Lemma}[1][]{\refstepcounter{cnt:Lemma}#1 {\bf Lemma \cntLemma}}

\newcounter{cnt:Assumption}
\newcommand{\cntAssumption}{\arabic{cnt:Assumption}}
\newcommand{\Assumption}[1][]{\refstepcounter{cnt:Assumption}#1 {\bf Assumption \cntAssumption}}

\newcounter{cnt:Proposition}
\newcommand{\cntProposition}{\arabic{cnt:Proposition}}
\newcommand{\Proposition}[1][]{\refstepcounter{cnt:Proposition}#1 {\bf Proposition \cntProposition}}

\newcounter{cnt:Corollary}
\newcommand{\cntCorollary}{\arabic{cnt:Corollary}}
\newcommand{\Corollary}[1][]{\refstepcounter{cnt:Corollary}#1 {\bf Corollary \cntCorollary}}

\newcounter{cnt:Observation}
\newcommand{\cntObservation}{\arabic{cnt:Observation}}
\newcommand{\Observation}[1][]{\refstepcounter{cnt:Observation}#1 {\bf Observation \cntObservation}}

\newcounter{cnt:Definition}
\newcommand{\cntDefinition}{\arabic{cnt:Definition}}
\newcommand{\Definition}[1][]{\refstepcounter{cnt:Definition}#1 {\bf Definition \cntDefinition}}

\newcounter{cnt:Theorem}
\newcommand{\cntTheorem}{\arabic{cnt:Theorem}}
\newcommand{\Theorem}[1][]{\refstepcounter{cnt:Theorem}#1 {\bf Theorem \cntTheorem}}

\begin{document}

\title{Fluctuation Theorem for Many-Body Pure Quantum States}

\author[1]{Eiki Iyoda}
\affil[1]{
Department of Applied Physics, The University of Tokyo,
7-3-1 Hongo, Bunkyo-ku, Tokyo 113-8656, Japan
}
\author[2]{Kazuya Kaneko}
\affil[2]{
Department of Basic Science, The University of Tokyo,
3-8-1 Komaba, Meguro-ku, Tokyo 153-8902, Japan
}
\author[1]{Takahiro Sagawa}

\maketitle
\begin{abstract} 
We prove the second law of thermodynamics and the nonequilibirum fluctuation theorem for pure quantum states.
The entire system obeys reversible unitary dynamics, 
where the initial state of the heat bath is not the canonical distribution but is a single energy-eigenstate that satisfies the eigenstate-thermalization hypothesis (ETH).
Our result is mathematically rigorous and based on the Lieb-Robinson bound, 
which gives the upper bound of the velocity of information propagation in many-body quantum systems.
The entanglement entropy of a subsystem is shown connected to thermodynamic heat, highlighting the foundation of the information-thermodynamics link.
We confirmed our theory by numerical simulation of hard-core bosons, and observed dynamical crossover from thermal fluctuations to bare quantum fluctuations.
Our result reveals a universal scenario that the second law emerges from quantum mechanics, 
and can experimentally be tested by artificial isolated quantum systems such as ultracold atoms.
\end{abstract}

\textit{Introduction.}
Although the microscopic laws of physics do not prefer a particular direction of time, the macroscopic world exhibits inevitable irreversibility represented by the second law of thermodynamics.  
Modern researches has revealed that even a pure quantum state, described by a single wave function without any genuine thermal fluctuation, can relax to macroscopic thermal equilibrium by a reversible unitary evolution~\cite{Neumann1929,Jensen1985,Tasaki1998,Reimann2008,Rigol2008,Linden2009,Polkovnikov2011,Linden2012,Goldstein2013,Gogolin2016,Tasaki2016Typ}. 
Thermalization of isolated quantum systems, which is relevant to the zeroth law of thermodynamics, is now a very active area of researches in theory~\cite{Neumann1929,Tasaki1998,Reimann2008,Linden2009,Linden2012,Goldstein2013}, numerics~\cite{Jensen1985,Rigol2007,Rigol2008,Biroli2010,Cassidy2011,Calabrese2011,Mallayya2017}, and experiments~\cite{Kinoshita2006,Hofferberth2007,Gring2012,Trotzky2012,Langen2015,Clos2016,Kaufman2016}.
Especially, the concepts of typicality~\cite{Popescu2006,Goldstein2006,Sugita2007,Tasaki2016Typ} and the eigenstate thermalization hypothesis (ETH)~\cite{Berry1977,Peres1984,Jensen1985,Srednicki1994,Rigol2008,Rigol2009,Biroli2010,Kim2014,Alba2015,Beugeling2014,Garrison2015,DAlessio2016,Mori2016} have played significant roles.

However, the second law of thermodynamics, which states that the thermodynamic entropy increases in isolated systems, has not been fully addressed in this context.  
We would emphasize that the informational entropy (i.e., the von Neumann entropy) of such a genuine quantum system never increases, but is always zero~\cite{Nielsen2000}.
In this sense, a fundamental gap between the microscopic and macroscopic worlds has not yet been bridged:
How does the second law emerge from pure quantum states?

In a rather different context,  a general theory of the second law and its connection to information  has recently been developed even out of equilibrium~\cite{Parrondo2015,Sagawa2008,Sagawa2010,Funo2013}, which has also been experimentally  verified in laboratories~\cite{Toyabe2010,Berut2012,Koski2014,Vidrighin2016}. This has revealed that  information contents and thermodynamic quantities  can be treated  on an equal footing, as originally illustrated by Szilard and Landauer in the context of Maxwell's demon~\cite{Landauer1961,Leff2003}.  
This research direction invokes a crucial assumption that the heat bath is, at least in the initial time, in the canonical distribution~\cite{Sagawa2012}; this special initial condition effectively breaks the time-reversal symmetry and leads to the second law of thermodynamics.  
The same assumption has been employed in various modern researches on thermodynamics, such as the nonequilibrium fluctuation theorem
\cite{Jarzynski1997,Crooks1999,Jarzynski2000,Kurchan2000,Tasaki2000FT,Esposito2009,Campisi2011,Sagawa2012,Collin2005,Batalhao2015,An2015} and the thermodynamic resource theory~\cite{Horodecki2013,Brandao2015}.

Based on these streams of researches, in this Letter we rigorously derive the second law of thermodynamics for isolated quantum systems in pure states.
We consider a small system and a large heat bath, where the bath is initially in a single energy-eigenstate.
Such an eigenstate is a pure quantum state, and does not include any statistical mixture as is the case for the canonical distribution.
The second law that we show here is formulated with the von Neumann entropy of the system, 
ensuring the information-thermodynamics link, 
which is a characteristic of our study in contrast to previous approaches~\cite{Tasaki2000,Ikeda2015,DAlessio2016}.  
Furthermore, we prove the integral fluctuation theorem~\cite{Jarzynski1997,Tasaki2000FT,Esposito2009,Jin2016}, a universal relation in nonequilibrium statistical mechanics, which expresses the second law as an equality rather than an inequality.

The key of our theory is combining the following two fundamental concepts.
One is the Lieb-Robinson bound~\cite{Lieb1972,Hastings2006}, 
which characterizes the finite group velocity of information propagation in quantum many-body systems with local interaction.
The other is the ETH, 
which states that even a single energy-eigenstate can behave as thermal~\cite{Berry1977,Peres1984,Jensen1985,Srednicki1994,Rigol2008,Rigol2009,Biroli2010,Kim2014,Alba2015,Beugeling2014,Garrison2015,DAlessio2016,Mori2016}.
In this Letter, we newly prove a variant of the ETH~\cite{Biroli2010,Mori2016}, 
which is referred to as the {\textit weak} ETH and states that most of the energy eigenstate satisfies the ETH, 
if an eigenstate is randomly sampled from the microcanonical energy shell.

Our theory provides a rigorous scenario of the emergence of the second law from quantum mechanics, which is relevant to experiments of artificial isolated quantum systems.
Furthermore, our approach to the second law would be applicable to quite a broad class of modern researches of thermodynamics, 
from thermalization in ultracold atoms~\cite{Trotzky2012} to scrambling in black holes~\cite{Hayden2007,Sekino2008,Maldacena2015,Maldacena2016}.

\textit{Setup.}
We first formulate our setup with a heat bath in a pure state.
Suppose that the entire system consists of  system S and  bath B. 
We assume that bath B is a quantum many-body system on a $d$-dimensional hypercubic lattice with $N$ sites.
The Hamiltonian is given by
\begin{equation}
\hat{H}=\hat{H}_\rS+\hat{H}_\rI+\hat{H}_\rB,
\label{Main_total_Hamiltonian}
\end{equation}
where $\hat{H}_\rS$ and $\hat{H}_\rB$ are respectively the Hamiltonians of system S and bath B,
and $\hat{H}_\rI$ represents their interaction.
We assume that $\hat{H}_\rB$ is translation invariant with local interaction, 
and system S is locally in contact with bath B  (see Fig.~1(a)).
We also assume that the correlation functions in the canonical distribution with respect to $\hat{H}_\rB$ is exponentially decaying for any local observables,
which implies that bath B is not on a critical point.

The initial state of the total system is given by
\begin{equation}
\hat{\rho}(0) =\hat \rho_{\rm S} (0) \otimes | E_i \rangle \langle E_i |, 
\label{Main_initial}
\end{equation}
where $\hat \rho_{\rm S} (0)$ is the initial density operator of system S, 
and $| E_i \rangle$ is the initial energy eigenstate of bath B.
We sample $| E_i \rangle$ from the set of the energy eigenstates in the microcanonical energy shell in a uniformly random way, 
as will be described in detail later.
We can then define temperature $T$ of $| E_i \rangle$ as the temperature of the corresponding energy shell.
We note that the initial correlation between S and B is assumed to be zero.  

The total system then obeys a unitary time evolution by the Hamiltonian: 
$\hat{\rho}(t)=\hat{U}\hat{\rho}(0)\hat{U}^\dag$ with $\hat U:= \exp (- {\rm i} \hat H t / \hbar)$.
Such a situation can experimentally be realized with ultracold atoms by quenching an external potential at time $0$.
Let  $\hat{\rho}_\rS(t):=\tr_\rB\bbra{\hat{\rho}(t)}$
and $\hat{\rho}_\rB(t):=\tr_\rS\bbra{\hat{\rho}(t)}$
be the density operators of system S and bath B at time $t$, respectively.
The change in the von Neumann entropy of S is given by $\Delta S_\rS:=S_\rS(t)-S_\rS(0)$ with $S_\rS (t) :=-\tr_\rS\bbra{\hat{\rho}_\rS (t) \ln \hat{\rho}_\rS (t) }$.
We also define the heat emitted from bath B  by 
$Q:=
-\mathrm{tr}_\rB[
\hat{H}_\mathrm{B}(
\hat{\rho}_\mathrm{B}(t)
-
\hat{\rho}_\mathrm{B}(0)
)
]$.

If  the initial state of system S is pure (i.e., $\hat \rho_{\rm S} (0) = | \psi \rangle \langle \psi |$), the total system is also pure, whose von Neumann entropy vanishes.  In such a case,  the final state $\hat \rho (t)$ remains in a pure state because of the unitarity, but is entangled.  
Correspondingly, the final state of S is mixed and has non-zero von Neumann entropy, 
which is regarded as the entanglement entropy.

\textit{Second law.}
We now discuss our first main result.
If  $| E_i \rangle$ is a typical energy eigenstate, that satisfies the ETH,  
the second law of thermodynamics is shown to hold within a small error:
\begin{align}
\Delta S_\rS - \beta Q \geq - \varepsilon_{\mathrm{2nd}},
\label{Main_Clausius}
\end{align}
where $\varepsilon_\mathrm{2nd}$ is a positive error term.  
We can rigorously prove that $\varepsilon_\mathrm{2nd}$ can be arbitrarily small 
if bath B is sufficiently large.
The left-hand side of inequality (\ref{Main_Clausius})  is regarded as the average entropy production $\langle \sigma \rangle :=\Delta S_\rS - \beta Q$, where $\langle \cdots \rangle$ describes the ensemble average, 
and $\sigma$ is the stochastic entropy production that will be introduced later.
We note that, if the initial state of bath B is not pure but in the canonical distribution
$\hat{\rho}_\rB^{\mathrm{can}}:=e^{-\beta\hat{H}_\rB}/\mathrm{tr}[e^{-\beta\hat{H}_\rB}]$, 
inequality (\ref{Main_Clausius}) exactly holds without any error~\cite{Sagawa2012}.

The second law (\ref{Main_Clausius}) implies that the information-thermodynamics link emerges in genuine quantum systems, 
if we look at the informational entropy of subsystem S, though that of the total system remains unchanged.
A significant consequence of inequality (\ref{Main_Clausius}) is the Landauer erasure principle~\cite{Landauer1961}. 
Suppose that the initial state of S stores one bit of information such that  $S_\rS (0) = \ln 2$, 
and it is erased in the final state: $S_{\rm S} (t) = 0$.  
We then have $\Delta S_{\rm S} = - \ln 2$, and the heat emission from S, represented by $-Q$, 
is bounded by $k_{\rm B}T \ln 2$ within a small error.  
While the Landauer principle and its generalizations have been derived in various ways~\cite{Parrondo2015,Sagawa2014,Shizume1995,Piechocinska2000,Esposito2011,Sagawa2011,Reeb2014}, 
we here showed that it emerges in the presence of a pure quantum bath.
 
We will prove inequality (\ref{Main_Clausius})  in Supplemental Material in a mathematically rigorous way.  
Here We only discuss the essentials of the proof, 
where the key ingredients are the Lieb-Robinson bound~\cite{Lieb1972,Hastings2006} and the weak ETH~\cite{Biroli2010,Mori2016}. 

\textit{Lieb-Robinson bound.}
The Lieb-Robinson bound gives an upper bound of the velocity of information propagation,
and is applicable to any system on a generic lattice with local interaction.
To apply the Lieb-Robinson bound, 
we divide bath B into $\rB_1$ and $\rB_2$,
such that $\rB_1$ is near system S and $\rB_2$ is far from S (see Fig.~1(b)).
Then, the Lieb-Robinson bound~\cite{Lieb1972,Hastings2006} sets the shortest time $\tau$,
at which information about $\rB_2$ reaches S across $\rB_1$. 
We refer to $\tau$ as the Lieb-Robinson time. 

The detailed formulation of the Lieb-Robinson bound is the following.
Let $\tilde{\rS}$ be the union of S and the support of $\hat{H}_\rI$.
Let $\hat{A}_\mathrm{\tilde{S}}$ and $\hat{B}_{\partial \rB_1}$ 
be arbitrary operators with the supports $\tilde{\rS}$ and $\partial \rB_1$, respectively,
where $\partial \rB_1$ is the boundary between $\rB_1$ and $\rB_2$.
Let $\mathrm{dist}(\mathrm{\tilde{S}},\partial\mathrm{B}_1)$ be the spatial distance between $\tilde{\rS}$ and $\partial \rB_1$ on the lattice, and let $| \rm \tilde{S} |$ and $| \partial \rB_1 |$ be the numbers of the sites in $\tilde{\rS}$ and $\partial \rB_1$ respectively.
The Lieb-Robinson bound is formulated in terms of the operator norm $\| \cdot \| $ as 
\begin{align}
\label{Main_eq:LRB}
\frac{
\|
[
\hat{A}_\mathrm{\tilde{S}}(t),\hat{B}_{\partial \rB_1}
]
\|
}
{
\|\hat{A}_\mathrm{\tilde{S}}\|
\|\hat{B}_{\partial \rB_1}\|
}
\leq
C
|\tilde{\rS}||\partial \rB_1|
e^{-\mu\mathrm{dist}(\tilde{\rS}, \partial \rB_1)}
(e^{v|t|}-1),
\end{align}
where $\hat{A}_\mathrm{\tilde{S}}(t):=\hat{U}^\dag\hat{A}_\mathrm{\tilde{S}}\hat{U}$ 
represents the time evolution in the Heisenberg picture,
and $C$, $v$, $\mu$ are positive constants.
In particular, $v/\mu$ represents the velocity of information propagation.
The Lieb-Robinson time is then given by $\tau : = \mu \dist(\tilde{S},\partial\rB_1) / v$.

\textit{Weak ETH.}
We next consider the concept of the weak ETH.
Let $D$ be the dimension of the Hilbert space of the microcanonical energy shell of bath B,
which is exponentially large with respect to $N$, and let $\{ | E_i \rangle \}_{i=1}^D$ be an orthonormal set of the energy eigenstates of $\hat H_\rB$ in the energy shell.
Suppose that we choose $| E_i \rangle$ from  $\{ | E_i \rangle \}_{i=1}^D$ 
in a uniformly random way.
As proved in Supplemental Material, if $\rB_2$ is sufficiently larger than $\rB_1$, we typically have that 
\begin{equation}
\tr_{\rB}\bbra{\hat{O}_{\rB_1}| E_i \rangle \langle E_i | }
\simeq
\tr_{\rB}\bbra{\hat{O}_{\rB_1}\hat{\rho}_\rB^{\mathrm{can}}},
\label{Main_wETH}
\end{equation}
which implies that $| E_i \rangle$ is indistinguishable from $\hat{\rho}_\rB^{\mathrm{can}}$ if we only look at any operator $\hat{O}_{\rB_1}$ on $\rB_1$ with $\|\hat{O}_{\rB_1}\|=1$.
We refer to this theorem as the weak ETH, which is a variant  of a theorem shown in Ref.~\cite{Biroli2010,Mori2016}. 
We note that the equivalence of the canonical and the microcanonical ensembles for reduced density operators~\cite{Tasaki2016, Brandao2015EOE} 
plays an important role here.

We note that the weak ETH  is true even if bath B is an integrable system~\cite{Kim2014,Alba2015}. 
However, it has been shown that atypical states that do not satisfy Eq.~(\ref{Main_wETH}) have large weights after quantum quench in the case of integrable systems~\cite{Biroli2010}, 
and the steady values of macroscopic observables are consistent with the generalized Gibbs ensemble (GGE)~\cite{Rigol2006,Rigol2007,Vidmar2016} rather than the microcanonical ensemble.
In this sense, the weak ETH is physically significant only for nonintegrable systems, though it is mathematically true even for integrable systems.
The reason why the weak ETH is called ``weak'' is that there is another concept called the ``strong'' ETH (or just ETH) that is believed to be true only for nonintegrable systems,
where every energy eigenstate satisfies Eq.~(\ref{Main_wETH}) without exception~\cite{Kim2014}.

\textit{Outline of the proof of (\ref{Main_Clausius}).}
We are now in the position to discuss the outline of the proof of the second law~(\ref{Main_Clausius}).
In the short time regime $t\ll\tau$,
system S cannot feel the existence of $\rB_2$, and the heat bath effectively reduces to $\rB_1$.
From the weak ETH, if $\rB_2$ is sufficiently larger, 
the initial state $|E_i\rangle$ of bath B is typically indistinguishable from the canonical distribution,
if we only look at any operator in $\rB_1$.
Thus, the reduced density operators of system S at time $t\ll\tau$ are almost the same for the initial energy eigenstate and the initial canonical distribution.
Therefore, the conventional proof of the second law with the canonical bath approximately applies to the present situation, 
leading to inequality (\ref{Main_Clausius}).

\textit{Integral fluctuation theorem.}
We next discuss the IFT for the case that bath B is initially in an energy eigenstate, which is our second main result.
Let  $\sigma$ be  the stochastic entropy production defined as follows.
Suppose that one performs projection measurements of 
$\hat{\sigma} (t) :=-\ln\hat{\rho}_\rS(t)+\beta\hat{H}_\rB$ at the initial and the final times, where the first and the second terms on the right-hand side respectively represent the informational and the energetic terms, corresponding to the first and the second terms on the left-hand side of inequality (\ref{Main_Clausius}).
Then, $\sigma$ is given by the difference between the obtained outcomes of these measurements.
The average of the stochastic entropy production is equivalent to the aforementioned average entropy production: 
$\tbra{\sigma} =\Delta S_\rS-\beta Q$.

The conventional IFT states that, if the initial state of bath B is the canonical distribution, 
\begin{equation}
\langle e^{-\sigma} \rangle = 1.
\end{equation}
We note that the IFT holds even when system S is far from equilibrium.
A crucial feature of the IFT is that it reproduces the second law $\langle \sigma \rangle \geq 0$ from the Jensen inequality $\langle e^{-\sigma} \rangle \geq e^{- \langle \sigma \rangle}$.
Furthermore, the IFT can reproduce the fluctuation-dissipation theorem~\cite{Esposito2009}.

Our result is the IFT in the case that bath B is initially in a typical energy eigenstate (i.e., with initial condition (\ref{Main_initial})):
\begin{align}
| \langle e^{-\sigma} \rangle -1 |\leq \varepsilon_{\mathrm{FT}},
\label{Main_FT_pure}
\end{align}
where $\varepsilon_{\mathrm{FT}}$ is a positive error term.
We can rigorously prove that $\varepsilon_{\mathrm{FT}}$ can be arbitrarily small if bath B is sufficiently large.
We note that  the detailed fluctuation theorem~\cite{Esposito2009} also holds with an initial typical eigenstate, 
from which we can prove the IFT as a corollary  (see Supplemental Material for details).

The central idea of the proof of the IFT~(\ref{Main_FT_pure}) is almost the same as that of the second law, which is outlined above.
On the other hand, we need to make an additional assumption to prove inequality~(\ref{Main_FT_pure}) that
\begin{align}
\label{Main_eq:assump_FT}
[\hat{H}_\rS+\hat{H}_\rB,\hat{H}_\rI]\simeq 0,
\end{align}
which means that the sum of the energies of S and B is conserved at the level of fluctuations, 
and does not necessarily mean that $\hat H_\rI$ itself is small.
We note that assumption (\ref{Main_eq:assump_FT}) is consistent with the concept of  ``thermal operation" 
in the thermodynamic resource theory~\cite{Horodecki2013,Brandao2015},
where the left-hand side of Eq.~(\ref{Main_eq:assump_FT}) is assumed to be exactly zero.
The rigorous meaning of ``$\simeq$" is discussed in Supplemental Material.
If the left-hand side of Eq.~(\ref{Main_eq:assump_FT}) is nonzero but small,
a small positive error term  $\varepsilon_{\rm I}$ should be added to the right-hand side of inequality~(\ref{Main_FT_pure}),
which cannot be arbitrarily small even in the large-bath limit.
However, we will numerically show later that this additional term is negligible in practice.

\textit{Estimation of the error terms.}
We evaluate the error terms in inequalities (\ref{Main_Clausius}) and (\ref{Main_FT_pure}) with respect to the size $N$ of bath B.
We set $| \rB_1 | = \mathcal O (N^\alpha)$ with $0<\alpha < 1/2$.
The error from the weak ETH is bounded by 
$\mathcal{O}(N^{-(1-2\alpha)/4+\delta})+\mathcal{O}(N^{-(1-2\alpha)/8+\delta/2}/\sqrt{\tilde{\varepsilon}})$,
where $\delta>0$ is an unimportant constant that can be arbitrarily small,
and $\tilde{\varepsilon}$ is the fraction of atypical eigenstates in the weak ETH.
The error from the Lieb-Robinson bound is bounded by $e^{-\mu\dist(\tilde{\rS},\partial\rB_1)}(e^{vt}-vt-1)$,
which is negligible compared with the error term from the weak ETH for sufficiently large $N$,
but increases in time with $\mathcal{O}(t^2)$ up to $t\simeq 1/v$.

\textit{Numerical simulation.}
We performed numerical simulation of hard-core bosons with nearest-neighbor repulsion by exact diagonalization.
System S is on a single site and bath B is on a two-dimensional lattice (see the inset of Fig.~2).
The annihilation (creation) operator of a boson at site $i$ is written as $\hat{c}_i$ ($\hat{c}_i^\dag$),
which satisfies the commutation relations
$[\hat{c}_i,\hat{c}_j]=[\hat{c}^\dag_i,\hat{c}^\dag_j]=[\hat{c}_i,\hat{c}^\dag_j]=0$ for $i\neq j$, 
$\{\hat{c}_i,\hat{c}_i\}=\{\hat{c}^\dag_i,\hat{c}^\dag_i\}=0
$, and $
\{\hat{c}_i,\hat{c}^\dag_i\}=1$.
The occupation number is defined as $\hat{n}_i:=\hat{c}^\dag_i\hat{c}_i$.
Let $i=0$ be the index of the site of system S.
The Hamiltonians in Eq.~(\ref{Main_total_Hamiltonian}) are then given by 
\begin{align}
\label{Main_eq_NS_Ham1}
\hat{H}_\rS
&=
\omega\hat n_0
,
\quad
\hat{H}_\rI
=
-\gamma^\prime
\sum_{<0,j>}
(\hat{c}^\dag_0 \hat{c}_j+\hat{c}^\dag_j \hat{c}_0),\\
\label{Main_eq_NS_Ham2}
\hat{H}_\rB
&=
\omega
\sum_i \hat n_i
-
\gamma
\sum_{<i,j>}
(\hat{c}^\dag_i \hat{c}_j+\hat{c}^\dag_j \hat{c}_i)
+
g
\sum_{<i,j>}
\hat{n}_i \hat{n}_j,
\end{align}
where $\omega>0$ is the on-site potential,
$-\gamma$ is the hopping rate in bath B,
$-\gamma^\prime$ is the hopping rate between system S and bath B,
and $g>0$ is the repulsion energy.
The initial state of system S is given as $\ket{\psi}:=\hat{c}_0^\dag\ket{0}$.
We set the size of bath B by $16=4\times 4$, and the initial number of bosons in bath B by $4$.
To evaluate the Lieb-Robinson time, we set $\dist(\tilde{S},\partial \rB_1)=1$.
We can then evaluate that $v \simeq \gamma $ and $\mu=1$ if $g\ll \gamma$,
and therefore the Lieb-Robinson time is given by $\tau\simeq\gamma^{-1}$.
We set the inverse temperature of the initial eigenstate as $\beta=0.1$.

Figure 2 shows the time dependence of $\langle \sigma \rangle$, which implies that the second law indeed holds.
The average entropy production gradually increases up to $t \simeq \tau$, 
and then saturates with some oscillations around the time average.
We note that the oscillation for $\gamma^\prime=4\omega$ is the Rabi oscillation between system S and a part of B.
Remarkably, we observed that the second law holds even in a much longer time regime $t \gg \tau$,  
though our theorem ensures the second law only up to $t \simeq \tau$.
This implies that the second law is so robust against bare quantum fluctuations of pure quantum states.

We next confirmed the IFT~(\ref{Main_FT_pure}).
As shown in Fig.~3, $\langle e^{-\sigma}\rangle$ is very close to unity up to $t \simeq \tau$, as predicted by our theorem.
After $t \simeq \tau$, however, the deviation of  $\langle e^{-\sigma} \rangle$  from unity becomes significant, which is consistent with our theorem.
This deviation comes from the effect of bare quantum fluctuations of the initial state, 
because if the initial state was the canonical distribution, such deviation would never be observed.
As time increases, system S more and more feels the effect of bare quantum fluctuations, and the deviation becomes larger. 
This is regarded as a dynamical crossover from emergent thermal fluctuations to bare quantum fluctuations across the Lieb-Robinson time $\tau$; The IFT holds only for the former.
Such a crossover is not clearly observed in the second law (Fig.~2), because the second law only concerns the average of the stochastic entropy production, while its fluctuations play a significant role in the IFT.
Our numerical result also clarifies that our theory indeed accounts for the validity of IFT in the short time regime, because the numerically observed time scale of the breakdown of the IFT is consistent with our theory.

As shown in the inset of Fig.~3,  
the error term of the IFT is proportional to $t^2$ up to $t \simeq 1/v \simeq \tau$ in our numerical simulation. 
In fact, our error evaluation based on the Lieb-Robinson bound predicts that 
an error term increases in time with $t^2$-dependence as discussed before, 
if the additional error term $\varepsilon_{\rm I}$, which could also increase in time, is zero (or equivalently, the  left-hand side of (\ref{Main_eq:assump_FT}) is zero).
Therefore, our numerical result clarifies that the contribution from the left-hand side of (\ref{Main_eq:assump_FT}) is negligible in our setup, though it is not exactly zero in our Hamiltonians (\ref{Main_eq_NS_Ham1}) and (\ref{Main_eq_NS_Ham2}).

\textit{Concluding remarks.}
In this Letter, we have established 
the second law~(\ref{Main_Clausius}) and the IFT~(\ref{Main_FT_pure}) for unitary dynamics 
in the presence of a heat bath that is initially in a typical energy eigenstate.
Our result implies that thermal fluctuations can emerge from quantum fluctuations in a short time regime, 
and the former crosses over to the latter in time.
Our rigorous mathematical proof is based on the Lieb-Robinson bound (4) and the weak ETH (5).
We also performed numerical simulation of two-dimensional hard-core bosons, 
and confirmed our theoretical results.
 
We remark that inequality~(\ref{Main_Clausius}) only ensures the entropy increase between the initial and final times, and does not imply the monotonic entropy increase in continuous time.
The extension of our result to the monotonic second law in continuous time could be possible with controlled approximations such as the Born-Markov approximation~\cite{Breuer2002},
as is the case for the standard master equation approach~\cite{Horowitz2013,Hekking2013}.
This is a future direction of our work, though it would be technically nontrivial.

Our results can experimentally be tested with artificial isolated quantum systems such as ultracold atoms on an optical lattice, 
by employing the technique of single-site addressing~\cite{Cheneau2012}.  
Another candidate of experimentally relevant systems is superconducting qubits, 
where fully-controlled dynamics of thermalization can be observed~\cite{Pekola2015}.
To examine the relevance of our theory to non-artificial complex materials in noisy open environment is an future issue.

\vspace{2mm}
\noindent
\underbar{{\bf Acknowledgement}}
The authors are most grateful to Hal Tasaki for valuable discussions,
especially on the equivalence of ensembles.
The authors also thank Takashi Mori and Jae Dong Noh for valuable discussions.
E.I. and T.S. are supported by JSPS KAKENHI Grant Number JP16H02211.
E.I. is also supported by JSPS KAKENHI Grant Number 15K20944.
T.S. is also supported by JSPS KAKENHI Grant Number JP25103003.


\begin{figure}[t]
\begin{center}
\includegraphics[width=0.8\linewidth]{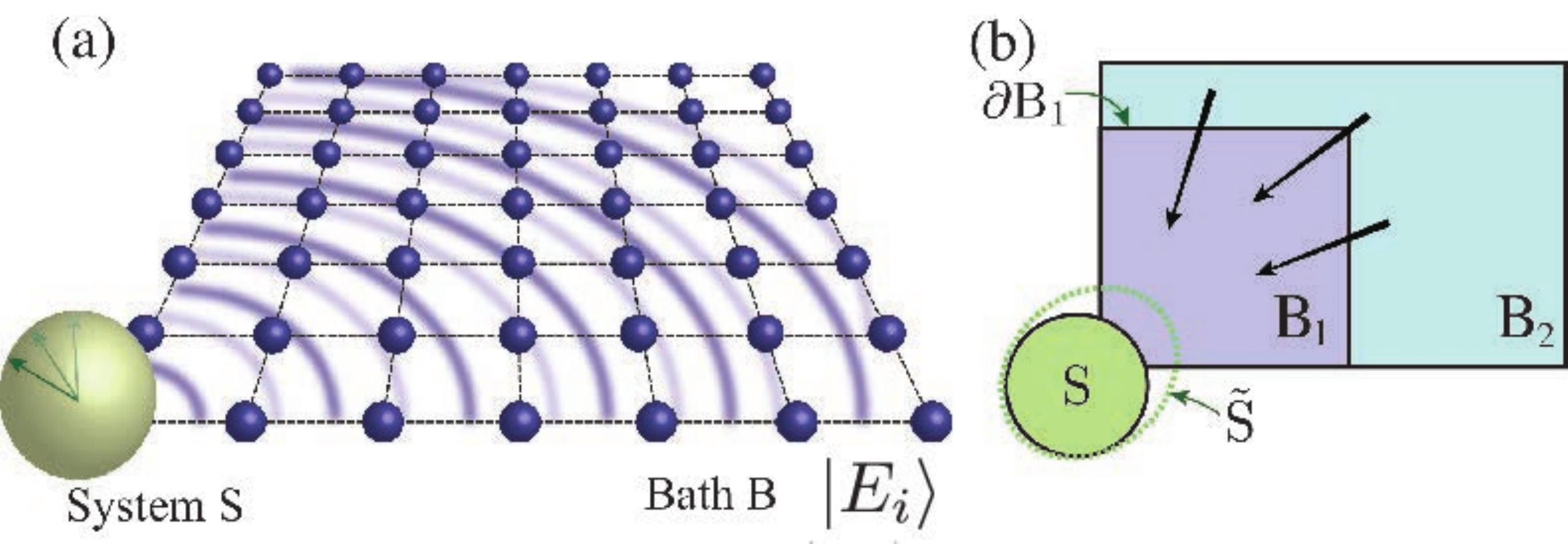}
\end{center}
\label{Main_fig1}
\caption{
Schematics of our setup.
(a) The total system, where the initial state of B is an energy eigenstate $| E_i \rangle$.
(b) The key idea of our proof. 
Bath B is divided into $\rB_1$ and  $\rB_2$, 
where only  $\rB_1$ is attached to system S.
The boundary between $\rB_1$ and  $\rB_2$ is denoted by $\partial \rB_1$.
A slightly extended region of S, including the support of $\hat{H}_\rI$, is denoted by $\tilde \rS$.
}
\end{figure}

\begin{figure}[t]
\begin{center}
\includegraphics[width=0.7\linewidth]{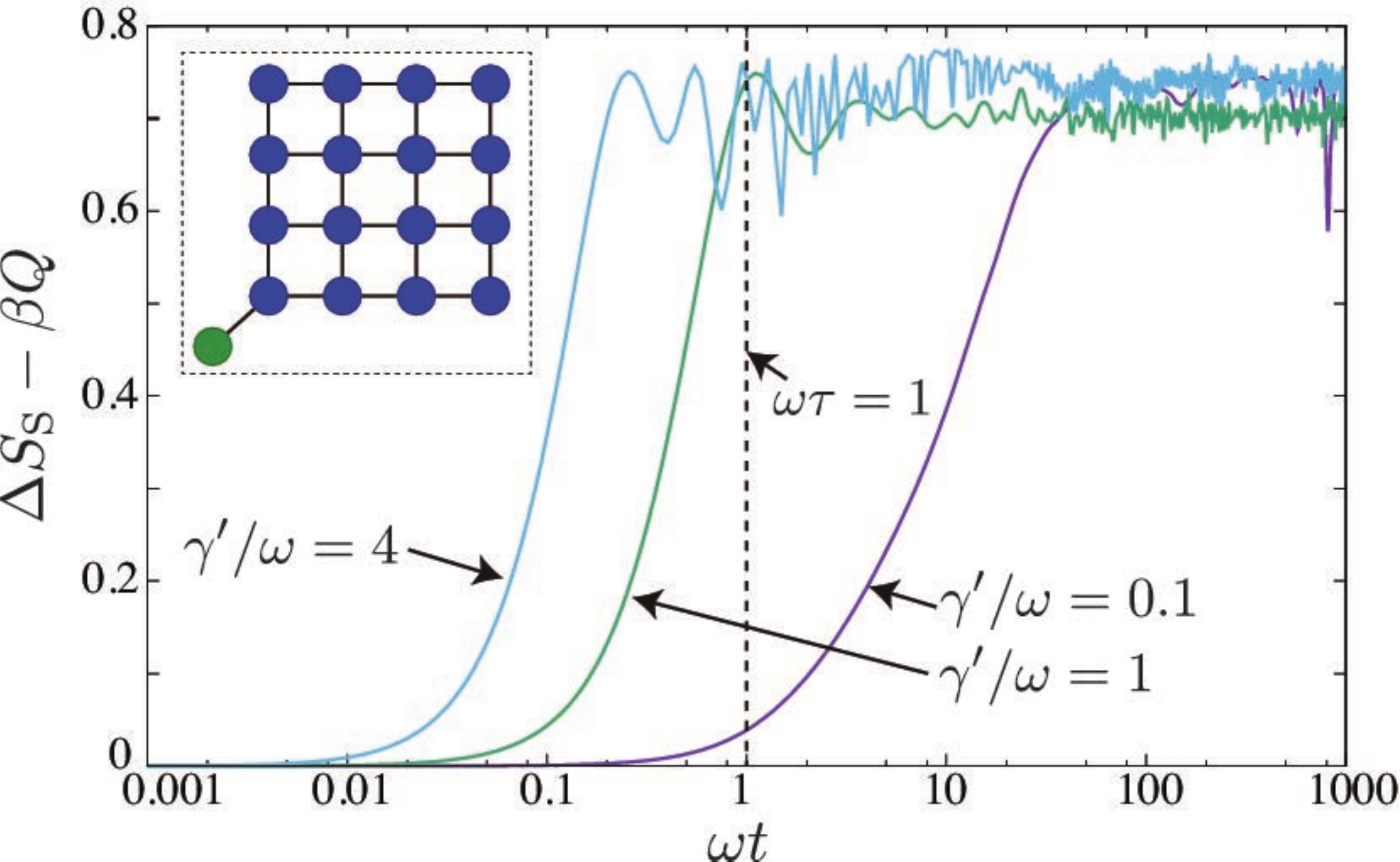}
\end{center}
\label{Main_fig2}
\caption{
Time dependence of average entropy production $\langle\sigma\rangle$ 
with parameters $\gamma^\prime/\omega=0.1$, $1$, $4$, $\gamma/\omega=1$, and $g/\omega=0.1$.
(Inset) The lattice structure for the numerical calculation.
System S is on a single site, and bath B is on a two-dimensional square lattice with $16=4\times 4$ sites with $4$ bosons.
}
\end{figure}

\begin{figure}[t]
\begin{center}
\includegraphics[width=0.7\linewidth]{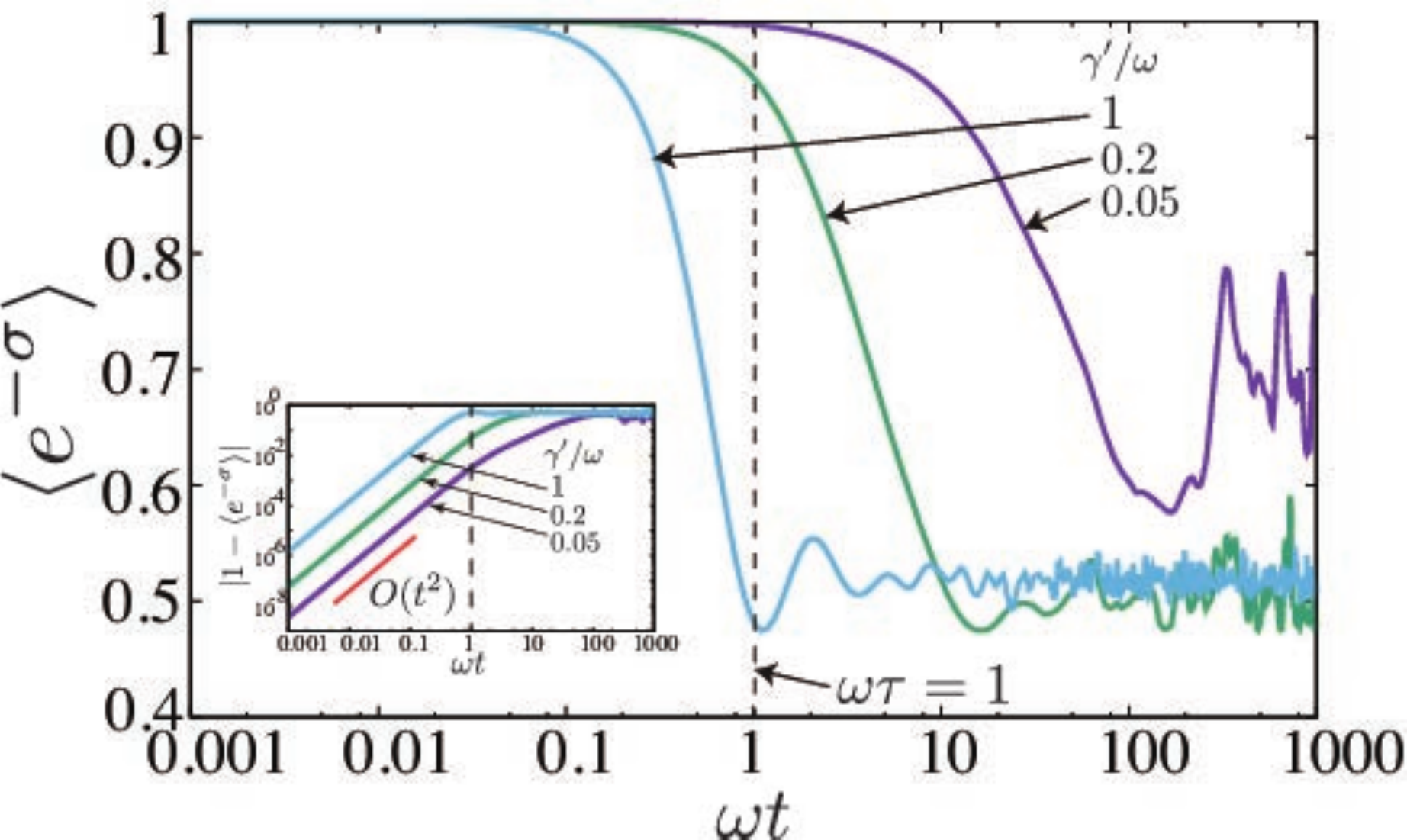}
\end{center}
\label{Main_fig3}
\caption{
Time dependence of $\langle e^{-\sigma} \rangle$ that numerically confirms the IFT.
The structure of S and B is the same as that in the inset of Fig.~2. 
The parameters are given by $\gamma^\prime/\omega=0.05,0.2,1$, $\gamma/\omega=1$, and $g/\omega=0.1$.
(Inset) Time dependence of the deviation of  $\langle e^{-\sigma} \rangle$ from unity. 
}
\end{figure}

\clearpage

{\Large Supplemental Material : Fluctuation Theorem for Many-Body Pure Quantum States}

\

{\large Eiki Iyoda$^{1}$, Kazuya Kaneko$^{2}$, Takahiro Sagawa$^{1}$}

\

[1] Department of Applied Physics, The University of Tokyo, 7-3-1 Hongo, Bunkyo-ku, Tokyo 113-8656, Japan

[2] Department of Basic Science, The University of Tokyo, 3-8-1 Komaba, Meguro-ku, Tokyo 153-8902, Japan


\vspace{10mm}

\def\theequation{S\arabic{equation}}
\def\thefigure{S\arabic{figure}}
\setcounter{equation}{0}
\setcounter{page}{1}
\setcounter{figure}{0}

\tableofcontents


\

In this Supplemental Material, 
we prove the second law of thermodynamics and the fluctuation theorem for a system 
in contact with a heat bath in a pure state [inequalities~(3) and (7) in the main text]  in a mathematically rigorous way.
For this purpose, we need several technical assumptions that will be discussed in detail.  
However, the bare essentials of the proof are the same as illustrated in the main text.
In addition, we show supplementary results of numerical simulation.

The organization of Supplemental Material is as follows.  In Sec.~\ref{sec:review},
we briefly review the established proof of the conventional second law and the fluctuation theorem 
with a heat bath that is initially in the canonical distribution.
In Sec.~\ref{sec:inequality}, we remark on the operator norm and the trace norm.
In Sec.~\ref{sec:LRB}, we formulate the Lieb-Robinson bound that is a key of our proof.
In Sec.~\ref{sec:MicroCanonical}, we define the canonical ensemble and the microcanonical ensemble.  
In Sec.~\ref{sec:weakETH}, we show our theorems on the weak eigenstate-thermalization hypothesis (ETH).
In Sec.~\ref{sec:setup}, we describe our basic setup of the proof of the second law and the fluctuation theorem.
In Sec.~\ref{sec:locality}, we prove two important lemmas. In particular, 
Lemma \ref{lemma:Reference} plays a key role in the proof of our main results.
In Secs.~\ref{sec:2nd} and \ref{sec:FT}, we prove our main results: 
the second  law (Theorem 1) and the fluctuation theorem (Theorem 2), respectively.
In Sec.~\ref{sec:typicality}, we discuss the typicality in the Hilbert space.
In Sec.~\ref{sec:Discussion}, we remark on some assumptions in our setup.
In Sec.~\ref{sec:Numerical}, we show the details of our numerical calculation and supplementary numerical results.

Throughout Supplemental Material,   we set $\hbar=1$ and $k_{\rm B} = 1$ for simplicity.
We will only consider finite-dimensional Hilbert spaces.

\section{The conventional second law and the fluctuation theorem}
\label{sec:review}
In this section, 
we review the conventional proof of the second law and the fluctuation theorem 
for a system in contact with a heat bath that is initially in the canonical distribution.

\subsection{Setup}
\begin{figure}[t]
\begin{center}
\includegraphics[width=0.3\linewidth]{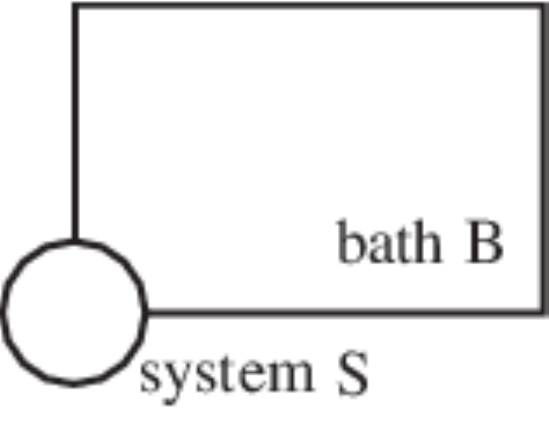}
\end{center}
\caption{
Schematic of the setup for the conventional proof of the second law and the fluctuation theorem.
The composite system consists of system S and bath B, where the initial state of bath B is  the canonical distribution.
}
\label{fig:fig_s_1}
\end{figure}

Suppose that the total system consists of  system S and  bath B as shown in Fig.~\ref{fig:fig_s_1}.
The Hamiltonian of the composite system is given by
\begin{equation}
\hat{H}
:=
\hat{H}_\mathrm{S}
+
\hat{H}_{\mathrm{I}}
+
\hat{H}_{\mathrm{B}},
\end{equation}
where 
$\hat{H}_\mathrm{S}$ and $\hat{H}_\mathrm{B}$ are the Hamiltonian of system S and bath B, respectively.
The interaction Hamiltonian between system S and bath B is denoted by $\hat{H}_\mathrm{I}$.
We note that there is no additional assumption on $\hat{H}$ 
(such as locality) in this section.
The coupling between system S and bath B can either be weak or strong.

The initial state is represented by a product state without any correlation between S and B: 
\begin{equation}
\hat{\rho}(0)=\hat{\rho}_\rS(0)\otimes\hat{\rho}_\rB^\can.
\label{canonical_initial}
\end{equation}
The initial state of system S is arbitrary,
and the initial state of bath B is described by the canonical distribution
$\hat{\rho}_\rB^{\mathrm{can}}:=\exp(-\beta\hat{H}_\rB)/Z_\rB$,
where $Z_\rB:=\tr_\rB[\exp(-\beta\hat{H}_\rB)]$ is the partition function and $\beta$ is the inverse temperature.

The composite system obeys the unitary evolution described by $\hat{U}:=\exp(-i\hat{H}t)$, and the final state of the system is given by 
$\hat{\rho}(t)=\hat{U}\hat{\rho}(0)\hat{U}^\dag$.
The reduced density operators for the final state of system S and bath B are defined as
$\hat{\rho}_\rS(t):=\tr_{\rB}[\hat{\rho}(t)]$
and 
$\hat{\rho}_\rB(t):=\tr_{\rS}[\hat{\rho}(t)]$,
respectively.

\subsection{The second law of thermodynamics}
We first note that  the von Neumann entropy of a density operator $\hat \rho$ is given by
\begin{equation}
S(\hat{\rho}):=-\mathrm{tr}\bbra{\hat{\rho} \ln\hat{\rho}}.
\end{equation}
The change in the von Neumann entropy of system S is represented as
$\Delta S_\rS:= S_\rS(t)-S_\rS(0):=S(\hat{\rho}_\rS(t))-S(\hat{\rho}_\rS(0))$.
We also define the heat  as
$Q:=
-\mathrm{tr}_\rB[
\hat{H}_\mathrm{B}(
\hat{\rho}_\mathrm{B}(t)
-
\hat{\rho}_\mathrm{B}(0)
)
]$.
The second law of thermodynamics in this setup is then stated as follows:
\begin{itembox}[l]
{\Proposition[\label{prop:2nd}] (Second law of thermodynamics with the canonical bath~\cite{S_Sagawa2012})}
\begin{equation}
\label{eq:conv_2nd}
\Delta S_\rS- \beta Q\geq 0.
\end{equation}
\end{itembox}

\noindent
{\bf Proof.}
The von Neumann entropy of the composite system in the initial and the final states are given by
\begin{equation}
\begin{split}
S(\hat{\rho}(0))
&=
-\mathrm{tr}_\rS\bbra{\hat{\rho}_\rS(0)\ln \hat{\rho}_\rS(0)}
-\mathrm{tr}_\rB\bbra{\hat{\rho}_\rB^{\mathrm{can}}\ln \hat{\rho}_{\mathrm{B}}^{\mathrm{can}}}
=
S_{\mathrm{S}}(0)
+\beta\mathrm{tr}_\rB[\hat{\rho}_\rB^{\mathrm{can}}
\hat{H}_\rB]
+\ln Z_\rB,
\\
S(\hat{\rho}(t))
&=
-
\tr_{\rS\cup\rB}
[
\hat{\rho}(t)\ln\hat{\rho}(t)
]
\leq
-
\tr_{\rS\cup\rB}
\bbra{
\hat{\rho}(t)\ln(\hat{\rho}_{\mathrm{S}}(t)\otimes \hat{\rho}_\mathrm{B}^{\mathrm{can}})
}
=
S_{\mathrm{S}}(t)+\beta\mathrm{tr}_\rB[\hat{\rho}_\rB(t)\hat{H}_\rB]
+\ln Z_\rB,
\end{split}
\end{equation}
where we used the Klein inequality:
$\mathrm{tr}\bbra{\hat{\rho}\ln\hat{\rho}}
\geq
\mathrm{tr}\bbra{\hat{\rho}\ln\hat{\sigma}}$.
Since the von Neumann entropy is invariant under unitary dynamics $S(\hat{\rho}(0)) = S(\hat{\rho}(t))$, we obtain inequality~(\ref{eq:conv_2nd}).
\hspace{\fill}$\Box$

\subsection{The fluctuation theorem}
\label{subsec:FT}

We consider the concept of stochastic entropy production~\cite{S_Esposito2009}.
We first define
\begin{equation}
\hat{\sigma}(t)\defL \beta \hat{H}_\mathrm{B}-\ln\hat{\rho}_S(t)
\end{equation}
in the Schr\"{o}dinger picture,
and consider the projection measurements of $\hat \sigma (t)$ at time $0$ and $t$.
Let $\sigma_{\rm i}$ and $\sigma_{\rm f}$ be the measurement outcomes at time $0$ and $t$, respectively.
The joint probability distribution of  $\sigma_{\rm i}$ and $\sigma_{\rm f}$ is given by
\begin{equation}
P_\rF[\sigma_{\rm f}; \sigma_{\rm i}]
:=
\tr
[
\hat{P}_{\sigma_{\rm f}}
\hat{U}
\hat{P}_{\sigma_{\rm i}}
\hat{\rho}(0)
\hat{P}_{\sigma_{\rm i}}
\hat{U}^\dag
\hat{P}_{\sigma_{\rm f}}
],
\end{equation}
where $\hat{P}_{\sigma_{\rm i}}$ and $\hat{P}_{\sigma_{\rm f}}$ are the projection operators corresponding to eigenvalues $\sigma_{\rm i}$ and $\sigma_{\rm f}$ of $\hat \sigma (0)$ and $\hat \sigma (t)$, respectively.
We note that $\hat \rho (0)$ and $\hat \sigma (0)$ are, and thus $\hat \rho (0)$ and $\hat{P}_{\sigma_{\rm i}}$ are, commutable.
The index $\rF$ in $P_\rF[\sigma_{\rm f}; \sigma_{\rm i}]$  means the forward time evolution.
The stochastic entropy production is then defined as the difference between the two measurement outcomes:
\begin{equation}
\sigma := \sigma_{\rm f} - \sigma_{\rm i}.
\end{equation}
The probability distribution of the stochastic entropy production is written as $P_\rF(\sigma)$, which is given by
\begin{equation}
P_\rF(\sigma)
:=
\sum_{\sigma_{\rm f}, \sigma_{\rm i}} \delta(\sigma-(\sigma_{\rm f}-\sigma_{\rm i}))
P_\rF[\sigma_{\rm f}; \sigma_{\rm i}],
\end{equation}
where $\delta (\cdot )$ is the delta function.
We note that in the proof of our main results in the subsequent sections, we will not use the concept of the delta function in order to avoid introducing advanced mathematical tools for the rigorous argument.

The characteristic function of the stochastic entropy production is given by the Fourier transformation of $P_\rF(\sigma)$~\cite{S_Esposito2009}:
\begin{equation}
\label{eq:def_Gu}
G_\mathrm{F}(u)
:=
\int_{- \infty}^{+\infty} d\sigma e^{iu\sigma} 
P_\rF(\sigma),
\end{equation}
where  $u$ is referred to as the counting field.
The $n$th differential of the characteristic function gives 
the $n$th order moment of the stochastic entropy production:
\begin{equation}
\tbra{\hat{\sigma}^n}
=
\left.
\frac{\partial^n G_\mathrm{F}(u)}{\partial(iu)^n}
\right|_{u=0}.
\end{equation}

We note that the ensemble average of the entropy production is given by
\begin{equation}
\langle \sigma \rangle =\Delta S_\rS-\beta Q,
\end{equation}
which is the left-hand side of the second law~(\ref{eq:conv_2nd}).
Therefore, we can rewrite the second law~(\ref{eq:conv_2nd}) in terms of the entropy production by
\begin{equation}
\langle \sigma \rangle \geq 0.
\label{conv_2nd_sigma}
\end{equation}

By noting that $[\hat{\sigma}(0),\hat{\rho}(0)]=0$, the characteristic function is rewritten as
\begin{equation}
\label{eq:CF_Gu}
G_\mathrm{F}(u)
=
\tr
[
\hat{U}
e^{-iu\hat{\sigma}(0)}
\hat{\sigma}(0)
\hat{U}^\dag
e^{iu\hat{\sigma}(t)}
],
\end{equation}
which can be regarded as the definition of $G_\mathrm{F}(u)$ in the case that  we do not use the delta function.

We next consider the reversed time evolution that obeys unitary evolution $\hat U^\dagger$.
The initial state of the reversed process is given by the product state with the final state of system S 
and the canonical distribution of bath $\rB$:
$
\hat{\rho}_\rR (0)
:=
\hat{\rho}_\rS(t)
\otimes
\hat{\rho}_\rB^{\mathrm{can}}$, where index $\rR$ represents the reversed processes.
The density operator at time $t$ of the reversed process is given by $\hat \rho_\rR (t) = \hat U^\dagger \hat{\rho}_\rR (0) \hat U$.

We then define the stochastic entropy production in the reversed processes by the difference between the measurement outcomes of $\hat \sigma (t)$ and $\hat \sigma (0)$ at the initial and the final times of the reversed processes, respectively.
Let $P_\rR(\sigma)$ be the probability distribution of the stochastic entropy production in the reversed processes, and $G_\rR(u)$ be the corresponding characteristic function. By noting that $[\hat{\sigma}(t),\hat{\rho}_\rR(0)]=0$, we have 
\begin{equation}
G_\mathrm{R}(u)
=
\tr
[
\hat{U}^\dag
e^{-iu\hat{\sigma}(t)}
\hat{\rho}_\rR (0)
\hat{U}
e^{iu\hat{\sigma}(0)}
].
\end{equation}

We note that ${\rm tr}_\rB [\hat \rho_\rR(t)] \neq \hat \rho_\rS (0)$ in general.  Therefore, the entropy production defined above is not necessarily related to the change in the von Neumann entropy of system S in the reversed process.  In this sense, the physical interpretation of the entropy production is not very clear in the reversed process.   If ${\rm tr}_\rB [\hat \rho_\rR(t)] = \hat \rho_\rS (0)$ happens to be true, the entropy production in the reversed process has a clear interpretation related to the von Neumann entropy of the reversed process.  Such a situation is physically realizable if the initial and the final states of system S are in thermal equilibrium in both of the forward and the reversed processes.

Apart from its physical interpretation, the entropy production in the reversed process is always mathematically well-defined, and the following argument including the reversed process is true for any case.  In addition, the entropy production in the reversed process is a useful concept to prove some theorems that only include quantities in the forward process, such as the integral fluctuation theorem.  Therefore, we will not go into details of the physical interpretation of the reversed processes in the following argument.

We now discuss the fluctuation theorem.
The detailed fluctuation theorem characterizes a universal relationship between the probabilities of the entropy changes in the forward and the reversed processes.
In the following argument, it is convenient to  formulate the detailed fluctuation theorem in terms of the characteristic functions:
\begin{itembox}[l]
{\Proposition[\label{prop:FT}] (Detailed fluctuation theorem with the canonical bath~\cite{S_Esposito2009})}
\begin{equation}
\label{eq:FT_Gu}
G_\mathrm{F}(u)=G_\mathrm{R}(-u+i).
\end{equation}
\end{itembox}

\noindent
{\bf Proof.}
Letting $v:=-u+i$, we have
\begin{equation}
\begin{split}
G_\mathrm{F}(u)
&=
\tr
[
\hat{U}
e^{-i(-v+i)\hat{\sigma}(0)}
\hat{\rho}(0)
\hat{U}^\dag
e^{i(-v+i)\hat{\sigma}(t)}
]
\\
&=
\tr
[
\hat{U}
e^{iv\hat{\sigma}(0)}
e^{\hat{\sigma}(0)}
\hat{\rho}(0)
\hat{U}^\dag
e^{-iv\hat{\sigma}(t)}
e^{-\hat{\sigma}(t)}
]
\\
&=
\tr
[
\hat{U}
e^{iv\hat{\sigma}(0)}
\hat{U}^\dag
e^{-iv\hat{\sigma}(t)}
\hat{\rho}_{\mathrm{S}}(t)
\otimes
\hat{\rho}_\mathrm{B}^\mathrm{can}
]
\\
&=
\tr
[
\hat{U}^\dag
e^{-iv\hat{\sigma}(t)}
\hat{\rho}_{\mathrm{S}}(t)
\otimes
\hat{\rho}_\mathrm{B}^\mathrm{can}
\hat{U}
e^{iv\hat{\sigma}(0)}
]
\\
&=
G_\mathrm{R}(-u+i),
\end{split}
\end{equation}
which implies Eq.~(\ref{eq:FT_Gu}).
\hspace{\fill}$\Box$

\

We next discuss the integral fluctuation theorem (IFT)~\cite{S_Kurchan2000,S_Tasaki2000}.  We consider the following quantity:
\begin{equation}
\tbra{e^{-\sigma}}
:=
\int_{- \infty}^{+\infty} d\sigma
P_\rF(\sigma)e^{-\sigma},
\end{equation}
or equivalently
\begin{equation}
\tbra{e^{-\sigma}}
:=
\sum_{\sigma_{\rm f}, \sigma_{\rm i}} 
P_\rF[\sigma_{\rm f}; \sigma_{\rm i}] e^{-(\sigma_{\rm f} -\sigma_{\rm i} )}.
\end{equation}
We then have the following corollary.
\begin{itembox}[l]
{\Corollary[\label{cor:IFT}] (Integral fluctuation theorem with the canonical bath)}
\begin{equation}
\label{integralFT}
\tbra{e^{-\sigma}} = 1.
\end{equation}
\end{itembox}

\noindent
{\bf Proof.}
We note that $\tbra{e^{-\sigma}} = G_{\rm F}(i)$ and $G_{\rm R} (0) =1$.  By substituting $u=i$ to the detailed fluctuation theorem~(\ref{eq:FT_Gu}), we obtain the IFT~(\ref{integralFT}).
\hspace{\fill}$\Box$

\

We make some remarks on the fluctuation theorem.
Applying the inverse Fourier transform to Eq.~(\ref{eq:FT_Gu}),
we obtain the detailed fluctuation theorem in terms of the probability distributions:
\begin{equation}
\label{eq:FT_conv_P}
\frac{P_\rF(\sigma)}{P_\rR(-\sigma)}=e^{\sigma}.
\end{equation}
By integrating the both-hand sides of $P_\rF(\sigma)e^{-\sigma} = P_\rR(-\sigma)$, which is equivalent to Eq.~(\ref{eq:FT_conv_P}),
we again obtain the IFT:
\begin{equation}
\tbra{e^{-\sigma}}
=
\int_{- \infty}^{+\infty} d\sigma
P_\rF(\sigma)e^{-\sigma}
=
\int_{- \infty}^{+\infty} d\sigma
P_\rR(-\sigma)
=
1.
\end{equation}
By using the Jensen inequality to Eq.~(\ref{integralFT}),
we obtain $\tbra{e^{-\sigma}}\geq e^{-\tbra{\sigma}}$, which reproduces the second law~(\ref{conv_2nd_sigma}).

\section{Norms of operators}
\label{sec:inequality}

As a preliminary, we  briefly review the operator norm and the trace norm for  finite-dimensional Hilbert spaces.
Let $| \psi \rangle$ be a vector of a Hilbert space, and $\|\ket{\psi}\|$ be its norm. 

We consider operators on the Hilbert space, which is not necessarily Hermitian.
First, the operator norm of an operator $\hat{X}$ is defined as
\begin{equation}
\|\hat{X}\|
:= 
\sup_{\|\ket{\Psi}\|=1}
\sqrt{\langle \Psi |
\hat X^\dagger 
\hat{X}
| \Psi \rangle
},
\end{equation}
which is equal to the largest singular value of $\hat X$.  
Second, the trace norm of an operator $\hat{X}$ is defined as
\begin{equation}
\|\hat{X}\|_1:=  {\rm tr}\left[ \sqrt{ \hat{X}^\dagger \hat X } \right],
\end{equation}
which is the sum of the singular values of $\hat X$.
If  $\hat X$ is Hermitian, the above definition reduces to
\begin{equation}
\|\hat{X}\|_1\defL {\rm tr}[ | \hat{X} | ].
\end{equation}
We  note some useful (and well-known) properties of the norms:
\begin{itembox}[l]
{\Proposition{\label{prop:IneqNorm}}}
For any operators,
\begin{equation}
| {\rm tr} [ \hat X ] | \leq \| \hat X \|_1,
\end{equation}
\begin{equation}
\| \hat X \hat Y \|_1 \leq \| \hat X \| \| \hat Y \|_1.
\end{equation}
\end{itembox}

\section{The Lieb-Robinson bound}
\label{sec:LRB}

In this section, 
we review the Lieb-Robinson bound \cite{S_Lieb1972,S_Hastings2006} that plays a key role in our study.
The Lieb-Robinson bound  gives an upper bound of the velocity of information propagation in a quantum many-body system on a general graph, which is formulated in terms of the operator norm of the commutator of two operators in spatially distinct regions on the graph.

We consider a general setup for the Lieb-Robinson bound as follows.
The system is defined on a finite graph, written as $\Lambda:=(\Lambda_\mathrm{v},\Lambda_\mathrm{e})$, where the sets of all vertices (sites) and all edges (bonds) are denoted by $\Lambda_\mathrm{v}$ and $\Lambda_\mathrm{e}$, respectively.
A region X in the graph is defined as a set of sites: $\rX \subset \Lambda_\mathrm{v}$.
We write the number of elements of a set as $|\cdot|$. For example, $| \Lambda_\mathrm{v} |$ describes the number of all the sites in the graph.

We next define the spatial distance on the graph.
The distance between two sites ${\rm x}, {\rm y} \in \Lambda_{\rm v}$ is defined as the number of bonds 
 in the shortest path that connects  x and y, which we denote as $\dist(\mathrm{x},\mathrm{y})$.
The distance between two regions $\rX$ and $\rY$ is correspondingly defined as 
\begin{equation}
\dist(\rX,\rY):=\min_{\rx\in\rX,\ry\in\rY}\dist(\rx,\ry).
\end{equation}

The Hamiltonian in our setup is expressed as the sum of local Hamiltonians:
\begin{equation}
\hat{H}=\sum_{\rZ\subset\Lambda_\mathrm{v}} \hat{h}_\rZ,
\label{local_Hamiltonian}
\end{equation}
where $\hat{h}_\rZ$ is a local Hamiltonian on a bounded support $\rZ$.
The sum on the right-hand side of Eq.~(\ref{local_Hamiltonian}) is over a particular set of bounded supports.
We then make the following assumption.
\begin{itembox}[l]
{\Assumption{\label{assump:LRB}} (Conditions on the local Hamiltonians and the graph~\cite{S_Lieb1972,S_Hastings2006})}
There exist  $\lambda_0 >0$, $\mu > 0$, and $p_0 > 0$, such that for any $\rx,\ry\in\Lambda_\mathrm{v}$,
\begin{equation}
\sum_{\mathrm{Z}\ni \rx,\ry}
\|\hat{h}_\mathrm{Z}\|
\leq
\lambda_0\exp[-\mu\mathrm{dist}(\rx,\ry)],
\end{equation}
\begin{equation}
\label{eq:LRB_assump2}
\sum_{z\in\Lambda_\mathrm{v}}
\exp[-\mu(\mathrm{dist}(\rx,\rz)+\mathrm{dist}(\rz,\ry))]
\leq
p_0
\exp[-\mu\mathrm{dist}(\rx,\ry)].
\end{equation}
\end{itembox}
The constants $\lambda_0$ and $\mu$ are determined by the Hamiltonian and the graph,
which represent the interaction strength and the interaction range, respectively.  
For example, if we consider the case of the nearest-neighbor interaction on a hypercubic lattice,
$\lambda_0$ and $\mu$ are determined to satisfy $\max_{\rZ}\|\hat{h}_\rZ\|\leq \lambda_0 e^{-\mu}$.
If the interaction is bounded as $\|\hat{h}_\rZ\|\leq J$ with $J>0$ being a constant, 
$\lambda_0$ is given by $\lambda_0=J e^{\mu}$ with arbitrary $\mu$.
The constant $p_0$ is determined by the graph structure and $\mu$.
For the case of nearest-neighbor interaction on the two-dimensional square lattice, $p_0$ is given by $2$.

The Lieb-Robinson bound states that there exists an upper bound of the velocity of information propagation for a general Hamiltonian which satisfies Assumption \ref{assump:LRB}.
In Sec.~\ref{sec:setup}, we will formulate a more specific setup for our study,
where  Assumption \ref{assump:LRB} is also assumed and 
the Lieb-Robinson bound is applicable.

We now state the Lieb-Robinson bound as follows.
\begin{itembox}[l]
{\Proposition{\label{prop:LRB}} (Lieb-Robinson bound~\cite{S_Lieb1972,S_Hastings2006})}
Let $\hat{A}$ and $\hat{A}^\prime$ be operators with supports A and $\rA^\prime$, respectively.  
We consider the Heisenberg picture of $\hat{A}$, defined as $\hat{A}(t):=\hat{U}^\dag\hat{A}\hat{U}$ with  $\hat{U}:=\exp(-i\hat{H}t)$.
Under Assumption \ref{assump:LRB}, defining  $C:=2/p_0$ and $v:=\lambda_0p_0$, the Lieb-Robinson bound states that
\begin{equation}
\label{eq:LRB}
\|
[
\hat{A}(t),\hat{A}^\prime
]
\|
\leq
C
\|\hat{A}\|
\|\hat{A}^\prime\|
|\rA||\rA^\prime|
\exp[-\mu\mathrm{dist}(\rA,\rA^\prime)]
(e^{v|t|}-1).
\end{equation}
\end{itembox}
The constants $C$ and $v$ depend on the interaction and the graph structure through $\lambda_0$ and $p_0$.
Here,  
we refer to $v/\mu$ as the Lieb-Robinson velocity.
We then define a characteristic time as
\begin{equation}
\tau:= \frac{\mu\dist(\rA,\rA^\prime)}{v},
\end{equation}
to which we refer as the Lieb-Robinson time.
If $t\ll\tau$, the right-hand side of inequality~(\ref{eq:LRB}) becomes exponentially small with respect to the distance between $\rA$ and $\rA^\prime$.

In the case of nearest-neighbor interaction on the two-dimensional square lattice, 
we have $C=1/2$, $v=2Je^\mu$, and $\tau=\mu\dist(\rA,\rA^\prime) / (2e^\mu J)$.
Since $\mu$ is arbitrary,
we can choose it to maximize the Lieb-Robinson time, and $\tau$ is maximized when $\mu=1$, where  $v=2eJ$ and $\tau=\dist(\rX,\rY) / (2eJ)$.

\section{The canonical and the microcanonical ensembles}
\label{sec:MicroCanonical}

In this section, we rigorously define the canonical and the microcanonical ensembles, 
and discuss their equivalence in line with Ref.~\cite{S_Tasaki2016}.
We consider a quantum many-body system on a lattice, to which we refer as ``bath B",
as it will play a role of a heat bath in our main theorems in Secs.~\ref{sec:setup}-\ref{sec:typicality}.
We assume that bath B is on a $d$-dimensional hypercubic lattice with the periodic boundary condition.
We denote the set of sites of bath B by the same notation, B.
We denote the number of sites in bath B by $N:=|\rB|$, and the side length of $\rB$ by $L$ such that  $L^d=N$.
Let  $\mathcal{H}_\mathrm{i}$ be the local Hilbert space on site $\ri \in \rB$.
The total Hilbert space of B is denoted by $\mathcal H_\rB:=\otimes_{\ri\in \rB}\mathcal{H}_\mathrm{i}$.
The dimension of $\mathcal{H}_\rB$ is denoted by $D_\rB$.

We assume that the Hamiltonian of bath B is represented as the sum of local Hamiltonians:
\begin{equation}
\hat{H}_\mathrm{B}
=
\sum_{\mathrm{Z}\subset \mathrm{B}} \hat{h}_{\mathrm{Z}},
\label{l_H}
\end{equation}
where the sum is taken over a particular set of bounded region Z.
We write the spectrum decomposition of $\hat H_\rB$ as
\begin{equation}
\hat H_\rB := \sum_{i=1}^{D_\rB} E_i | E_i \rangle \langle E_i |,
\end{equation}
where 
$E_i$ is an energy eigenvalue, and $| E_i \rangle$ is the corresponding eigenstate.
We assume the locality of the interaction on the lattice, which implies that the interaction range of the Hamiltonian is independent of the size of bath B:
\begin{itembox}[l]
{\Assumption{\label{assump:LocalInt}} (Locality of the interaction)}
There exists an integer $k>0$ that is independent of $N$,  such that  for any $\hat h_{\rm Z}$ with support Z, $\dist(\ri,\rj)\leq k$ holds for  $\ri, \rj \in \rZ$.
We refer to $k$ as the interaction range inside bath B.
In addition, we  assume that $\| \hat h_{\rZ} \|$ is independent of $N$ for any $\hat{h}_\rZ$. 
\end{itembox}
We also assume that bath $\rB$ is translation invariant:
\begin{itembox}[l]
{\Assumption{\label{assump:TransInv}} (Translation invariance)}
The local Hilbert spaces $\mathcal{H}_\mathrm{i}$ with $\ri \in \rB$ are identical, 
and the local Hamiltonians $\hat h_{\rm Z}$ are the same for any Z in Eq.~(\ref{l_H}),
where $\rZ$'s are defined over the lattice in a translation invariant way.
Let $D_{\mathrm{loc}}$ be the dimension of $\mathcal{H}_{\ri}$.
\end{itembox}

\

We now define the canonical ensemble.
\begin{itembox}[l]
{\Definition[\label{def:Canonical}] (Canonical ensemble)}
For a given inverse temperature $\beta$,
the density operator of the canonical ensemble is defined as
\begin{equation}
\hat{\rho}^{\mathrm{can}}_{\rB}
:=
\frac{1}{Z(\beta)}e^{-\beta \hat{H}_\rB},
\end{equation}
where 
$Z(\beta):=\tr[e^{-\beta\hat{H}_\rB}]$ is the partition function.
\end{itembox}
We also define the average energy density of the canonical ensemble at inverse temperature $\beta$ by
\begin{equation}
u^\mathrm{can}(\beta ) :=\frac{1}{N} \tr[\hat{H}_\rB\hat{\rho}_\rB^{\can}].
\label{u_beta}
\end{equation}

\

To define the microcanonical ensemble,
we employ the framework of Ref.~\cite{S_Tasaki2016},
where the equivalence of the canonical and the microcanonical ensembles has
rigorously been proved.
Let $\Delta>0$ be the width of the microcanonical energy shell,
which can depend on $N$
but is bounded from below by $N$-independent constant $\delta^\prime>0$.
This implies that 
$\Delta$ can be in the order of $\mathcal{O}(1) \leq \Delta \leq \mathcal{O}(N)$.
\begin{itembox}[l]
{\Definition[\label{def:EnergyShell}] (Microcanonical energy shell)}
The energy shell with energy $U$ and width $\Delta$ is defined as the following set of indices of energy eigenvalues of $\hat H_\rB$:
\begin{equation}
M_{U,\Delta}:=
\{
i:E_i\in (U-\Delta,U]
\}.
\end{equation}
We also define the Hilbert space $\mathcal H_{U,\Delta}$ 
that is spanned by the energy eigenstates 
$\{ | E_i \rangle \}_{i\in M_{U,\Delta}}$
in the energy shell.
We denote the dimension of the Hilbert space of the energy shell as $D(U,\Delta):=|M_{U,\Delta}|$.
\end{itembox}

To relate the microcanonical and the canonical ensembles, we define energy $U(\beta)$ that is determined by a given inverse temperature $\beta$.
In line with Ref.~\cite{S_Tasaki2016}, we employ the following definition.
\begin{itembox}[l]
{\Definition[\label{def:NDepEnergy}] }
For a given inverse temperature $\beta$,
$U(\beta)$ is defined as
\begin{equation}
U(\beta)
:=
\delta^\prime
\argmax_{\nu\in\mathbb{Z}}
D(\nu\delta^\prime,\delta^\prime)
e^{-\beta \nu\delta^\prime}.
\end{equation}
\end{itembox}
The above definition is motivated by  the Legendre transformation from $\beta$ to $U$, and is  roughly rewritten  as
\begin{equation}
U(\beta) := \argmin_U \left( \beta U - \ln D(U,\delta^\prime) \right),
\end{equation}\
where $\ln D(U,\delta^\prime) $ is the thermodynamic entropy.

We now define the microcanonical ensemble for a given inverse temperature $\beta$.
\begin{itembox}[l]
{\Definition[\label{def:Microcanonical}] (Microcanonical ensemble)}
The density operator of the microcanonical ensemble corresponding to the energy shell $M_{U(\beta),\Delta}$ is defined as
\begin{equation}
\hat{\rho}^{\mathrm{MC}}_{\rB}
:=
\frac{1}{D(U(\beta),\Delta)}\sum_{i\in M_{U(\beta),\Delta}} \ket{E_i}\bra{E_i}.
\end{equation}
\end{itembox}
In the following, we write $D:=D(U(\beta),\Delta)$ for simplicity.

\

We next discuss two fundamental assumptions that are required to prove the equivalence of the ensembles.
We first consider the correlation function.
\begin{itembox}[l]
{\Definition[\label{def:Correlation}] (Correlation function)}
Let $\hat{\rho}$ be an arbitrary density operator.
The correlation function of arbitrary two operators $\hat{A}$ and $\hat{A}^\prime$
is defined as
\begin{equation}
\mathrm{cor}_{\hat{\rho}}(\hat{A},\hat{A}^\prime)
:=
\tr[\hat{\rho} \hat{A}\hat{A}^\prime]
-
\tr[\hat{\rho} \hat{A}]\tr[\hat{\rho} \hat{A}^\prime].
\end{equation}
\end{itembox}
We then assume the following property of the correlation functions for the canonical ensemble.
\begin{itembox}[l]
{\Assumption[\label{assump:ExpDecayT}] (Exponential decay of correlation functions: Assumption I of \cite{S_Tasaki2016})}
For a given inverse temperature $\beta$,
there exist positive constants $\xi$ and $\gamma_1$ that satisfy the following.
Let $\hat{A}$ and $\hat{A}^\prime$ be arbitrary operators with supports $\rA,\rA^\prime\subset \rB$.
For any $N$, we have 
\begin{equation}
|\mathrm{cor}_{\hat{\rho}^\mathrm{can}_{\rB}}(\hat{A},\hat{A}^\prime)|
\leq
C(\hat{A},\hat{A}^\prime)
\exp\left[
-\dist(\rA,\rA^\prime)/\xi
\right],
\end{equation}
where
\begin{equation}
C(\hat{A},\hat{A}^\prime)
:=
\gamma_1
\|\hat{A}\|
\|\hat{A}^\prime\|
|\rA|
|\rA^\prime|.
\end{equation}
\end{itembox}
Assumption \ref{assump:ExpDecayT} can be rigorously proved for any $d$ and sufficiently small $\beta$,
by using the cluster expansion method~\cite{S_Park1982,S_Frolich2014}.
We note that a slightly different version of the exponential decaying of correlation functions have been proved in Ref.~\cite{S_Kliesch2014} for sufficiently small $\beta$.

\

We next  consider the Massieu function $\varphi(\beta)$, which is related to the free energy by $f(\beta)=-\varphi(\beta)/\beta$.
\begin{itembox}[l]
{\Definition[\label{def:Massieu}] (Massieu function)}
We define the Massieu function of $N$ and $\beta$ as
\begin{equation}
\varphi_N(\beta)
:=
\frac{1}{N}
\log Z(\beta).
\end{equation}
It is known rigorously in general \cite{S_Ruelle1999} that the following limit exists:
\begin{equation}
\varphi(\beta)
:=
\lim_{N\rightarrow\infty}
\varphi_N(\beta).
\end{equation}
\end{itembox}
We then assume the following properties of the Massieu function.
\begin{itembox}[l]
{\Assumption[\label{assump:Massieu}] (Properties of the Massieu function: Assumption II of \cite{S_Tasaki2016})}
For a given inverse temperature $\beta$,
there exist $\beta_1$ and $\beta_2$ such that $\beta_1<\beta<\beta_2$,
and the following two properties are valid.
First, there exists $\gamma_0$ such that
\begin{equation}
|
\varphi_N(\beta^\prime)
-
\varphi(\beta^\prime)
|
\leq
\frac{\gamma_0}{N}
\end{equation}
for any $\beta^\prime\in[\beta_1,\beta_2]$ and $N$.
Second, the Massieu function $\varphi(\beta)$ is twice continuously differentiable,
and satisfies $\varphi^{\prime\prime}(\beta)\geq c_0$
with a constant $c_0>0$ in interval $[\beta_1,\beta_2]$.
\end{itembox}
Assumption \ref{assump:Massieu} can also be proved for any $d$ and sufficiently small $\beta$,
by using the cluster expansion method~\cite{S_Park1982,S_Frolich2014}.

\

We now  define the thermodynamic limit with the inverse temperature being fixed:
\begin{itembox}[l]
{\Definition[\label{def:TDlimit}] (Thermodynamic limit)}
The thermodynamic limit is given by $N \to \infty$ with $\beta$ being fixed.  
In the thermodynamic limit, the interaction range $k$ in Assumption \ref{assump:LocalInt} is kept constant.
\end{itembox}
In the following argument, the phrase of ``sufficiently large $N$'' will be used in the sense of the above thermodynamic limit.

\

We are now in the position to state the equivalence of ensembles in terms of the reduced density operators of a subsystem.
\begin{itembox}[l]
{\Proposition[\label{prop:EquiEns}] (Equivalence of ensembles: Main theorem in \cite{S_Tasaki2016})}
Let $\rB_1$ be a hypercube or a pair of identical hypercubes in B,
whose side length is $l=L^\alpha$ with $0\leq\alpha<1/2$.
Under Assumptions \ref{assump:LocalInt}, \ref{assump:TransInv}, \ref{assump:ExpDecayT}, and \ref{assump:Massieu},
for a given inverse temperature $\beta>0$ and
for any $\varepsilon_2>0$, 
\begin{equation}
\label{ineq:equi_ens}
\|
\mathrm{tr}_{\rB\backslash\rB_1}[\hat{\rho}_{\rB}^\mathrm{can}]
-
\mathrm{tr}_{\rB\backslash\rB_1}[\hat{\rho}_{\rB}^{\mathrm{MC}}]
\|_1
\leq
\varepsilon_2
\end{equation}
holds for sufficiently large $N$.
More precisely, the left-hand side above is bounded as
\begin{equation}
\label{ineq:equi_ens_error}
\|
\mathrm{tr}_{\rB\backslash\rB_1}[\hat{\rho}_\rB^\mathrm{MC}]
-
\mathrm{tr}_{\rB\backslash\rB_1}[\hat{\rho}^{\mathrm{can}}_\mathrm{B}]
\|_1
\leq
\mathcal{O} \sbra{N^{-(1-2\alpha)/4+\delta}},
\end{equation}
where $\delta>0$ is an arbitrarily small constant.
\end{itembox}
The above theorem implies that the canonical and the microcanonical ensembles are locally indistinguishable, when we only look at $\rB_1$ that is not too large.

Strictly speaking,  $l$ should be an integer given by $\lfloor L^\alpha \rfloor $ (i.e., the maximum integer that is not larger than $L^\alpha$).
To avoid too much complications of notations, however, we just omit  $\lfloor \cdots \rfloor $ throughout this paper, 
and do not go into a technically strict argument that distinguishes whether $L^\alpha$ is an integer or not.

We note that Proposition \ref{prop:EquiEns} in Ref.~\cite{S_Tasaki2016} is a variant of Theorem 1 in Ref.~\cite{S_Brandao2015}.
We here employ the equivalence of the ensembles in the form of Ref.~\cite{S_Tasaki2016} that is applicable up to $\alpha < 1/2$, while $\alpha < 1/(d+1)$ in  Ref.~\cite{S_Brandao2015}.

\section{Weak eigenstate-thermalization hypothesis (ETH)}
\label{sec:weakETH}

In this section, we formulate and prove the weak ETH in the setup introduced in the previous section.
Let $\mathcal{M}_{U(\beta),\Delta} := \{ | E_i \rangle \}_{i=1}^D$ be the set of the eigenstates of $\hat{H}_\rB$ in the energy shell $M_{U(\beta),\Delta}$.
The goal of this section is to prove Eq.~(5) in the main text, which states that
\begin{equation}
\label{weak_ETH_main}
\tr[\hat{O}_{\rB_1}
| E_i \rangle \langle E_i |]
\simeq 
\tr[
\hat{O}_{\rB_1}
\hat{\rho}_\rB^{\mathrm{can}}]
\end{equation}
holds for a typical choice of $| E_i \rangle$ from $\mathcal{M}_{U(\beta),\Delta}$,
where 
$\rB_1$ is a hypercube in B with  side length $l=L^\alpha$ as is the case for Proposition \ref{prop:EquiEns},
and
$\hat{O}_{\rB_1}$ is any bounded operator on $\rB_1$.
We first note that 
\begin{equation}
\label{eq:wETH_split}
|
\tr[\hat{O}_{\rB_1}
(
| E_i \rangle \langle E_i |
-
\hat{\rho}_\rB^{\mathrm{can}}
)
]
|
\leq
|
\tr[\hat{O}_{\rB_1}
(
| E_i \rangle \langle E_i |
-
\hat{\rho}_\rB^{\mathrm{MC}}
)
]
|
+
|
\tr[\hat{O}_{\rB_1}
(
\hat{\rho}_\rB^{\mathrm{MC}}
-
\hat{\rho}_\rB^{\mathrm{can}}
)
]
|
.
\end{equation}
The second term on the right-hand side can be evaluated from the equivalence of the ensembles (i.e., Proposition \ref{prop:EquiEns}).  Therefore, we will focus on the first term on the right-hand side of inequality (\ref{eq:wETH_split}).

\subsection{Weak ETH for operators}

We consider any operator $\hat O$ that is defined on the Hilbert space $\mathcal H_\rB$.
We define the expectation value of $\hat O$ for $| E_i \rangle \in \mathcal{M}_{U(\beta),\Delta}$ as
\begin{equation}
O_i:=\bra{E_i}\hat{O}\ket{E_i}.
\end{equation}
The microcanonical average of $\hat O$ is then rewritten as
\begin{equation}
\overline{O}:= {\rm tr}[\hat O \hat{\rho}_{\rB}^\mathrm{MC}] 
= \frac{1}{D}\sum_{i\in M_{U(\beta),\Delta}} O_i.
\end{equation}
We consider the fluctuation of $O_i$'s around $\overline{O}$, which can be quantified as
\begin{equation}
(\Delta O)^2
:=
\frac{1}{D}
\sum_{i\in M_{U(\beta),\Delta}} (O_i)^2
-
\left( \overline{O} \right)^2.
\label{O_fluctuation}
\end{equation}
If this fluctuation is very small, almost all the energy eigenstates (i.e., typical energy eigenstates) satisfy $O_i \simeq \overline{O}$,
which is the main idea of the weak ETH:
\begin{itembox}[l]
{\Definition[\label{def:WeakETH}] (Weak ETH for operators~\cite{S_Biroli2010})}
We say that an observable $\hat{O}$ satisfies the weak ETH, 
if for any $\varepsilon>0$, 
\begin{equation}
(\Delta O)^2<\varepsilon
\label{eq:weakETH_def}
\end{equation}
holds for sufficiently large $N$.
\end{itembox}

Biroli \textit{et al.}~\cite{S_Biroli2010} discussed the weak ETH for local operators, 
while their proof was not rigorous.
We now make their argument rigorous, and extend it to quasi-local operators, 
where their argument for local operators is the case of $\alpha=0$ in the following lemma.

\begin{itembox}[l]
{\Lemma[\label{lemma:WeakETH}] (Weak ETH for quasi-local operators)}
Let $\rB_1$ be a hypercube in $\rB$, whose side length is $l=L^\alpha$ with $0\leq \alpha< 1/2$.
Under Assumptions \ref{assump:LocalInt}, \ref{assump:TransInv}, \ref{assump:ExpDecayT}, and \ref{assump:Massieu},
the weak ETH holds for any operator $\hat{O}_{\rB_1}$ with support $\rB_1$ and normalized as  
$\|\hat{O}_{\rB_1}\|=1$.
\end{itembox}

\begin{figure}[t]
\begin{center}
\includegraphics[width=0.7\linewidth]{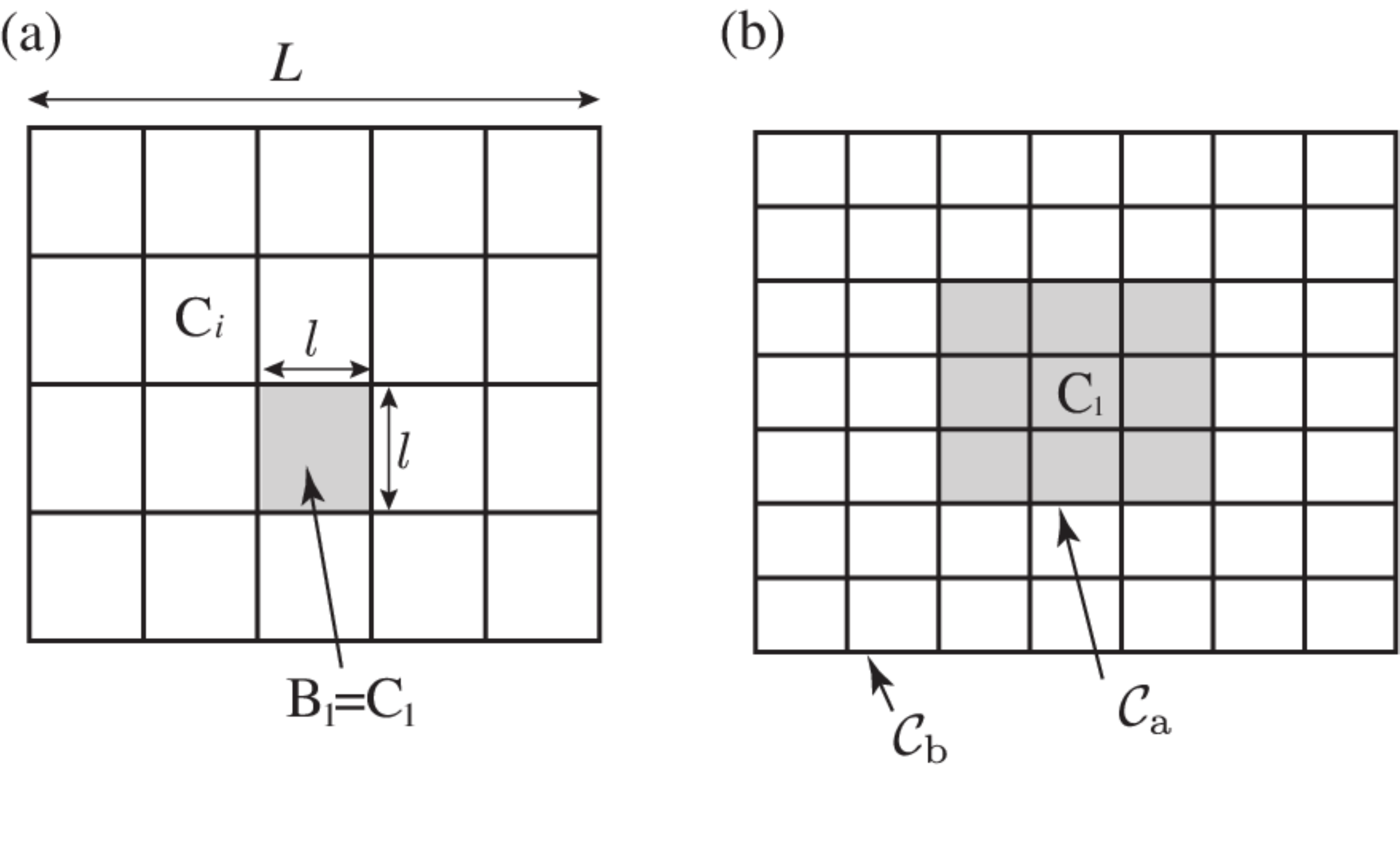}
\end{center}
\caption{
Schematics for the proof of Lemma \ref{lemma:WeakETH}.
(a)
Bath B is divided into hypercubes $\rC_i$,
which are identical copies of $\rC_1$.
(b) 
The set of blocks $\mathcal{C}$ is divided into $\mathcal{C}_\ra$ and $\mathcal{C}_\rb$
such that $\mathcal{C}_\ra$ consists of $\rC_1$ and its neighbors.
}
\label{fig:fig_s_blocks}
\end{figure}

\noindent
{\bf Proof.}
Let $n:=|\rB_1|$.
We divide bath B into $N/n$ blocks, which are identical copies of $\rB_1$.
We label them as $\rC_j$ with $j=1,\cdots,N/n$,
where $\rC_1$ is identified to $\rB_1$ itself (see Fig.~\ref{fig:fig_s_blocks}(a)).
Correspondingly, $\hat{O}_{\rB_1}$ is rewritten as $\hat{O}_{\rC_1}$.
Let $\mathcal{C}$ be the set of the blocks 
$\mathcal{C}:=\{\rC_1,\rC_2,\cdots,\rC_{N/n}\}$.
We also define $\hat{O}_{\rC_j}~(j=2,3,\cdots, | \mathcal C|)$, 
which are translational copies of $\hat{O}_{\rC_1}$ defined on $\rC_j$.
From Assumption \ref{assump:TransInv} (translation invariance),
the expectation values of $\hat{O}_{\rC_j}$'s are the same: For any $| E_i \rangle$ ($i=1,2,\cdots, D$) and any $j$ ($=1,2,\cdots, | \mathcal C|$),
\begin{equation}
\bra{E_i}
\hat{O}_{\rC_j}
\ket{E_i}
=
\bra{E_i}
\hat{O}_{\rC_1}
\ket{E_i}.
\end{equation}
Therefore, for any $| E_i \rangle$,
\begin{equation}
\bra{E_i}
\hat{O}_{\rB_1}
\ket{E_i}
=
\frac{1}{|\mathcal{C}|}
\sum_{\rC_j\in\mathcal{C}}
\bra{E_i}
\hat{O}_{\rC_j}
\ket{E_i}.
\end{equation}
By defining
\begin{equation}
\hat O := \frac{1}{|\mathcal{C}|}
\sum_{\rC_j\in\mathcal{C}}
\hat{O}_{\rC_j},
\end{equation}
we obtain
\begin{equation}
\bra{E_i}
\hat{O}_{\rC_1}
\ket{E_i}
=
\bra{E_i}
\hat{O}
\ket{E_i}.
\end{equation}
We will then consider $\langle E_i | \hat O | E_i \rangle$ instead of $\langle E_i | \hat{O}_{\rC_1} | E_i \rangle$ in the following.

We evaluate the fluctuation of $\langle E_i | \hat O | E_i \rangle$, 
quantified by $(\Delta O)^2$ in Eq.~(\ref{O_fluctuation}), 
in line with Ref.~\cite{S_Biroli2010}.
From the Cauchy-Schwarz inequality,
 $(\Delta O)^2$ is bounded from above as
\begin{equation}
(\Delta O)^2
\leq
\frac{1}{D}
\sum_{i\in M_{U(\beta),\Delta}}(O^2)_i
-
(\overline{O})^2
=
\tbra{
\delta O^2
}_\mathrm{MC},
\end{equation}
where $(O^2)_i := \langle E_i | \hat O^2 | E_i \rangle$ and
\begin{equation}
\tbra{
\delta O^2
}_\mathrm{MC}
:=
{\rm tr} [
\hat O^2 
\hat{\rho}^\mathrm{MC}_{\rB}
] - \left( {\rm tr} [\hat O \hat \rho_\rB^{\rm MC}] \right)^2.
\end{equation}
Since $\hat{O}$ is the sum of $\hat{O}_{\rC_j}$'s, we can expand $\tbra{\delta O^2}_\mathrm{MC}$ as
\begin{equation}
\begin{split}
\tbra{
\delta O^2
}_\mathrm{MC}
&=
\frac{1}{|\mathcal{C}|^2}
\sum_{\rC_j,\rC_k\in\mathcal{C}}
\sbra{
\langle
\hat{O}_{\rC_j}\hat{O}_{\rC_k}
\rangle_\mathrm{MC}
-
\langle
\hat{O}_{\rC_j}
\rangle_\mathrm{MC}
\langle
\hat{O}_{\rC_k}
\rangle_\mathrm{MC}
}
\\
&=
\frac{1}{|\mathcal{C}|}
\sum_{\rC_j\in\mathcal{C}}
\sbra{
\langle
\hat{O}_{\rC_j}\hat{O}_{\rC_1}
\rangle_\mathrm{MC}
-
\langle
\hat{O}_{\rC_j}
\rangle_\mathrm{MC}
\langle
\hat{O}_{\rC_1}
\rangle_\mathrm{MC}
}
\\
&=
\frac{1}{|\mathcal{C}|}
\sum_{\rC_j\in\mathcal{C}}
\mathrm{cor}_{\hat{\rho}_\rB^\mathrm{MC}}
(
\hat{O}_{\rC_j},\hat{O}_{\rC_1}
),
\end{split}
\label{eq:deltaOmc}
\end{equation}
where $\mathrm{cor}_{\hat{\rho}_\rB^\mathrm{MC}}(\hat{O}_{\rC_j},\hat{O}_{\rC_1}) := \langle
\hat{O}_{\rC_j}\hat{O}_{\rC_1}
\rangle_\mathrm{MC}
-
\langle
\hat{O}_{\rC_j}
\rangle_\mathrm{MC}
\langle
\hat{O}_{\rC_1}
\rangle_\mathrm{MC}$, and we used translation invariance to obtain the second line.

The correlation functions on the right-hand side of Eq.~(\ref{eq:deltaOmc}) are bounded as
\begin{equation}
\begin{split}
&
\frac{1}{|\mathcal{C}|}
\sum_{\rC_j\in\mathcal{C}}
\mathrm{cor}_{\hat{\rho}_\rB^\mathrm{MC}}
(
\hat{O}_{\rC_j},\hat{O}_{\rC_1}
)
\\
\leq&
\frac{1}{|\mathcal{C}|}
\sum_{\rC_j\in\mathcal{C}}
\mathrm{cor}_{\hat{\rho}_\rB^\mathrm{can}}
(
\hat{O}_{\rC_j},\hat{O}_{\rC_1}
)
+
\frac{1}{|\mathcal{C}|}
\sum_{\rC_j\in\mathcal{C}}
\abra{
\mathrm{cor}_{\hat{\rho}_\rB^\mathrm{MC}}
(
\hat{O}_{\rC_j},\hat{O}_{\rC_1}
)
-
\mathrm{cor}_{\hat{\rho}_\rB^\mathrm{can}}
(
\hat{O}_{\rC_j},\hat{O}_{\rC_1}
)
},
\end{split}
\label{eq:deltaOmc2}
\end{equation}
where $\mathrm{cor}_{\hat{\rho}_\rB^\mathrm{can}}$ is the correlation function for the canonical ensemble.

To evaluate the first term on the right-hand side of inequality (\ref{eq:deltaOmc2}),
we devide $\mathcal{C}$ into $\mathcal{C}_\ra$ and $\mathcal{C}_\rb$ such that
$\rC_1$ and its neighbors are in $\mathcal{C}_\ra$ (see Fig.~\ref{fig:fig_s_blocks}(b)).
We note that $|\mathcal{C}_\ra|=3^d$.
The first term on the right-hand side of inequality~(\ref{eq:deltaOmc2}) is written as
\begin{equation}
\begin{split}
\frac{1}{|\mathcal{C}|}
\sum_{\rC_j\in\mathcal{C}}
\mathrm{cor}_{\hat{\rho}_\rB^\mathrm{can}}
(
\hat{O}_{\rC_j},\hat{O}_{\rC_1}
)
&=
\frac{1}{|\mathcal{C}|}
\sum_{\rC_j\in\mathcal{C}_\ra}
\mathrm{cor}_{\hat{\rho}_\rB^\mathrm{can}}
(
\hat{O}_{\rC_1},\hat{O}_{\rC_1}
)
+
\frac{1}{|\mathcal{C}|}
\sum_{\rC_j\in\mathcal{C}_\rb}
\mathrm{cor}_{\hat{\rho}_\rB^\mathrm{can}}
(
\hat{O}_{\rC_j},\hat{O}_{\rC_1}
).
\end{split}
\label{eq:CorCanSplit}
\end{equation}
From
$|\mathrm{cor}_{\hat{\rho}_\rB^\mathrm{can}}
(
\hat{O}_{\rC_j},\hat{O}_{\rC_1}
)|
\leq
2\|\hat{O}_{\rC_j}\|\|\hat{O}_{\rC_1}\|=2$,
the first term on the right-hand side of Eq.~(\ref{eq:CorCanSplit})
is evaluated as
\begin{equation}
\begin{split}
\frac{1}{|\mathcal{C}|}
\sum_{\rC_j\in\mathcal{C}_\ra}
\mathrm{cor}_{\hat{\rho}_\rB^\mathrm{can}}
(
\hat{O}_{\rC_j},\hat{O}_{\rC_1}
)
&\leq 
2\frac{|\mathcal{C}_\ra|}{|\mathcal{C}|}.
\end{split}
\end{equation}
To evaluate the second term on the right-hand side of Eq.~(\ref{eq:CorCanSplit}),
we use Assumption \ref{assump:ExpDecayT} (exponential decay of correlation functions)
and obtain
\begin{equation}
|\mathrm{cor}_{\hat{\rho}^\mathrm{can}_{\rB}}(\hat{O}_{\rC_j},\hat{O}_{\rC_1})|
\leq
\gamma_1
N^{2\alpha}
\exp\left[
-\dist(\rC_j,\rC_1)/\xi
\right],
\end{equation}
where $\gamma_1$ is a positive constant.
If $\alpha=0$, 
the sum in the second term on the right-hand side of Eq.~(\ref{eq:CorCanSplit}) converges 
from Assumption \ref{assump:ExpDecayT},
and therefore, there exists a positive constant $\gamma_2$ such that
\begin{equation}
\sum_{\rC_j\in\mathcal{C}_\rb}
\mathrm{cor}_{\hat{\rho}_\rB^\mathrm{can}}
(
\hat{O}_{\rC_j},\hat{O}_{\rC_1}
)
\leq
\gamma_2.
\end{equation}
If $0<\alpha<1/2$, the sum in the second term on the right-hand side of Eq.~(\ref{eq:CorCanSplit}) 
converges to $0$ in the limit of $N\rightarrow\infty$:
\begin{equation}
\begin{split}
\sum_{\rC_j\in\mathcal{C}_\rb}
\mathrm{cor}_{\hat{\rho}_\rB^\mathrm{can}}
(
\hat{O}_{\rC_j},\hat{O}_{\rC_1}
)
&\leq 
\gamma_1
|\mathcal{C}_\rb|
N^{2\alpha}
\exp
\bbra{-\min_{k\in\mathcal{C}_\rb}(\dist(\rC_k,\rC_1))/\xi}
\\
&\leq
\gamma_1
N^{1-\alpha}
N^{2\alpha}
\exp
\bbra{-\min_{k\in\mathcal{C}_\rb}(\dist(\rC_k,\rC_1))/\xi}
\rightarrow 0.
\end{split}
\end{equation}
Thus, the sum in the first term on the right-hand side of inequality~(\ref{eq:deltaOmc2}) is bounded 
with a positive constant $\gamma_2$ as
\begin{equation}
\begin{split}
\label{eq:wETH_Op_1}
\sum_{\rC_j\in\mathcal{C}}
\mathrm{cor}_{\hat{\rho}_\rB^\mathrm{can}}
(
\hat{O}_{\rC_j},\hat{O}_{\rC_1}
)
&\leq 
\gamma_2
\end{split}
\end{equation}
for $0\leq\alpha<1/2$.

The second term on the right-hand side of inequality~(\ref{eq:deltaOmc2}) is bounded from 
Proposition \ref{prop:EquiEns} (the equivalence of the ensembles):
There exists a positive constant $\gamma_3$ such that
\begin{equation}
\begin{split}
\label{eq:wETH_Op_2}
\frac{1}{|\mathcal{C}|}
\sum_{\rC_j\in\mathcal{C}}
\abra{
\mathrm{cor}_{\hat{\rho}_\rB^\mathrm{MC}}
(
\hat{O}_{\rC_j},\hat{O}_{\rC_1}
)
-
\mathrm{cor}_{\hat{\rho}_\rB^\mathrm{can}}
(
\hat{O}_{\rC_j},\hat{O}_{\rC_1}
)
}
\leq&
\frac{1}{|\mathcal{C}|}
\sum_{\rC_j\in\mathcal{C}}
\gamma_3 N^{-(1-2\alpha)/4+\delta}
\\
=&
\gamma_3 N^{-(1-2\alpha)/4+\delta}
\end{split}
\end{equation}
holds for sufficiently large $N$.

By combining 
inequalities (\ref{eq:deltaOmc2}), (\ref{eq:wETH_Op_1}), and (\ref{eq:wETH_Op_2}),
we finally obtain
\begin{equation}
\tbra{
\delta O^2
}_\mathrm{MC}
\leq
\gamma_2
 N^{-(1-\alpha)}
+
\gamma_3 N^{-(1-2\alpha)/4+\delta}
\end{equation}
for sufficiently large $N$.
The right-hand side above can be arbitrarily small if $N$ is sufficiently large, which proves the lemma.
\hspace{\fill}$\Box$

\

From the proof of Lemma \ref{lemma:WeakETH},
we obtain
\begin{equation}
(\Delta O)^2\leq  
\mathcal{O}\sbra{N^{-(1-\alpha)}}
+
\mathcal{O}\sbra{N^{-(1-2\alpha)/4+\delta}},
\end{equation}
where the dominant term on the right-hand side is $\mathcal{O}\sbra{N^{-(1-2\alpha)/4+\delta}}$.

\subsection{Weak ETH with eigenstate-typicality}
From Lemma \ref{lemma:WeakETH}, 
almost all the energy eigenstates give approximately the same expectation values of a quasi-local operator as that of the microcanonical ensemble.
In this subsection, we formulate this in terms of typicality in the set of energy eigenstates.

Let ${\bf X}$ be a statement about the energy eigenstates in $\mathcal{M}_{U(\beta),\Delta}$ corresponding to the energy shell $M_{U(\beta),\Delta}$.
We define the probability that $\bf X$ is true with respect to the uniform distribution on $\mathcal{M}_{U(\beta),\Delta}$:
\begin{align}
P_{\mathcal{M}_{U(\beta),\Delta}}[{\bf X}]:=\frac{n_{\mathrm{true}}}{D},
\end{align}
where $n_{\mathrm{true}}$ is the number of the energy eigenstates in $\mathcal{M}_{U(\beta),\Delta}$ for which statement ${\bf X}$ is true.
We then define the concept of the eigenstate typicality as follows.
\begin{itembox}[l]
{\Definition[\label{def:wETHTypical}] ($\tilde{\varepsilon}$-eigenstate-typical statement)}
Let $\tilde{\varepsilon}>0$.
We say that statement $\bf X$ about the energy eigenstates in $\mathcal{M}_{U(\beta),\Delta}$ holds $\tilde{\varepsilon}$-eigenstate-typically,
if $P_{\mathcal{M}_{U(\beta),\Delta}}[{\bf X}]>1-\tilde{\varepsilon}$ holds.
\end{itembox}

We now state the weak ETH for quasi-local observables in terms of the eigenstate typicality.
\begin{itembox}[l]
{\Lemma[\label{lemma:EigenMC}] (Weak ETH with eigenstate typicality)}
Let $\rB_1$ be a hypercube in $\rB$, whose side length is $l=L^\alpha$ with $0\leq\alpha<1/2$.
Let $\hat{O}$ be any operator on $\rB_1$ with $\|\hat{O}\|=1$.
Under Assumptions \ref{assump:LocalInt}, \ref{assump:TransInv}, \ref{assump:ExpDecayT}, and \ref{assump:Massieu},
for any $\varepsilon_1>0$ and any $\tilde{\varepsilon}>0$, 
\begin{equation}
\label{eq:LemmaWeakETH}
|O_i-\overline{O}|
\leq
\varepsilon_1
\end{equation}
holds $\tilde{\varepsilon}$-eigenstate-typically for sufficiently large $N$.
\end{itembox}

\noindent
{\bf Proof.}
From  the Chebyshev inequality, we have for any $\varepsilon_1 > 0$,
\begin{align}
P_{\mathcal{M}_{U(\beta),\Delta}}[|O_i-\overline{O}|>\varepsilon_1]< 
\frac{(\Delta O)^2}{\varepsilon_1^2}.
\end{align}
From Lemma \ref{lemma:WeakETH}, for any $\varepsilon_1 > 0$ and any $\tilde \varepsilon > 0$, 
\begin{align}
\label{evaluate_N}
\frac{(\Delta O)^2}{\varepsilon_1^2} < \tilde \varepsilon
\end{align}
holds for sufficiently large $N$.
Therefore,
\begin{equation}
\label{eq:weakETH_prob}
P_{\mathcal{M}_{U(\beta),\Delta}}[|O_i-\overline{O}|>\varepsilon_1]<\tilde{\varepsilon}
\end{equation}
holds for sufficiently large $N$,
which proves the lemma.
\hspace{\fill}$\Box$

\

From inequality~(\ref{evaluate_N}) and the argument in  Lemma \ref{lemma:WeakETH}, 
inequality (\ref{eq:LemmaWeakETH}) holds $\tilde{\varepsilon}$-eigenstate-typically 
if $N$ satisfies 
$\mathcal{O}\sbra{N^{-(1-2\alpha)/4+\delta}} \leq \varepsilon_1^2 \tilde \varepsilon$.
In other words,
\begin{equation}
\label{ineq:LemmaWeakETH_error}
|O_i-\overline{O}|
\leq
\mathcal{O}\sbra{\sqrt{N^{-(1-2\alpha)/4+\delta}/\tilde{\varepsilon}}}
\end{equation}
holds $\tilde{\varepsilon}$-eigenstate-typically.

\

By combining the weak ETH and the equivalence of the ensembles discussed in Sec.~\ref{sec:MicroCanonical}, 
we obtain the weak ETH of the form~(\ref{weak_ETH_main}) 
(i.e., Eq.~(5) in the main text).
In fact, from Lemma \ref{lemma:EigenMC} and Propositions \ref{prop:EquiEns}, 
we have the following corollary.
\begin{itembox}[l]
{\Corollary[\label{cor:EigenCan}]}
Let $\rB_1$ be a hypercube in $\rB$, whose side length is $l=L^\alpha$ with $0\leq\alpha<1/2$.
Let $\hat{O}$ be any operator on $\rB_1$ with $\|\hat{O}\|=1$.
Under Assumptions \ref{assump:LocalInt}, \ref{assump:TransInv}, \ref{assump:ExpDecayT}, and \ref{assump:Massieu},
for any $\varepsilon_{12} >0$ and any $\tilde \varepsilon > 0$, 
\begin{equation}
|
O_i
-
\tr_\rB[
\hat{O}
\hat{\rho}^{\mathrm{can}}_\mathrm{B}
]
|
\leq
\varepsilon_{12}
\label{new_coro0}
\end{equation}
holds $\tilde{\varepsilon}$-eigenstate-typically for sufficiently large $N$.
\end{itembox}

\noindent
{\bf Proof.}
The left-hand side of inequality (\ref{new_coro0}) is bounded as
\begin{align}
|
O_i-
\tr_\rB[
\hat{O}
\hat{\rho}^{\mathrm{can}}_\mathrm{B}
]
|
\leq&
|
O_i
-\overline{O}
|
+
|
\overline{O}
-
\tr_\rB[
\hat{O}
\hat{\rho}^{\mathrm{can}}_\mathrm{B}
]
|.
\label{ineq:WeakETHCan_proof1}
\end{align}
From Lemma \ref{lemma:EigenMC},
for any $\varepsilon_{12}>0$ and any $\tilde{\varepsilon}>0$,
\begin{align}
|
O_i
-\overline{O}
|
\leq
\varepsilon_{12}/2
\end{align}
holds $\tilde{\varepsilon}$-eigenstate-typically for sufficiently large $N$.
From Propositions \ref{prop:EquiEns},
for any $\varepsilon_{12}>0$,
\begin{align}
|
\overline{O}
-
\tr_\rB[
\hat{O}
\hat{\rho}^{\mathrm{can}}_\mathrm{B}
]
|
\leq&
\|\hat{O}\|
\|
\tr_{\rB_2}[\hat{\rho}^{\mathrm{MC}}_\mathrm{B}]
-
\tr_{\rB_2}[\hat{\rho}^{\mathrm{can}}_\mathrm{B}]
]
\|_1
\leq
\varepsilon_{12}/2
\end{align}
holds for sufficiently large $N$.
By summing up the above inequalities, we prove the corollary.
\hspace{\fill}$\Box$

\

From inequality (\ref{ineq:equi_ens_error}) in Proposition \ref{prop:EquiEns} and inequality (\ref{ineq:LemmaWeakETH_error}) in Lemma \ref{lemma:EigenMC},
inequality~(\ref{new_coro0}) in Corollary \ref{cor:EigenCan} holds if $N$ satisfies
$\mathcal{O}\sbra{N^{-(1-2\alpha)/4+\delta}}\leq \varepsilon_{12} /2$ and 
$\mathcal{O}\sbra{N^{-(1-2\alpha)/4+\delta}} \leq \varepsilon_{12}^2 \tilde \varepsilon / 4$.
In other words,
\begin{align}
|
O_i
-
\tr_\rB[
\hat{O}
\hat{\rho}^{\mathrm{can}}_\mathrm{B}
]
|
\leq
\mathcal{O}\sbra{N^{-(1-2\alpha)/4+\delta}}
+
\mathcal{O}\sbra{\sqrt{N^{-(1-2\alpha)/4+\delta}/\tilde{\varepsilon}}}
\end{align}
holds $\tilde{\varepsilon}$-eigenstate-typically.

\


If $\alpha=0$,
Corollary \ref{cor:EigenCan}
can be described in terms of the trace norm of the density operators of a subsystem.
\begin{itembox}[l]
{\Corollary[\label{cor:EigenCan_TraceNorm}]}
Let $\rB_1$ be a hypercube in $\rB$ with side length $l$ that is independent of the size $N$ of bath B.
Under Assumptions \ref{assump:LocalInt}, \ref{assump:TransInv}, \ref{assump:ExpDecayT}, and \ref{assump:Massieu},
for any $\varepsilon_{14} >0$ and any $\tilde \varepsilon > 0$, 
\begin{equation}
\|
\tr_{\rB\backslash\rB_1}[\ket{E_i}\bra{E_i}]
-
\tr_{\rB\backslash\rB_1}[\hat{\rho}^{\mathrm{can}}_\mathrm{B}]
\|_1
\leq
\varepsilon_{14}
\label{new_coro0_tracenorm}
\end{equation}
holds $\tilde{\varepsilon}$-eigenstate-typically for sufficiently large $N$.
\end{itembox}

{\bf Proof.}
Let $D_{\rB_1}$ be the dimension of the Hilbert space of $\rB_1$.
Let $\{\hat{O}_{k}\}$ ($k=1,\dots, 2D_{\rB_1}^2$ and $\|\hat{O}_{k}\|=1$) be a linear basis of the space of linear operators on $\rB_1$.
From Corollary \ref{cor:EigenCan},
for each $k$,
for any $\varepsilon_{14,1}>0$ and any $\tilde{\varepsilon}>0$, 
we find that 
\begin{align}
\left|\mathrm{tr}_{\rB_1}
\bbra{
\hat{O}_{k}
\sbra{
\tr_{\rB\backslash\rB_1}[\ket{E_i}\bra{E_i}]
-
\tr_{\rB\backslash\rB_1}[\hat{\rho}^{\mathrm{can}}_\mathrm{B}]
}
}
\right|
&\leq
\varepsilon_{14,1}
\end{align}
holds $\tilde{\varepsilon}/(2D_{\rB_1}^2)$-eigenstate-typically for sufficiently large $N$.
Because $D_{\rB_1}$ is independent of $N$,
for any $\varepsilon_{14,1}>0$ and any $\tilde{\varepsilon}>0$, 
we find that 
\begin{align}
\left|\mathrm{tr}_{\rB_1}
\bbra{
\hat{O}_{k}
\sbra{
\tr_{\rB\backslash\rB_1}[\ket{E_i}\bra{E_i}]
-
\tr_{\rB\backslash\rB_1}[\hat{\rho}^{\mathrm{can}}_\mathrm{B}]
}
}
\right|
&\leq
\varepsilon_{14,1}
\label{eq:cor:EigenCan_TN}
\end{align}
holds for all $k=1,\cdots,2D_{\rB_1}^2$, $\tilde{\varepsilon}$-eigenstate-typically for sufficiently large $N$.

Let $\hat{O}$ be any operator with support $\rB_1$ and with $\|\hat{O}\|=1$.
We can expand $\hat{O}$ as
$\hat{O}=\sum_{k=1}^{2D_{\rB_1}^2} c_k \hat{O}_{k}$ 
with $c_k\in\mathbb{C}$.
We have
\begin{align}
\begin{split}
\abra{
\mathrm{tr}_{\rB_1}
\bbra{
\hat{O}
\sbra{
\tr_{\rB\backslash\rB_1}[\ket{E_i}\bra{E_i}]
-
\tr_{\rB\backslash\rB_1}[\hat{\rho}^{\mathrm{can}}_\mathrm{B}]
}
}
}
&=
\abra{\mathrm{tr}_{\rB_1}
\bbra{
\sum_{k=1}^{2D_{\rB_1}^2}
c_k
\hat{O}_{k}
\sbra{
\tr_{\rB\backslash\rB_1}[\ket{E_i}\bra{E_i}]
-
\tr_{\rB\backslash\rB_1}[\hat{\rho}^{\mathrm{can}}_\mathrm{B}]
}
}
}
\\
&\leq
\sum_{k=1}^{2D_{\rB_1}^2}
|c_k|
\abra{
\mathrm{tr}_{\rB_1}
\bbra{
\hat{O}_{k}
\sbra{
\tr_{\rB\backslash\rB_1}[\ket{E_i}\bra{E_i}]
-
\tr_{\rB\backslash\rB_1}[\hat{\rho}^{\mathrm{can}}_\mathrm{B}]
}
}
}.
\end{split}
\end{align}
From inequality (\ref{eq:cor:EigenCan_TN}),
for any $\varepsilon_{14,1}>0$ and any $\tilde{\varepsilon}>0$,
\begin{align}
\begin{split}
\abra{
\mathrm{tr}_{\rB_1}
\bbra{
\hat{O}
\sbra{
\tr_{\rB\backslash\rB_1}[\ket{E_i}\bra{E_i}]
-
\tr_{\rB\backslash\rB_1}[\hat{\rho}^{\mathrm{can}}_\mathrm{B}]
}
}
}
\leq
\sum_{k=1}^{2D_{\rB_1}^2}
|c_k|
\varepsilon_{14,1}
\end{split}
\label{eq:cor:EigenCan_TN2}
\end{align}
holds $\tilde{\varepsilon}$-eigenstate-typically for sufficiently large $N$.
Because the set of $\hat{O}_k$ with $\|\hat{O}_k\|=1$ is compact,
\begin{align}
c:=\max_{\|\hat{O}_{\rB_1}\|=1}\sbra{\sum_{k=1}^{2D_{\rB_1}^2}|c_k|}
\end{align}
is finite, which is independent of $N$.
Therefore, the right-hand side of inequality (\ref{eq:cor:EigenCan_TN2}) is bounded as 
$\sum_{k=1}^{2D_{\rB_1}^2}
|c_k|
\varepsilon_{14,1}
\leq
\varepsilon_{14,1}c.
$
By letting $\varepsilon_{14} := \varepsilon_{14,1}c$, we prove the corollary.
\hspace{\fill}$\Box$

\section{Our setup}
\label{sec:setup}

We now formulate the setup of our main theorems on the second law and the fluctuation theorem.
We also introduce some definitions that are crucial in the proof of our main results. 

\begin{figure}[t]
\begin{center}
\includegraphics[width=0.3\linewidth]{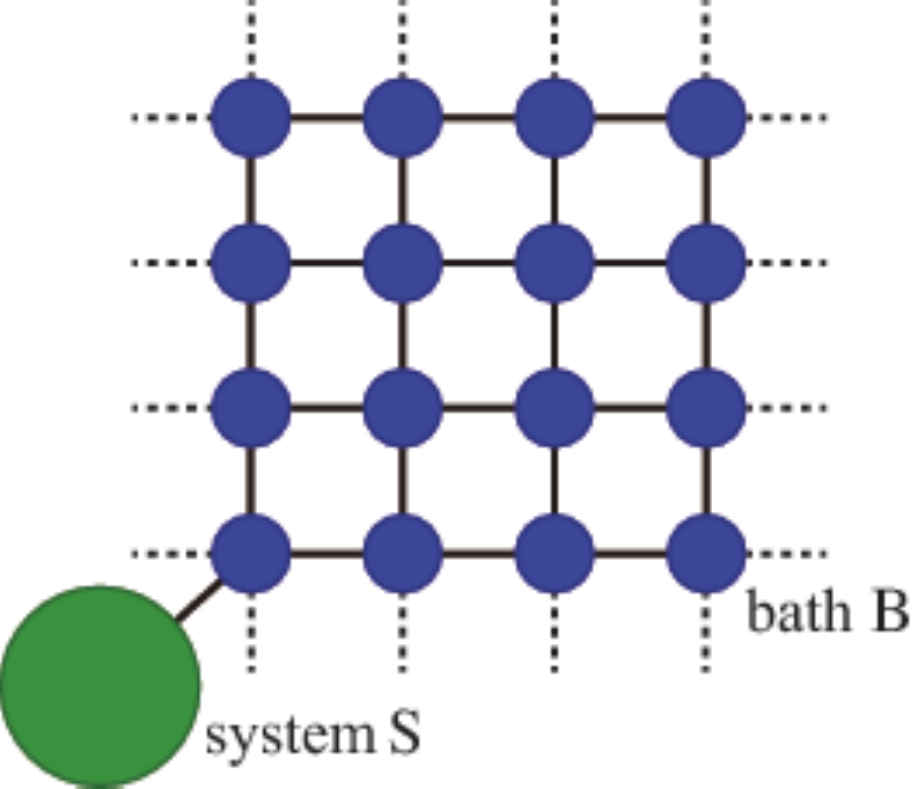}
\end{center}
\caption{
Schematic of our setup.
The total system consists of system S and bath B.
System S is locally attached to a part of bath B.
Bath B is a quantum many-body system on the $d$-dimensional hypercubic lattice with the periodic boundary condition and translation invariance.
}
\label{fig:fig_s_2a}
\end{figure}

We consider a composite system that consists of system S and bath B ~(see Fig.~\ref{fig:fig_s_2a} for a schematic).
Let  $\Lambda=(\Lambda_\mathrm{v},\Lambda_\mathrm{e})$ be the graph of the composite system, where $\Lambda_{\rm v}$ is the set of vertices (sites) and $\Lambda_{\rm e}$ is the set of edges (bonds), as is the case for Sec.~\ref{sec:LRB}.
The sets of sites of system S and bath B are denoted by the same notations: $\rS \subset \Lambda_{\rv}$ and $\rB \subset \Lambda_{\rv}$, respectively.
We note that $\rS \cup \rB = \Lambda_{\rv}$ and $\rS \cap \rB = \phi$ with $\phi$ being the empty set.
In the same manner as in Sec.~\ref{sec:MicroCanonical},
we assume that B is the $d$-dimensional hypercubic lattice with the periodic boundary condition, where $N:=|\rB|$. 
On the other hand, S can be an arbitrary graph.

The Hamiltonian of the total system is written as
\begin{equation}
\hat{H}
:=
\hat{H}_\mathrm{S}
+
\hat{H}_{\mathrm{I}}
+
\hat{H}_{\mathrm{B}},
\label{eq:Hamiltonian_total}
\end{equation}
where $\hat{H}_{\mathrm{S}}$ and $\hat{H}_{\mathrm{B}}$
are the Hamiltonians of system S and bath B, whose supports are  $\rS \subset \Lambda_{\rv}$ and $\rB \subset \Lambda_{\rv}$, respectively.
The interaction between system S and bath $\rB$ is represented by $\hat{H}_\mathrm{I}$.
We write the support of $\hat{H}_\mathrm{I}$ in bath B as $\rI \subset \Lambda_{\rm v}$:
$\rI:=\rB\cap \supp(\hat{H}_\rI)$.
We define $\tilde{\rS}:=\rS\cup\rI$.
The Hilbert space of $\tilde{\rS}$ is denoted by  $\mathcal H_{\tilde{\rS}}:=\otimes_{\ri\in {\tilde \rS}}\mathcal{H}_\mathrm{i}$
and the dimension of $\mathcal H_{\tilde{\rS}}$ is denoted by $D_{\tilde{\rS}}$.
We also define $\tilde{\rB}:= \rB \backslash \rI =(\rS\cup\rB)\backslash \tilde{\rS}$.

We assume that
the Hamiltonian of bath B is represented as the sum of local Hamiltonians,
as in Eq.~(\ref{l_H}).
We further assume all of the foregoing Assumptions
\ref{assump:LRB},
\ref{assump:LocalInt}, \ref{assump:TransInv}, \ref{assump:ExpDecayT}, and \ref{assump:Massieu},
such that all of the foregoing Propositions and Lemmas are applicable to bath B.
We also assume that the interaction between S and B is local:
\begin{itembox}[l]
{\Assumption{\label{assump:LocalIntS}} (Locality of the interaction between S and B)}
We assume that  $\rm I$ and $\hat{H}_\rI$ are independent of $N$.
We refer to $k' :=\dist (\rS, \tilde{\rB})$ as the interaction range between system S and bath B.
\end{itembox}
In the present setup with system S and bath B,
the thermodynamic limit is defined as follows.
\begin{itembox}[l]
{{\bf Definition \ref{def:TDlimit}'} (Thermodynamic limit)}
The thermodynamic limit is given by $N \to \infty$ with $\beta$ being fixed.  
In the thermodynamic limit, interaction ranges $k$ (in Assumption \ref{assump:LocalInt}) and $k'$ (in Assumption \ref{assump:LocalIntS}) are kept constant, and the graph structure and the Hamiltonian of system S do not change.
We note that  $D_\rS$ and $D_{\tilde{\rS}}$ do not change in the thermodynamic limit.
\end{itembox}

\

\begin{figure}[t]
\begin{center}
\includegraphics[width=0.5\linewidth]{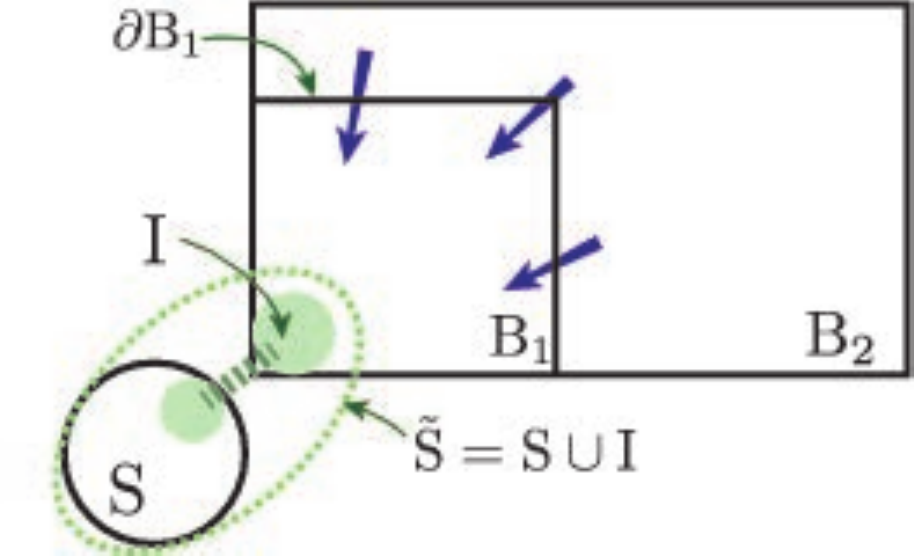}
\end{center}
\caption{
Schematic of regions on the graph.
Bath B is divided into two regions: $\rB_1$ and $\rB_2$.  
We also define $\tilde{\rS}:= \rS \cup \rI$, where $\tilde \rS$ is localized around S. 
We assume that $\rI \subset \rB_1$.
The information propagation from  $\rB_2$ is illustrated by the blue arrows, which does not reach $\tilde \rS$ in a sufficiently short time regime.
Before the information about $\rB_2$ reaches $\tilde \rS$, $\tilde \rS$ does not feel the existence of  $\rB_2$.
System S cannot distinguish whether the state of bath B is a typical energy eigenstate or the canonical distribution in the sufficiently short time regime.
}
\label{fig:fig_s_3}
\end{figure}

As discussed in the main text, the initial state of the composite system is given by
\begin{equation}
\hat \rho (0) := \hat \rho_\rS (0) \otimes \hat \rho_\rB (0), 
\label{setup_initial}
\end{equation}
where $\hat \rho_\rS (0) $ is an arbitrary state of system S.
On the other hand,
\begin{equation}
\hat{\rho}_\rB (0) := | E_i \rangle \langle E_i |
\end{equation}
is a typical energy eigenstate in the energy shell $M_{U(\beta),\Delta}$ of bath B,
in the sense of the weak ETH as discussed in Sec.~\ref{sec:weakETH}.

For the proof of our main results, 
we divide bath B into two regions: $\mathrm{B}_1$ and  $\mathrm{B}_2$ with $\rB=\rB_1\cup \rB_2$ and $\rB_1\cap\rB_2=\phi$.
We assume that $\rB_1$ is a hypercube in $\rB$, whose side length is $l=L^\alpha$ with $0<\alpha<1/2$.
As shown in Fig.~\ref{fig:fig_s_3},
$\mathrm{B}_1$ is near S, 
and $\mathrm{B}_2$ is far from S.
By noting that $\rI$ is localized around S because of the local interaction between S and B, we assume that $\rI \subset \rB_1$.
The Hamiltonians of these regions are defined as 
\begin{equation}
\hat{H}_\mathrm{X}
=
\sum_{\mathrm{Z}\subset \mathrm{X}} \hat{h}_{\mathrm{Z}}
\quad(\mathrm{X}=\rB_1,\rB_2).
\end{equation}
We then define the interaction Hamiltonian between $\rB_1$ and $\rB_2$ as
\begin{equation}
\hat{H}_{\partial\mathrm{B}_{1}}
:=
\hat{H}_\rB
-
\hat{H}_{\mathrm{B}_1}
-
\hat{H}_{\mathrm{B}_2},
\end{equation}
and define the boundary of the regions as the support of the interaction Hamiltonian:
\begin{equation}
\partial\mathrm{B}_{1}
:=
\supp(\hat{H}_{\partial\mathrm{B}_{1}}).
\end{equation}

\section{Truncated dynamics around system S}
\label{sec:locality}

We now go into the main part of our proof.  
In this section, we introduce a {\it reference dynamics} with the initial canonical ensemble,
to which the conventional proof of the second law and the fluctuation theorem is applicable.
We then show that the dynamics around system S is almost the same for both the actual dynamics and the reference dynamics.
In the following, we assume all of the foregoing Assumptions
\ref{assump:LRB},
\ref{assump:LocalInt}, \ref{assump:TransInv}, \ref{assump:ExpDecayT}, \ref{assump:Massieu}, and
\ref{assump:LocalIntS}.

\subsection{Truncation of the dynamics}
First of all, we compare the dynamics 
with the same initial condition $\hat{\rho}(0)$ as in Eq.~(\ref{setup_initial}),
but with different Hamiltonians:
the actual Hamiltonian $\hat H$ and a truncated Hamiltonian 
\begin{equation}
\hat{H}_\trunc:=\hat{H}_\mathrm{S}+\hat{H}_{\mathrm{I}}+\hat{H}_{\mathrm{B}_1}.
\end{equation}
The subscript $\trunc$ represents that the dynamics is ``truncated''.
The density operators at time $t$ under unitary evolutions with these Hamiltonians are given by
\begin{equation}
\begin{split}
\hat{\rho}(t)
&\defL
e^{-i\hat{H}t}
\hat{\rho}(0)
e^{i\hat{H}t},
\\
\hat{\rho}_\trunc(t)
&\defL
e^{-i\hat{H}_\trunc t}
\hat{\rho}(0)
e^{i\hat{H}_\trunc t}.
\end{split}
\end{equation}

We focus on the dynamics around system S (i.e., $\tilde{\rS}$ defined in Sec.~\ref{sec:setup}).
The corresponding reduced density operators are defined as  
$\hat{\rho}_{\tilde{\rS}}(t):=\mathrm{tr}_{\tilde{\rB}}[\hat{\rho}(t)]$
and
$\hat{\rho}_{\tilde{\rS},\trunc}(t):=\mathrm{tr}_{\tilde{\rB}}[\hat{\rho}_\trunc(t)]$.
We then have the following lemma.
\begin{itembox}[l]
{\Lemma[\label{lemma:TruncDynamics}]}
For any $\varepsilon_4>0$ and  $t> 0$, 
\begin{equation}
\label{eq:LRB_DM1}
\|
\hat{\rho}_{\tilde{\rS}}(t)
-
\hat{\rho}_{{\tilde{\rS}},\trunc}(t)
\|_1
\leq
\varepsilon_4
\end{equation}
holds for sufficiently large $N$.
\end{itembox}

\noindent
{\bf Proof.}
Let $\hat{O}_{\tilde{\rS}}$ be an arbitrary operator with support ${\tilde{\rS}}$ and with operator norm $\|\hat{O}_\mathrm{\tilde{\rS}}\|=1$.
We have
\begin{equation}
\label{eq:LRB_Lemma1_eq1}
\begin{split}
&
|\mathrm{tr}_{\tilde{\rS}}
[
\hat{O}_{\tilde{\rS}}
(\hat{\rho}_{\tilde{\rS}}(t)-\hat{\rho}_{{\tilde{\rS}},\trunc}(t)
)
]
|
\\
=&
|\mathrm{tr}_{\rS\cup\rB}
[
(
e^{i\hat{H}t}
\hat{O}_{\tilde{\rS}}
e^{-i\hat{H}t}
-
e^{i\hat{H}_\trunc t}
\hat{O}_{\tilde{\rS}}
e^{-i\hat{H}_\trunc t}
)
\hat{\rho}_{{\tilde{\rS}}}(0)
]
|
\\
\leq&
\|
(
e^{i\hat{H}t}
\hat{O}_{\tilde{\rS}}
e^{-i\hat{H}t}
-
e^{i\hat{H}_\trunc t}
\hat{O}_{\tilde{\rS}}
e^{-i\hat{H}_\trunc t}
)
\|
\cdot
\|
\hat{\rho}_{\tilde{\rS}}(0)
\|_1
\\
=&
\|
(
e^{i\hat{H}t}
\hat{O}_{\tilde{\rS}}
e^{-i\hat{H}t}
-
e^{i\hat{H}_\trunc t}
\hat{O}_{\tilde{\rS}}
e^{-i\hat{H}_\trunc t}
)
\|,
\end{split}
\end{equation}
where we used Proposition \ref{prop:IneqNorm} from the second to the third line.
To evaluate the final line of (\ref{eq:LRB_Lemma1_eq1}), we calculate as
\begin{equation}
\begin{split}
&
e^{i\hat{H}t}
\hat{O}_{\tilde{\rS}}
e^{-i\hat{H}t}
-
e^{i\hat{H}_\trunc t}
\hat{O}_{\tilde{\rS}}
e^{-i\hat{H}_\trunc t}
\\
=&
\int_0^t ds\frac{d}{ds}
\sbra{
e^{i\hat{H}s}
e^{i\hat{H}_\trunc(t-s)}
\hat{O}_{\tilde{\rS}}
e^{-i\hat{H}_\trunc(t-s)}
e^{-i\hat{H}s}
}
\\
=&
i
\int_0^t
ds
\{
e^{i\hat{H}s}
(\hat{H}-\hat{H}_\trunc)
e^{i\hat{H}_\trunc(t-s)}
\hat{O}_{\tilde{\rS}}
e^{-i\hat{H}_\trunc(t-s)}
e^{-i\hat{H}s}
\\
&\qquad
+
e^{i\hat{H}s}
e^{i\hat{H}_\trunc(t-s)}
\hat{O}_{\tilde{\rS}}
(\hat{H}_\trunc-\hat{H})
e^{-i\hat{H}_\trunc(t-s)}
e^{-i\hat{H}s}\}
\\
=&
i
\int_0^t
ds
\{
e^{i\hat{H}s}
(\hat{H}_{\partial\mathrm{B}_{1}}+\hat{H}_{\mathrm{B}_2})
e^{i\hat{H}_\trunc(t-s)}
\hat{O}_{\tilde{\rS}}
e^{-i\hat{H}_\trunc(t-s)}
e^{-i\hat{H}s}
\\
&\qquad
-
e^{i\hat{H}s}
e^{i\hat{H}_\trunc(t-s)}
\hat{O}_{\tilde{\rS}}
e^{-i\hat{H}_\trunc(t-s)}
(\hat{H}_{\partial\mathrm{B}_{1}}+\hat{H}_{\mathrm{B}_2})
e^{-i\hat{H}s}
\}
\\
=&
-i
\int_0^t
ds
e^{i\hat{H}s}
[
e^{i\hat{H}_\trunc(t-s)}
\hat{O}_{\tilde{\rS}}
e^{-i\hat{H}_\trunc(t-s)}
,
\hat{H}_{\partial\mathrm{B}_{1}}
]
e^{-i\hat{H}s},
\end{split}
\end{equation}
where we used $[\hat{H}_{\mathrm{B}_2},\hat{O}_{\tilde{\rS}}]=0$ and $[\hat{H}_{\mathrm{B}_2},\hat{H}_\trunc]=0$.
We thus obtain
\begin{equation}
\label{eq:norm_diff_short_S2}
\|
e^{i\hat{H}t}
\hat{O}_{\tilde{\rS}}
e^{-i\hat{H}t}
-
e^{i\hat{H}_\trunc t}
\hat{O}_{\tilde{\rS}}
e^{-i\hat{H}_\trunc t}
\|
\leq
\int_0^t
ds
\|
[
e^{i\hat{H}_\trunc(t-s)}
\hat{O}_{\tilde{\rS}}
e^{-i\hat{H}_\trunc(t-s)}
,
\hat{H}_{\partial\mathrm{B}_{1}}
]
\|.
\end{equation}
The Lieb-Robinson bound (Proposition \ref{prop:LRB}) is now applicable to the right-hand side of inequality (\ref{eq:norm_diff_short_S2}).
Let us set in Proposition \ref{prop:LRB} $\hat{A}(t) := \hat{O}_{\tilde{\rS}}(t-s)$ and $\hat{A}^\prime := \hat{H}_{\partial\rB_{1}}$, and identify $t$ to $t-s$, 
A to $\tilde \rS$, and $\rA^\prime$ to $\partial \rB_1$.
By noting that the Hamiltonian is given by $\hat{H}_\trunc$, we obtain
\begin{equation}
\label{eq:LRB_S_dynamics}
\|
[
e^{i\hat{H}_\trunc(t-s)}
\hat{O}_{\tilde{\rS}}
e^{-i\hat{H}_\trunc(t-s)}
,
\hat{H}_{\partial\mathrm{B}_{1}}
]
\|
\leq
C
\|
\hat{O}_{\tilde{\rS}}
\|
\cdot
\|
\hat{H}_{\partial\mathrm{B}_{1}}
\|
\cdot
|{\tilde{\rS}}|
\cdot
|\partial\mathrm{B}_{1}|
\cdot
e^{-\mu \mathrm{dist}({\tilde{\rS}},\partial\mathrm{B}_{1})}
(e^{v|t-s|}-1),
\end{equation}
where $C$ is the constant defined in Proposition \ref{prop:LRB}.
By substituting this to inequality~(\ref{eq:norm_diff_short_S2}),
we obtain 
\begin{equation}
\label{LR_S1}
\|
e^{i\hat{H}t}
\hat{O}_{\tilde{\rS}}
e^{-i\hat{H}t}
-
e^{i\hat{H}_\trunc t}
\hat{O}_{\tilde{\rS}}
e^{-i\hat{H}_\trunc t}
\|
\leq 
C
\|
\hat{O}_{\tilde{\rS}}
\|
\cdot
\|
\hat{H}_{\partial\mathrm{B}_{1}}
\|
\cdot
|{\tilde{\rS}}|
\cdot
|\partial\mathrm{B}_{1}|
\cdot
e^{-\mu\mathrm{dist}({\tilde{\rS}},\partial\mathrm{B}_{1})}
(e^{vt}-vt-1).
\end{equation}
By noting that $\| \hat{O}_{\tilde{\rS}}
\| = 1$, we finally obtain, from inequalities (\ref{eq:LRB_Lemma1_eq1}) and (\ref{LR_S1})
\begin{equation}
\label{eq:LRB_DM2}
\|
\hat{\rho}_{\tilde{\rS}}(t)
-
\hat{\rho}_{{\tilde{\rS}},\trunc}(t)
\|_1
\leq
C
\frac{
\|
\hat{H}_{\partial\mathrm{B}_{1}}
\|
}{v}
\cdot
|{\tilde{\rS}}|
\cdot
|\partial\mathrm{B}_{1}|
\cdot
e^{-\mu\mathrm{dist}({\tilde{\rS}},\partial\mathrm{B}_{1})}
(e^{vt}-vt-1).
\end{equation}
The right-hand side of inequality (\ref{eq:LRB_DM2}) exponentially decreases with respect to the distance between $\tilde{\rS}$ and $\partial\rB_{1}$.
Therefore, 
the right-hand side of inequality (\ref{eq:LRB_DM2}) can be arbitrarily small for sufficiently large $N$, and the lemma is proved. \hspace{\fill}$\Box$

In our setup,
the Lieb-Robinson time $\tau$ is defined as
\begin{equation}
\tau:=\mu\dist(\tilde{\rS},\partial\rB_{1})/v
\simeq (\mu/v)N^{\alpha/d}
.
\end{equation} 
From the proof of Lemma \ref{lemma:TruncDynamics},
the left-hand side of inequality 
(\ref{eq:LRB_DM1})
is bounded by a term proportional to 
$
e^{-\mu N^{\alpha/d}}
(e^{vt}-vt-1).$
This term increases in time with $\mathcal{O}(t^2)$ up to $t\simeq 1/v$.


\subsection{The reference dynamics}
\label{sec:ref_dynamics}
We now introduce the \textit{reference dynamics}, written as $\hat \rho^\rR (t)$, 
where the initial  state is given by
\begin{equation}
\hat \rho^\rR (0) := \hat{\rho}_\rS(0)\otimes\hat{\rho}_{\rB}^\can,
\label{ref_initial}
\end{equation}
and the Hamiltonian is given by the actual one: $\hat{H}$.
We note that the inverse temperature of $\hat{\rho}^\mathrm{can}_\rB$ in $\hat{\rho}^\rR (0)$ is determined by that of the corresponding energy shell
defined in Definition \ref{def:Microcanonical}.

In the reference dynamics, 
the argument in Sec.~\ref{sec:review} (the conventional proof of the second law and the fluctuation theorem) is applicable.
In this subsection, we will prove Lemma \ref{lemma:Reference}, 
which states that the difference between the actual and the reference dynamics
can be arbitrarily small in the thermodynamic limit.
Based on Lemma \ref{lemma:Reference}, 
we will prove the second law and the fluctuation theorem in Secs.
\ref{sec:2nd} and \ref{sec:FT}, respectively.

\

Before going to Lemma \ref{lemma:Reference},
we show a straightforward extension of Corollary \ref{cor:EigenCan} to any operator in the composite system $\rS\cup \rB_1$.
\begin{itembox}[l]
{\Corollary[\label{cor:EigenCan2}]}
Let $\hat{\rho}_\rS$ be any density operator of system S.
Let $\hat{O}_{\rS\cup\rB_1}$ be any operator defined on $\rS\cup\rB_1$
with $\|\hat{O}_{\rS\cup\rB_1}\|=1$. 
For any $\varepsilon_{13} >0$ and any $\tilde \varepsilon > 0$, 
\begin{equation}
\abra{
\tr_{\rS\cup\rB}\bbra{
\hat{O}_{\rS\cup\rB_1}
\sbra{
\hat{\rho}_\rS \otimes \ket{E_i}\bra{E_i}
-
\hat{\rho}_\rS \otimes \hat{\rho}^{\mathrm{can}}_\mathrm{B}
}}}
\leq
\varepsilon_{13}
\label{new_coro3}
\end{equation}
holds $\tilde{\varepsilon}$-eigenstate-typically for sufficiently large $N$.
\end{itembox}

\noindent
{\bf Proof.}
From Corollary \ref{cor:EigenCan},
for any $\varepsilon_{13} >0$ and any $\tilde \varepsilon > 0$, 
\begin{equation}
\abra{
\tr_{\rS\cup\rB}\bbra{
\hat{O}_{\rS\cup\rB_1}
\sbra{
\hat{\rho}_\rS(0) \otimes \ket{E_i}\bra{E_i}
-
\hat{\rho}_\rS(0) \otimes \hat{\rho}^{\mathrm{can}}_\mathrm{B}
}}}
\leq
\|\tr_\rS[
\hat{O}_{\rS\cup\rB_1}\hat{\rho}_\rS(0)
]\|
\varepsilon_{13}
\label{eq:new_coro3_1}
\end{equation}
holds $\tilde{\varepsilon}$-eigenstate-typically for sufficiently large $N$.
Let
$\hat{\rho}_{\rS}=\sum_{k=1}^{D_{{\rS}}}\rho_{kk}
\ket{\varphi_k}
\bra{\varphi_k}$
be the spectrum decomposition of $\hat{\rho}_{\rS}$,
where $\{\ket{\varphi_k}\}$ is an orthonormal basis of $\mathcal{H}_\rS$.
By defining $\hat{O}_{kj}:=\bra{\varphi_k}\hat{O}_{\rS\cup\rB_1}\ket{\varphi_j}$,
we have
\begin{equation}
\begin{split}
\|\tr_\rS[
\hat{O}_{\rS\cup\rB_1}
\rho_\rS
]
\|
=&
\nbra{
\sum_{k=1}^{D_{{\rS}}}
\rho_{kk}
\hat{O}_{kk}
}
\leq 
\sum_{k=1}^{D_{{\rS}}}
\rho_{kk}
\|
\hat{O}_{kk}
\|
\leq
\sum_{k=1}^{D_{{\rS}}}
\rho_{kk}
\|
\hat{O}_{\rS\cup\rB_1}
\|
=
1,
\end{split}
\label{eq:new_coro3_2}
\end{equation}
where we used
$\|\hat{O}_{kk}\|
\leq 
\|\hat{O}_{\rS\cup\rB_1}\|
=1
$
and $\sum_{k=1}^{D_\rS}\rho_{kk}=1$.
By combining inequalities (\ref{eq:new_coro3_1}) and (\ref{eq:new_coro3_2}),
we prove the corollary.
\hspace{\fill}$\Box$

\

The reduced density operator of $\tilde{\rS}$ in the reference dynamics is defined as
$\hat{\rho}^\rR_{\tilde{\rS}}(t):=\tr_{\tilde{\rB}}[\hat{\rho}^\rR(t)]$.
The difference between 
$\hat{\rho}_{\tilde{\mathrm{S}}}(t)$ and
$\hat{\rho}^\rR_{\tilde{\rS}}(t)$ is then evaluated as follows.
\begin{itembox}[l]
{\Lemma[\label{lemma:Reference}]}
For any $\varepsilon_5>0$, any $\tilde{\varepsilon}>0$, and $t>0$, 
\begin{equation}
\label{eq:LRB_DM3}
\|
\hat{\rho}_{\tilde{\mathrm{S}}}(t)
-
\hat{\rho}^\rR_{\tilde{\mathrm{S}}}(t)
\|_1
\leq
\varepsilon_5
\end{equation}
holds $\tilde{\varepsilon}$-eigenstate-typically for sufficiently large $N$.
\end{itembox}
\noindent
{\bf Proof.}
Let $\{\hat{O}_{\tilde{\rS},k}\}$ ($k=1,\dots, 2D_{\tilde{\rS}}^2$ and $\|\hat{O}_{\tilde{\rS},k}\|=1$) be 
a linear basis of the space of linear operators of $\mathcal{H}_{\tilde{\rS}}$.
We first note that
\begin{align}
|\mathrm{tr}_{\tilde{\rS}}
[
\hat{O}_{\tilde{\rS},k}
(
\hat{\rho}_{\tilde{\rS}}(t)-\hat{\rho}^\rR_{\tilde{\rS}}(t)
)
]
|
&\leq
|\mathrm{tr}_{\tilde{\rS}}
[
\hat{O}_{\tilde{\rS},k}
(
\hat{\rho}_{\tilde{\rS}}(t)-\hat{\rho}_{\tilde{\rS},\trunc}(t)
)
]
|
+
|\mathrm{tr}_{\tilde{\rS}}
[
\hat{O}_{\tilde{\rS},k}
(
\hat{\rho}_{\tilde{\rS},\trunc}(t)-\hat{\rho}^\rR_{\tilde{\rS},\trunc}(t)
)
]
|
\nonumber
\\
&+
|\mathrm{tr}_{\tilde{\rS}}
[
\hat{O}_{\tilde{\rS},k}
(
\hat{\rho}^\rR_{\tilde{\rS},\trunc}(t)-\hat{\rho}^\rR_{\tilde{\rS}}(t)
)
]
|.
\label{eq:LRB2_S_tilde}
\end{align}
The first and the third terms on the right-hand side of inequality~(\ref{eq:LRB2_S_tilde}) 
are in the form of the first line of inequality~(\ref{eq:LRB_Lemma1_eq1}), to which we can apply Lemma \ref{lemma:TruncDynamics}.  
Therefore, for any $\varepsilon_{5,1}>0$ and any $\varepsilon_{5,2}>0$, 
\begin{align}
|\mathrm{tr}_{\tilde{\rS}}
[
\hat{O}_{\tilde{\rS},k}
(
\hat{\rho}_{\tilde{\rS}}(t)-\hat{\rho}_{\tilde{\rS},\trunc}(t)
)
]
|
\leq \varepsilon_{5,1},
\label{eq:Lem4_51}
\\
|\mathrm{tr}_{\tilde{\rS}}
[
\hat{O}_{\tilde{\rS},k}
(
\hat{\rho}^\rR_{\tilde{\rS},\trunc}(t)-\hat{\rho}^\rR_{\tilde{\rS},\trunc}(t)
)
]
|
\leq \varepsilon_{5,2}
\label{eq:Lem4_52}
\end{align}
holds for sufficiently large $N$.

The second term on the right-hand side of inequality~(\ref{eq:LRB2_S_tilde}) is evaluated as
\begin{equation}
\begin{split}
&|\mathrm{tr}_{\tilde{\rS}}
[
\hat{O}_{\tilde{\rS},k}
(
\hat{\rho}_{\tilde{\rS},\trunc}(t)-\hat{\rho}^\rR_{\tilde{\rS},\trunc}(t)
)
]
|
\\
=&
|\mathrm{tr}_{\rS \cup \rB_1}
[
\hat{O}_{\tilde{\rS},k}
(
 {\rm tr}_{\rB_2} [
e^{-i\hat{H}_\trunc t}
\hat{\rho}(0)
e^{i\hat{H}_\trunc t}
]
-
 {\rm tr}_{\rB_2} [
e^{-i\hat{H}_\trunc t}
\hat{\rho}^\rR(0)
e^{i\hat{H}_\trunc t}
]
)
]
|
\\
=&
|\mathrm{tr}_{\rS \cup \rB_1}
[
\hat{O}_{\tilde{\rS},k}
(e^{-i\hat{H}_\trunc t}\hat \rho_{\rS \cup \rB_1}(0)e^{i\hat{H}_\trunc t} 
-
e^{-i\hat{H}_\trunc t}
\hat{\rho}^\rR_{\rS \cup \rB_1}(0)
e^{i\hat{H}_\trunc t}
)
]
|
\\
=&
|\mathrm{tr}_{\rS \cup \rB_1}
[
e^{i\hat{H}_\trunc t}
\hat{O}_{\tilde{\rS},k}
e^{-i\hat{H}_\trunc t}
(\hat{\rho}_{\rS \cup \rB_1}(0)
-
\hat{\rho}^\rR_{\rS \cup \rB_1}(0)
)
]
|,
\end{split}
\label{eq:LRB2_S_tilde2}
\end{equation}
where we used $\mathrm{tr}_{\rB_2}[ e^{-i\hat{H}_\trunc t} \hat{\rho}(0) e^{i\hat{H}_\trunc t} ] = e^{-i\hat{H}_\trunc t}\hat \rho_{\rS \cup \rB_1}(0)e^{i\hat{H}_\trunc t} $ with  $\hat \rho_{\rS \cup \rB_1}(0) := {\rm tr}_{\rB_2} [\hat \rho (0)]$ from the second to the third line.
By applying Corollary \ref{cor:EigenCan2},
for any $\varepsilon_{5,3}>0$ and any $\tilde{\varepsilon}>0$, 
\begin{equation}
|\mathrm{tr}_{\rS \cup \rB_1}
[
e^{i\hat{H}_\trunc t}
\hat{O}_{\tilde{\rS}}
e^{-i\hat{H}_\trunc t}
(\hat{\rho}_{\rS \cup \rB_1}(0)
-
\hat{\rho}^\rR_{\rS \cup \rB_1}(0)
)
]
|
\leq 
\varepsilon_{5,3}
\end{equation}
holds $\tilde{\varepsilon}/(2D_{\tilde{\rS}}^2)$-eigenstate-typically for sufficiently large $N$.

By summarizing the foregoing argument, and by letting $\varepsilon_{5,4} := \varepsilon_{5,1}+\varepsilon_{5,2}+\varepsilon_{5,3}$, 
we find that
for each $k$,
for any $\varepsilon_{5,4}>0$ and any $\tilde{\varepsilon}>0$, 
\begin{align}
|\mathrm{tr}_{\tilde{\rS}}
[
\hat{O}_{\tilde{\rS},k}
(
\hat{\rho}_{\tilde{\rS}}(t)-\hat{\rho}^\rR_{\tilde{\rS}}(t)
)
]
|
&\leq
\varepsilon_{5,4}
\label{eq:Lemma:Reference1}
\end{align}
holds $\tilde{\varepsilon}/(2D_{\tilde{\rS}}^2)$-eigenstate-typically for sufficiently large $N$.
Because $D_{\tilde{\rS}}$ is independent of $N$,
for any $\varepsilon_{5,4}>0$ and any $\tilde{\varepsilon}>0$, 
\begin{align}
|\mathrm{tr}_{\tilde{\rS}}
[
\hat{O}_{\tilde{\rS},k}
(
\hat{\rho}_{\tilde{\rS}}(t)-\hat{\rho}^\rR_{\tilde{\rS}}(t)
)
]
|
&\leq
\varepsilon_{5,4}
\label{eq:Lemma:Reference2}
\end{align}
holds for all $k=1,\cdots,2D_{\tilde{\rS}}^2$, $\tilde{\varepsilon}$-eigenstate-typically for sufficiently large $N$.

Let $\hat{O}_\mathrm{\tilde{\rS}}$ be an arbitrary operator with support $\tilde{\rS}$ and with operator norm $\|\hat{O}_\mathrm{\tilde{\rS}}\|=1$.
We can expand $\hat{O}_\mathrm{\tilde{\rS}}$ as
$\hat{O}_\mathrm{\tilde{\rS}}=\sum_{k=1}^{2D_{\tilde{\rS}}^2} c_k \hat{O}_{\tilde{\rS},k}$ 
with $c_k\in\mathbb{C}$.
We note that
\begin{align}
\begin{split}
|\mathrm{tr}_{\tilde{\rS}}
[
\hat{O}_{\tilde{\rS}}
(
\hat{\rho}_{\tilde{\rS}}(t)-\hat{\rho}^\rR_{\tilde{\rS}}(t)
)
]
|
&=
\abra{\mathrm{tr}_{\tilde{\rS}}
\bbra{
\sum_{k=1}^{2D_{\tilde{\rS}}^2}
c_k
\hat{O}_{\tilde{\rS},k}
\sbra{
\hat{\rho}_{\tilde{\rS}}(t)-\hat{\rho}^\rR_{\tilde{\rS}}(t)
}
}
}
\\
&\leq
\sum_{k=1}^{2D_{\tilde{\rS}}^2}
|c_k|
|\mathrm{tr}_{\tilde{\rS}}
[
\hat{O}_{\tilde{\rS},k}
(
\hat{\rho}_{\tilde{\rS}}(t)-\hat{\rho}^\rR_{\tilde{\rS}}(t)
)
]
|
\end{split}
\end{align}
From inequality (\ref{eq:Lemma:Reference2}),
for any $\varepsilon_{5,4}>0$ and any $\tilde{\varepsilon}>0$,
\begin{align}
\begin{split}
|\mathrm{tr}_{\tilde{\rS}}
[
\hat{O}_{\tilde{\rS}}
(
\hat{\rho}_{\tilde{\rS}}(t)-\hat{\rho}^\rR_{\tilde{\rS}}(t)
)
]
|
\leq
\sum_{k=1}^{2D_{\tilde{\rS}}^2}
|c_k|
\varepsilon_{5,4}
\end{split}
\label{eq:Lemma:Reference3}
\end{align}
holds $\tilde{\varepsilon}$-eigenstate-typically for sufficiently large $N$.
Because the set of $\hat{O}_{\tilde{\rS}}$ with $\|\hat{O}_{\tilde{\rS}}\|=1$ is compact,
\begin{align}
c:=\max_{\|\hat{O}_{\tilde{\rS}}\|=1}\sbra{\sum_{k=1}^{2D_{\tilde{\rS}}^2}|c_k|}
\end{align}
is finite, which is independent of $N$ because $\tilde{\rS}$ does not change in the thermodynamic limit.
Therefore, the right-hand side of inequality (\ref{eq:Lemma:Reference3}) is bounded as 
$\sum_{k=1}^{2D_{\tilde{\rS}}^2}
|c_k|
\varepsilon_{5,4}
\leq
\varepsilon_{5,4}c.
$
By letting $\varepsilon_{5} := \varepsilon_{5,4}c$, the lemma is proved.
\hspace{\fill}$\Box$

\

In the above proof, we used the Lieb-Robinson bound (Lemma \ref{lemma:TruncDynamics})
for evaluating 
the first and the third terms on the right-hand side of inequality~(\ref{eq:LRB2_S_tilde}).
They are proportional to $
e^{-\mu\mathrm{dist}({\tilde{\rS}},\partial\mathrm{B}_{1})}
(e^{vt}-vt-1)
$,
which is exponentially small with respect to $\dist(\tilde{\rS},\partial \rB_1)\simeq N^{\alpha/d}$.
We can evaluate the second term on the right-hand side of inequality~(\ref{eq:LRB2_S_tilde}) by Corollary \ref{cor:EigenCan2} as 
\begin{align}
|\mathrm{tr}_{\tilde{\rS}}
[
\hat{O}_{\tilde{\rS},k}
(
\hat{\rho}_{\tilde{\rS},\trunc}(t)-\hat{\rho}^\rR_{\tilde{\rS},\trunc}(t)
)
]
|
\leq
\mathcal{O}\sbra{N^{-(1-2\alpha)/4+\delta}}
+
\mathcal{O}\sbra{\sqrt{N^{-(1-2\alpha)/4+\delta}/\tilde{\varepsilon}}}
\end{align}
for sufficiently large $N$.

We note that $0<\alpha<1/2$ can be arbitrarily chosen.
If $\alpha$ is smaller, $e^{-\mu N^{\alpha/d}}$ is larger but $N^{-(1-2\alpha)/4+\delta}$ is smaller.
For sufficiently large $N$, 
$e^{-\mu N^{\alpha/d}}$ is smaller than $N^{-(1-2\alpha)/4+\delta}$ for any $0<\alpha<1/2$.
Therefore, a good error evaluation for sufficiently large $N$ is obtained by setting $0<\alpha\ll 1/2$.

By summing up, we have the following observation:
\begin{itembox}[l]
{\Observation[\label{obs:SizeDep}] (Size and time dependence of Lemma \ref{lemma:Reference})}
Let $\tilde{\varepsilon}>0$ be the probability of atypical eigenstates.
The left-hand side of inequality (\ref{eq:LRB_DM3}) in Lemma \ref{lemma:Reference}
is then bounded as
\begin{align}
\label{eq:obs:SizeDep}
\|
\hat{\rho}_{\tilde{\mathrm{S}}}(t)
-
\hat{\rho}^\rR_{\tilde{\mathrm{S}}}(t)
\|_1
\leq
\mathcal{O}\sbra{N^{-(1-2\alpha)/4+\delta}}
+
\mathcal{O}\sbra{\sqrt{N^{-(1-2\alpha)/4+\delta}/\tilde{\varepsilon}}}
\end{align}
in the short time regime $t\ll\tau$ and for sufficiently large $N$,
where we neglected the error term proportionate to 
$e^{-\mu\mathrm{dist}({\tilde{\rS}},\partial\mathrm{B}_{1})}(e^{vt}-vt-1)$.
We note that the neglected term increases in time with $\mathcal{O}(t^2)$ up to $t\simeq 1/v$.
\end{itembox}

\section{Proof of the second law of thermodynamics}
\label{sec:2nd}

In this section, we prove the second law of thermodynamics  [inequality~(3) in the main text] in the foregoing setup.

\

Before going to the proof, we note the following general inequality:
\begin{itembox}[l]
{\Proposition[\label{prop:vNEntIneq}] (Inequality for the von Neumann entropy~\cite{S_Andenaert2007})}
Let $\hat{\rho}$, $\hat{\rho}^\prime$ be density operators on a finite-dimensional Hilbert space with dimension $D$.
They satisfy
\begin{equation}
\label{eq:ineq1}
|S(\hat{\rho})-S(\hat{\rho}^\prime)|
\leq
\delta_\rho\ln (D-1)+H(\delta_\rho),
\end{equation}
where $\delta_\rho\defL{\|\hat{\rho}-\hat{\rho}^\prime\|_1}/2$ and $H$ is the Shannon entropy:
\begin{equation}
H(x)
:=
-x\ln x-(1-x)\ln(1-x).
\end{equation}
\end{itembox}

\

We now go into the main argument.
We define the following quantities as in the main text:
\begin{itembox}[l]
{\Definition[\label{def:vNEntHeat}] (The change in the von Neumann entropy of system S and the heat)}
The change in the von Neumann entropy of system S and the heat absorbed by system S are respectively defined as
\begin{equation}
\begin{split}
\Delta S_\rS
&:=
S(\hat{\rho}_\rS(t))-S(\hat{\rho}_\rS(0)),
\\
Q
&:=
-\mathrm{tr}_\rB[
\hat{H}_\mathrm{B}(
\hat{\rho}_\mathrm{B}(t)
-
\hat{\rho}_\mathrm{B}(0)
)
].
\end{split}
\end{equation}
\end{itembox}

The second law of thermodynamics is now stated as follows.

\begin{itembox}[l]
{\Theorem[\label{th:2nd}] (Second law of thermodynamics: inequality (3) in the main text)}
For any $\varepsilon_{\mathrm{2nd}}>0$, any $\tilde{\varepsilon}>0$, and  $t>0$, 
\begin{equation}
\label{eq:2nd}
\Delta S_\rS-\beta {Q}\geq -\varepsilon_{\mathrm{2nd}}
\end{equation}
holds $\tilde{\varepsilon}$-eigenstate-typically for sufficiently large $N$.
\end{itembox}

\noindent
{\bf Proof.}
As in Lemma \ref{lemma:Reference}, we consider the reference dynamics introduced in Sec.~\ref{sec:locality}.
The proof is divided  into five parts.

\noindent\textit{1. Second law for the reference dynamics.}
From Proposition \ref{prop:2nd},
the second law holds without any error for the reference dynamics.
We define the heat in the reference dynamics as
$
Q^\rR
\defL
-\tr_{\rS\cup\rB}[
\hat{H}_{\rB}(\hat{\rho}^\rR(t)-\hat{\rho}^\rR(0))
],
$
and  define the change in the von Neumann entropy in the reference dynamics as 
$\Delta S^\rR_{\rS}:= S(\hat{\rho}^\rR_{\rS}(t))-S(\hat{\rho}^\rR_{\rS}(0))$.
The second law for the reference dynamics is then given by
\begin{equation}
\label{eq:2nd_law_R}
\Delta S^\rR_{\rS}-\beta Q^\rR\geq 0.
\end{equation}

\noindent\textit{2. Division  of the von Neumann entropy.}
The change in the von Neumann entropy can be written as
\begin{equation}
\begin{split}
\Delta S_\rS
&=
S(\hat{\rho}_\rS(t))-S(\hat{\rho}_\rS(0))
\\
&=
S(\hat{\rho}^\rR_{\rS}(t))-S(\hat{\rho}^\rR_{\rS}(0))
+S(\hat{\rho}_\rS(t))-S(\hat{\rho}^\rR_{\rS}(t))
-S(\hat{\rho}_\rS(0))+S(\hat{\rho}^\rR_{\rS}(0))
\\
&=
\Delta S^\rR_{\rS}
+\sbra{S(\hat{\rho}_\rS(t))-S(\hat{\rho}^\rR_{\rS}(t)},
\end{split}
\label{eq:diff_vonNeumann_S}
\end{equation}
where we used $\hat{\rho}_\rS(0) = \hat{\rho}^\rR_{\rS}(0)$ from the second to the third line.
The first term in the last line of Eq.~(\ref{eq:diff_vonNeumann_S}) satisfies the second law (\ref{eq:2nd_law_R}) of the reference dynamics.
We will then evaluate $S(\hat{\rho}_\rS(t))-S(\hat{\rho}^\rR_{\rS}(t))$ as follows.

\noindent\textit{3. Approximation of the von Neumann entropy.}
By using Proposition \ref{prop:vNEntIneq} [Eq. (\ref{eq:ineq1})], we have
\begin{equation}
\label{eq:2nd_2}
|
S(\hat{\rho}_\rS(t))-S(\hat{\rho}^\rR_{\rS}(t))
|
\leq
\delta_\rS\ln (D_\rS-1)
+
H(\delta_\rS)
,
\end{equation}
where 
$\delta_\rS:=\|\hat{\rho}_\rS(t)-\hat{\rho}^\rR_{\rS}(t)\|_1 /2$.
We have $\|\hat{\rho}_\rS(t)-\hat{\rho}^\rR_{\rS}(t)\|_1 \leq \|\hat{\rho}_{\tilde \rS}(t)-\hat{\rho}^\rR_{\tilde{\rS}}(t)\|_1$ from the monotonicity of the trace norm.
Therefore, Lemma \ref{lemma:Reference} is applicable to evaluate $\delta_\rS$.
In addition, the right-hand side of inequality~(\ref{eq:2nd_2}) is continuous with respect to $\delta_\rS$,
and is equal to zero if $\delta_\rS=0$.
Therefore, 
for  any $\varepsilon_{\mathrm{2nd}}>0$ and any $\tilde{\varepsilon}>0$, 
\begin{equation}
|
S(\hat{\rho}_\rS(t))-S(\hat{\rho}^\rR_{\rS}(t))
|\leq\varepsilon_{\mathrm{2nd}}/3
\end{equation} 
holds $\tilde{\varepsilon}/3$-eigenstate-typically for sufficiently large $N$.

\noindent\textit{4. Approximation of  the heat.}
We next compare ${Q^\rR}$ with ${Q}$.
The total average energy 
$\trtotal[ \hat{H} \hat \rho (t) ]$
 is conserved with Hamiltonian $\hat{H}$.
Therefore, we have
\begin{equation}
\label{eq:Q1}
\begin{split}
{Q^\rR}
&:=
\tr_{\rS\cup\rB}[
(\hat{H}_\rS+\hat{H}_\rI)(\hat{\rho}^\rR(t)-\hat{\rho}^\rR(0))
],
\\
{Q}
&:=
\trtotal[
(\hat{H}_\rS+\hat{H}_\rI)(\hat{\rho}(t)-\hat{\rho}(0))
].
\end{split}
\end{equation}
By noting that the support of $\hat{H}_\rS+\hat{H}_\mathrm{I}$ is $\tilde{\mathrm{S}}$, the deviation between ${Q}$ and ${Q^\rR}$ is evaluated as
\begin{equation}
\label{eq:diff_Q}
\begin{split}
|{Q}-{Q^\rR}| \leq
|\tr_{\tilde{\rS}}[
(\hat{H}_\rS+\hat{H}_\mathrm{I})
(\hat{\rho}_{\tilde \rS}(t)-\hat{\rho}_{\tilde \rS}^\rR(t))
]|
+
|\tr_{\tilde{\rS}}[
(\hat{H}_\rS+\hat{H}_\mathrm{I})
(\hat{\rho}_{\tilde \rS}(0)-\hat{\rho}_{\tilde \rS}^\rR(0))
]|.
\end{split}
\end{equation}

We evaluate the first term on the right-hand side of inequality (\ref{eq:diff_Q}) as
\begin{equation}
\begin{split}
|\tr_{\tilde{\rS}}[
(\hat{H}_\rS+\hat{H}_\mathrm{I})
(\hat{\rho}_{\tilde \rS}(t)-\hat{\rho}_{\tilde \rS}^\rR(t))
]| 
\leq
\| \hat{H}_\rS+\hat{H}_\mathrm{I} \|   \| \hat{\rho}_{\tilde \rS}(t)-\hat{\rho}_{\tilde \rS}^\rR(t) \|_1,
\end{split}
\end{equation}
to which Lemma \ref{lemma:Reference} is straightforwardly applicable.
We note that $\| \hat{H}_\rS+\hat{H}_\mathrm{I} \|$ is assumed to be independent of $N$.
Therefore, for any $\varepsilon_{\mathrm{2nd}}>0$ and any $\tilde{\varepsilon}>0$, 
\begin{equation}
|\tr_{\tilde{\rS}}[
(\hat{H}_\rS+\hat{H}_\mathrm{I})
(\hat{\rho}_{\tilde \rS}(t)-\hat{\rho}_{\tilde \rS}^\rR(t))
]| 
\leq\varepsilon_{\mathrm{2nd}}/3
\end{equation} 
holds $\tilde{\varepsilon}/3$-eigenstate-typically for sufficiently large $N$.

The second term on the right-hand side of~(\ref{eq:diff_Q})
can also be evaluated from Lemma \ref{cor:EigenCan2}.
Therefore, for any $\varepsilon_{\mathrm{2nd}}>0$ and any $\tilde{\varepsilon}>0$, 
\begin{equation}
|\tr_{\tilde{\rS}}[
(\hat{H}_\rS+\hat{H}_\mathrm{I})
(\hat{\rho}_{\tilde \rS}(0)-\hat{\rho}_{\tilde \rS}^\rR(0))
]| 
\leq \varepsilon_{\mathrm{2nd}}/3
\end{equation}
holds $\tilde{\varepsilon}/3$-eigenstate-typically for sufficiently large $N$.

\noindent\textit{5. Finishing the proof.}
By summarizing  the foregoing argument, for any $\varepsilon_{\mathrm{2nd}}>0$ and any $\tilde{\varepsilon}>0$, 
\begin{equation}
| (\Delta S_\rS-\beta {Q}) - ( \Delta S^\rR_{\rS}-\beta Q^\rR ) | \leq \varepsilon_{\mathrm{2nd}}
\label{second_f}
\end{equation}
holds $\tilde{\varepsilon}$-eigenstate-typically for sufficiently large $N$.
By combining inequality (\ref{second_f}) with inequality (\ref{eq:2nd_law_R}), 
we prove the theorem.
\hspace{\fill}$\Box$

\

The error term $\varepsilon_{\mathrm{2nd}}$ comes from 
the difference between the von Neumann entropies in system S,
$S(\hat{\rho}_\rS(t))-S(\hat{\rho}^\rR_{\rS}(t))$,
and the heat $Q-Q^\rR$
between the actual dynamics and the reference dynamics.
From Observation \ref{obs:SizeDep},
we evaluate the error as
\begin{equation}
| (\Delta S_\rS-\beta {Q}) - ( \Delta S^\rR_{\rS}-\beta Q^\rR ) | \leq 
\mathcal{O}\sbra{N^{-(1-2\alpha)/4+\delta}}
+
\mathcal{O}\sbra{\sqrt{N^{-(1-2\alpha)/4+\delta}/\tilde{\varepsilon}}}
\end{equation}
in the short time regime $t\ll \tau$ and for sufficienly large $N$.

\section{Proof of the fluctuation theorem}
\label{sec:FT}

In this section, we prove 
the integral fluctuation theorem with 
the heat bath in an energy eigenstate [inequality (7) in the main text] in the foregoing setup.
We first prove the detailed fluctuation theorem, and then obtain the integral fluctuation theorem as a corollary.
By noting that the initial state of bath B is an energy eigenstate, we can define the characteristic functions as follows:
\begin{itembox}[l]
{\Definition[\label{def:CharaFunc}] (Characteristic functions of the entropy production)}
The characteristic functions of the entropy production in the forward and the reversed processes are respectively defined as
\begin{equation}
\begin{split}
&
G_\rF(u)
:=
\tr_{\rS\cup\rB}[
\hat{U}
e^{-iu\beta\hat{H}_{\rB}}e^{iu\ln\hat{\rho}_\rS(0)}
\hat{\rho}_\rS(0)\otimes\hat{\rho}_\rB(0)
\hat{U}^\dag
e^{-iu\ln\hat{\rho_\rS}(t)}
e^{iu\beta\hat{H}_{\rB}}
],
\\
&
G_\rR(u)
:=
\tr_{\rS\cup\rB}[
\hat{U}^\dag
e^{-iu\beta\hat{H}_{\rB}}e^{iu\ln\hat{\rho}_\rS(t)}
\hat{\rho}_\rS(t)\otimes\hat{\rho}_\rB(0)
\hat{U}
e^{-iu\ln\hat{\rho_\rS}(0)}
e^{iu\beta\hat{H}_{\rB}}
],
\end{split}
\end{equation}
where $u \in \mathbb C$.
\end{itembox}

We next give an important, but rather technical definition.

\begin{itembox}[l]
{\Definition[\label{def:EpsilonI}]}
For $t>0$ and $u \in \mathbb C$, we define $\varepsilon_\rI  \geq 0$ by
\begin{equation}
\label{eq:G_UV_assump}
\frac{\varepsilon_\rI}{6}
:=
\limsup_{N \to \infty}
\|
e^{iu\hat{H}_\rB}\hat{U} e^{-iu\hat{H}_\rB}
-
e^{-iu\hat{H}_\rS}
\hat{U}
e^{iu\hat{H}_\rS}
\|,
\end{equation}
where $N \to \infty$ means the thermodynamic limit.
\end{itembox}

We discuss the meaning of $\varepsilon_\rI$.
In the main text (Eq.~(8)), we discussed an assumption on the interaction Hamiltonian:
\begin{equation}
[\hat{H}_\rS+\hat{H}_\rB,\hat{H}_\rI]\simeq 0.
\end{equation}
The precise meaning of the above assumption is now formulated as 
\begin{equation}
\varepsilon_{\rm I} \ll 1.
\end{equation}
In fact, if $[\hat{H}_\rS+\hat{H}_\rB,\hat{H}_\rI]$ is exactly zero, 
$\|
e^{iu\hat{H}_\rB}\hat{U} e^{-iu\hat{H}_\rB}
-
e^{-iu\hat{H}_\rS}
\hat{U}
e^{iu\hat{H}_\rS}
\|$ is also exactly zero.

Logically speaking, we do not need to assume  $\varepsilon_{\rm I} \ll 1$ in the following mathematical argument.  However, if $\varepsilon_{\rm I}$ is not small, 
the error term can become large in the fluctuation theorem; our main theorem is physically meaningless in such a situation.
Fortunately, our numerical simulation shown in the main text demonstrates  that the contribution from $\varepsilon_{\rm I}$ is indeed negligible in our model of numerical simulation.
In particular, the inset of Fig.~3 in the main text shows that the main contribution to the error term of the fluctuation theorem is not   $\varepsilon_{\rm I}$.

\

The fluctuation theorem is now stated as follows.

\begin{itembox}[l]
{\Theorem[\label{th:FT}] (Detailed fluctuation theorem)}
For any $\varepsilon_{\mathrm{FT}}>0$, any $\tilde{\varepsilon}>0$,  $t>0$ and $u \in \mathbb C$, 
\begin{equation}
\label{eq:FT}
|G_\rF(u)-G_\rR(-u+i)|
\leq 
\varepsilon_{\mathrm{FT}}+\varepsilon_\rI
\end{equation}
holds $\tilde{\varepsilon}$-eigenstate-typically for sufficiently large $N$.
\end{itembox}

\noindent
{\bf Proof.}
We again consider  the reference dynamics introduced in  Sec.~\ref{sec:locality}.
The  proof is similar to that of the second law, which is  divided into four parts.

\noindent\textit{1. Fluctuation theorem for the reference dynamics.}
We define the characteristic functions for the reference dynamics:
\begin{equation}
\begin{split}
G^\rR_{\rF}(u)
&:=
\tr_{\rS\cup\rB}[
\hat{U}
e^{-iu\beta\hat{H}_{{\rB}}}e^{iu\ln\hat{\rho}_{\rS}^\rR(0)}
\hat{\rho}_{\rS}^\rR(0)\otimes
\hat{\rho}^\can_{\rB}
\hat{U}^\dag
e^{-iu\ln\hat{\rho}_{\rS}^\rR(t)}
e^{iu\beta\hat{H}_{{\rB}}}
],
\\
G^\rR_{\rR}(u)
&:=
\tr_{\rS\cup\rB}[
\hat{U}^\dag
e^{-iu\beta\hat{H}_{{\rB}}}e^{iu\ln\hat{\rho}_{\rS}^\rR(t)}
\hat{\rho}_{\rS}^\rR(t)\otimes\hat{\rho}_{\rB}^{\mathrm{can}}
\hat{U}
e^{-iu\ln\hat{\rho}_{\rS}^\rR(0)}
e^{iu\beta\hat{H}_{{\rB}}}
].
\end{split}
\end{equation}
From Proposition \ref{prop:FT},
the fluctuation theorem holds without any error for the reference dynamics:
\begin{equation}
\label{eq:FT_Reference}
G_{\rF}^\rR(u)-G_{\rR}^\rR(-u+i)=0.
\end{equation}

\noindent\textit{2. Rewriting the characteristic functions.}
We define
\begin{equation}
\begin{split}
\delta_\rI
&:=
\|
e^{iu\hat{H}_\rB}\hat{U} e^{-iu\hat{H}_\rB}
-
e^{-iu\hat{H}_\rS}
\hat{U}
e^{iu\hat{H}_\rS}
\|.
\end{split}
\end{equation}
From Definition \ref{def:EpsilonI} and the property of the limit superior,  
we obtain
\begin{equation}
\delta_\rI \leq \varepsilon_\rI/4
\label{e_I}
\end{equation}
for sufficiently large $N$.
The characteristic functions can then be evaluated as
\begin{equation}
\begin{split}
|
G_\rF(u)
&-
\trtotal[
\hat{U}
\hat{V}_0
\hat{\rho}(0)
\hat{U}^\dag
\hat{V}_t^\dag
]
|
\leq \delta_\rI,
\\
|
G_{\rF}^\rR(u)
&-
\tr_{\rS\cup\rB}[
\hat{U}
\hat{V}_0
\hat{\rho}^\rR(0)
\hat{U}^\dag
\hat{V}_{t}^{\rR\dag}
]
|
\leq \delta_{\rI},
\end{split}
\end{equation}
where we defined $\hat{V}_t:=e^{iu\beta\hat{H}_\rS}e^{iu\ln\hat{\rho}_\rS(t)}$
and
$\hat{V}_{t}^\rR:=e^{iu\beta\hat{H}_\rS}e^{iu\ln\hat{\rho}_{\rS}^\rR(t)}$.
We can also evaluate $G_\rR(u)$ and $G_{\rR}^\rR(u)$ in the same manner.
The key point here is that we have rewritten the characteristic functions within a small error  in such a way that Lemma \ref{lemma:Reference} is applicable.

\noindent\textit{3. The Lieb-Robinson bound.}
We evaluate the difference between $G_{\rF}(u)$ and $G^\rR_{\rF}(u)$ as follows:
\begin{equation}
\begin{split}
|G_{\rF}(u)-G_{\rF}^\rR(u)|
\leq&
2\delta_\rI
\\
&+
|
\trtotal[\hat{U}\hat{V}_0\hat{\rho}(0)\hat{U}^\dag\hat{V}_t^\dag]
-
\tr_{\rS\cup\rB}[\hat{U}\hat{V}_0\hat{\rho}^\rR(0)\hat{U}^\dag\hat{V}_t^\dag]
|
\\
&+
|
\tr_{\rS\cup\rB}[\hat{U}\hat{V}_0\hat{\rho}^\rR(0)\hat{U}^\dag\hat{V}_t^\dag]
-
\tr_{\rS\cup\rB}[\hat{U}\hat{V}_0\hat{\rho}^\rR(0)\hat{U}^\dag\hat{V}_{t}^{\rR\dag}]
|.
\end{split}
\label{eq:FT_diff_proof}
\end{equation}
Since the support of $\hat{V}_t^\dag$ is S,
Lemma \ref{lemma:Reference} is applicable to the second term on the right-hand side of inequality~(\ref{eq:FT_diff_proof}),
by replacing
$\tilde \rS$ with S, $\hat{\rho}(0)$ with $\hat{V}_0\hat{\rho}(0)$, and $\hat{\rho}^\rR(0)$ with $\hat{V}_0\hat{\rho}^\rR(0)$.
Therefore, for any $\varepsilon_{\rm FT} > 0$ and any $\tilde \varepsilon >0$, 
\begin{equation}
|
\trtotal[\hat{U}\hat{V}_0\hat{\rho}(0)\hat{U}^\dag\hat{V}_t^\dag]
-
\tr_{\rS\cup\rB}[\hat{U}\hat{V}_0\hat{\rho}^\rR(0)\hat{U}^\dag\hat{V}_t^\dag]
|
\leq 
\varepsilon_{\rm FT} / 4
\label{F_proof1}
\end{equation}
holds $\tilde \varepsilon /4$-eigenstate-typically for sufficiently large $N$.

The third term  on the right-hand side of inequality~(\ref{eq:FT_diff_proof}) is evaluated  as
\begin{equation}
\begin{split}
&
|
\tr_{\rS\cup\rB}[\hat{U}\hat{V}_0\hat{\rho}^\rR(0)\hat{U}^\dag\hat{V}_t^\dag]
-
\tr_{\rS\cup\rB}[\hat{U}\hat{V}_0\hat{\rho}^\rR(0)\hat{U}^\dag\hat{V}_{t}^{\rR\dag}]
|
\\
\leq&
\|
e^{-iu\beta\hat{H}_\rS}\hat{U}\hat{V}_0\hat{\rho}^\rR(0)\hat{U}^\dag 
\|
\|
e^{iu\ln\hat{\rho}_\rS(t)}
-
e^{iu\ln\hat{\rho}_{\rS}^\rR(t)}
\|_1.
\end{split}
\label{eq:FT_diff_proof2}
\end{equation}
Since $e^{iu\ln x}$ is a continuous function of $x$, if 
$\|
\hat{\rho}_\rS(t)
-
\hat{\rho}_{\rS}^\rR(t)
\|_1$
is small, 
$\|
e^{iu\ln\hat{\rho}_\rS(t)}
-
e^{iu\ln\hat{\rho}_{\rS}^\rR(t)}
\|_1$
is also small.
By applying  Lemma \ref{lemma:Reference},
$\|
e^{iu\ln\hat{\rho}_\rS(t)}
-
e^{iu\ln\hat{\rho}_{\rS}^\rR(t)}
\|_1$
is bounded from above.
Therefore, for any $\varepsilon_{\rm FT} > 0$ and any $\tilde \varepsilon >0$, 
\begin{equation}
|
\tr_{\rS\cup\rB}[\hat{U}\hat{V}_0\hat{\rho}^\rR(0)\hat{U}^\dag\hat{V}_t^\dag]
-
\tr_{\rS\cup\rB}[\hat{U}\hat{V}_0\hat{\rho}^\rR(0)\hat{U}^\dag\hat{V}_{t}^{\rR\dag}]
|
\leq 
\varepsilon_{\rm FT} / 4
\label{F_proof2}
\end{equation}
holds $\tilde \varepsilon /4$-eigenstate-typically for sufficiently large $N$.

By summing up inequalities~(\ref{e_I}), (\ref{F_proof1}), and (\ref{F_proof2}), for any $\varepsilon_{\rm FT} > 0$ and any $\tilde \varepsilon >0$, 
\begin{equation}
| G_{\rF}(u)-G_{\rF}^\rR(u)|
\leq
\varepsilon_{\rm FT}/2 + \varepsilon_\rI/2
\end{equation}
holds $\tilde \varepsilon /2$-eigenstate-typically for sufficiently large $N$.
In the same manner, we can show that 
and for any $\varepsilon_{\rm FT} > 0$ and any $\tilde \varepsilon >0$, 
\begin{equation}
|G_{\rR}(-u+i)-G_{\rR}^\rR(-u+i)|
\leq
\varepsilon_{\rm FT}/2 + \varepsilon_\rI/2
\end{equation}
holds $\tilde \varepsilon /2$-eigenstate-typically for sufficiently large $N$.

\noindent\textit{4. Finishing the proof.}
By combining the above results in the forward and the reversed processes, we finally prove that 
for any $\varepsilon_{\mathrm{FT}}>0$ and any $\tilde{\varepsilon}>0$,
\begin{equation}
|G_\rF(u)-G_\rR(-u+i)|
\leq
\varepsilon_{\mathrm{FT}}+\varepsilon_\rI
\end{equation}
holds $\tilde{\varepsilon}$-eigenstate-typically for sufficiently large $N$.
\hspace{\fill}$\Box$

\

We note that the convergence of the left-hand side of inequality (\ref{eq:FT}) is not uniform with respect to $u$ and $t$.
As a corollary of Theorem \ref{th:FT}, 
the integral fluctuation theorem is obtained.
\begin{itembox}[l]
{\Corollary[\label{cor:IFT_pure}] (Integral fluctuation theorem: inequality~(7) in the main text)}
For any $\varepsilon_{\mathrm{FT}}>0$, any $\tilde{\varepsilon}>0$,  and $t>0$, 
\begin{equation}
\label{eq:IFT}
|\langle e^{-\sigma}\rangle-1|
\leq 
\varepsilon_{\mathrm{FT}}+\varepsilon_\rI
\end{equation}
holds $\tilde{\varepsilon}$-eigenstate-typically for sufficiently large $N$.
\end{itembox}
{\bf Proof.}
As discussed in Corollary \ref{cor:IFT},
$\tbra{e^{-\sigma}} = G_{\rm F}(i)$ and $G_{\rm R} (0) =1$.
By substituting $u=i$ to the detailed fluctuation theorem (\ref{eq:FT}), we obtain inequality~(\ref{eq:IFT}).
\hspace{\fill}$\Box$

\

The size dependence and the time dependence of the left-hand side of
inequalities (\ref{eq:FT}) or (\ref{eq:IFT}) can be evaluated by the same manner as Observation \ref{obs:SizeDep} in Sec.~\ref{sec:ref_dynamics},
but $\varepsilon_\rI$ should be added to the right-hand side of inequality~(\ref{eq:obs:SizeDep}).

\section{Typicality in the Hilbert space}
\label{sec:typicality}

In the foregoing sections, we have discussed the second law and the fluctuation theorem 
in the case that the initial state of bath B is an energy eigenstate.
In this section, 
we will discuss a slightly different setup in terms of the typicality in the whole Hilbert space
of the energy shell, and prove a similar lemma to Lemma \ref{lemma:Reference} in Sec.~\ref{sec:ref_dynamics}.
We note that we still assume that bath B is on a hypercubic lattice with the periodic boundary condition.

We suppose that the initial state of bath B is given by a pure state 
\begin{align}
\ket{\Psi}=\sum_{i\in M_{U(\beta),\Delta}}c_i \ket{E_i},
\label{eq:initial_Haar}
\end{align}
where $c_i\in\mathbb{C}$ and $\sum_{i\in M_{U(\beta),\Delta}}|c_i|^2=1$.
We note that $\ket{\Psi}$ is not necessarily a single energy eigenstate.
The initial state of the total system is given by 
\begin{align}
\hat{\rho}(0)
=
\hat{\rho}_\rS(0)\otimes\hat{\rho}_\rB(0),
\end{align}
where $\hat{\rho}_\rB(0):=|\Psi\rangle \langle \Psi |$.
The total system then obeys the unitary evolution with Hamiltonian $\hat{H}$ defined in Eq.~(\ref{eq:Hamiltonian_total}),
where all of the foregoing Assumptions
\ref{assump:LRB},
\ref{assump:LocalInt}, \ref{assump:TransInv}, \ref{assump:ExpDecayT}, \ref{assump:Massieu}, and
\ref{assump:LocalIntS} are satisfied.

Let $\mathcal H_{{U(\beta),\Delta}}^\prime$ be the set of  unit vectors in the Hilbert space  $\mathcal H_{{U(\beta),\Delta}}$ of the energy shell.
We consider the Haar measure (i.e., the uniform measure) of $\mathcal H_{{U(\beta),\Delta}}^\prime$.
We denote by $| \mathcal V |$ the volume of a subset $\mathcal V \subset \mathcal H_{{U(\beta),\Delta}}^\prime$ with the Haar measure.  We introduce the concept of $\tilde \varepsilon$-typical statement:
\begin{itembox}[l]
{\Definition[\label{def:Typical}] ($\tilde{\varepsilon}$-typical statement)}
Let $\tilde{\varepsilon}>0$.
We say that a statement $\bf X$ about a unit vector in $\mathcal H_{{U(\beta),\Delta}}^\prime$ holds $\tilde{\varepsilon}$-typically, if
there exists a subset $\mathcal V\subset \mathcal H_{{U(\beta),\Delta}}^\prime$ 
such that 
${|\mathcal V|}/{|\mathcal H_{{U(\beta),\Delta}}^\prime |}>1-\tilde{\varepsilon}$
and $\bf X$ holds for any 
$|\Psi\rangle\in\mathcal{V}$.
\end{itembox}

\

As discussed in Sec.~\ref{subsec:FT},
it is necessary to employ the projection measurements 
on the initial and the final states in order to define the stochastic entropy production $\sigma$.
If the initial state (\ref{eq:initial_Haar}) is not an energy eigenstate,
it is projected to the diagonal ensemble
\begin{equation}
\hat{\rho}_\rB^\mathrm{diag}:=\sum_{i\in M_{U(\beta),\Delta}} |c_i|^2 \ket{E_i}\bra{E_i}.
\label{eq:initial_diag}
\end{equation}
From Lemma \ref{lemma:EigenMC} in Sec.~\ref{sec:weakETH},
we obtain the following corollary.
\begin{itembox}[l]
{\Corollary[\label{cor:DiagMC}]}
Let $\hat{O}$ be any operator on $\rB_1$ with $\|\hat{O}\|=1$,
where $\rB_1$ is a hypercube in B with side length $l=L^\alpha$ with $0\leq \alpha<1/2$.
For any $\varepsilon_6>0$ and any $\tilde{\varepsilon}>0$,
\begin{equation}
|
\tr_\rB[
\hat{O}
\hat{\rho}_\rB^\mathrm{diag}
]
-
\overline{O}
|
\leq
\varepsilon_6
\label{eq:cor:DiagMC}
\end{equation}
holds $\tilde{\varepsilon}$-typically for sufficiently large $N$.
\end{itembox}

\noindent
{\bf Proof.}
By defining $O_i:=\langle E_i| \hat{O}|E_i\rangle$ with $|E_i\rangle\in\mathcal{M}_{U(\beta),\Delta}$, we have
\begin{equation}
\label{eq:rhodist_diagonal1}
\begin{split}
|
\tr_\rB[
\hat{O}
\hat{\rho}_\rB^\mathrm{diag}
]
-
\overline{O}
|
&=
\abra{
\sum_{i\in M_{U(\beta),\Delta}}
|c_i|^2
(
O_i-\overline{O}
)
}
\\
&\leq
\sum_{i\in M_{U(\beta),\Delta}}
|c_i|^2
|O_i-\overline{O}|.
\end{split}
\end{equation}
From Lemma \ref{lemma:EigenMC}, for any $\varepsilon_{6,1}>0$ and any $\tilde{\varepsilon}_6>0$, we have
\begin{equation}
\label{eq:diagonal_1}
|O_i-\overline{O}|
\leq
\varepsilon_{6,1}
\end{equation}
$\tilde{\varepsilon}_6$-eigenstate-typically for sufficiently large $N$.
We then
split $M_{U(\beta),\Delta}$ into a typical subset $M_\rt$ and an atypical subset $M_{\ra}:=M\setminus M_\rt$,
where inequality (\ref{eq:diagonal_1}) is satisfied 
if and only if $i\in M_\rt$.
We note that $|M_{\ra}|\leq\tilde{\varepsilon}_6 D$.
The right-hand side of inequality~(\ref{eq:rhodist_diagonal1}) is then evaluated as
\begin{equation}
\label{eq:rhodist_diagonal2}
\begin{split}
\sum_{i\in M_{U(\beta),\Delta}}
|c_i|^2
|O_i-\overline{O}|
=&
\sum_{i\in M_\rt}
|c_i|^2
|O_i-\overline{O}|
+
\sum_{i\in M_\ra}
|c_i|^2
|O_i-\overline{O}|
\\
\leq &
\varepsilon_{6,1}
+
\sum_{i\in M_\ra}
|c_i|^2
|O_i-\overline{O}|.
\end{split}
\end{equation}

From the property of the Haar measure of $\mathcal H_{{U(\beta),\Delta}}^\prime$~\cite{S_Sugita2007},
we have
\begin{equation}
\begin{split}
\overline{
|c_i|^2
}
&=\frac{1}{D},
\\
\overline{
|c_i|^4
}
&=\frac{2}{D(D+1)},
\\
\overline{
|c_i|^2|c_j|^2
}
&=\frac{1}{D(D+1)}\quad (i\neq j),
\end{split}
\end{equation}
where $\overline{\cdots}$ describes the average with respect to the Haar measure.
Therefore, we obtain
\begin{align}
\overline{
\sum_{i\in M_\ra}|c_i|^2
}
&=
\frac{|M_\ra|}{D},
\\
\overline{
\sbra{\sum_{i\in M_\ra}|c_i|^2}^2
}
-
\sbra{\overline{
\sum_{i\in M_\ra}|c_i|^2
}}^2
&=
\frac{1}{D}\frac{|M_\ra|(D-|M_\ra|)}{D(D+1)}
<
\frac{1}{D}.
\end{align}

From the Chebyshev inequality, we have that for any $\varepsilon_{6,2} > 0$,
\begin{align}
P\bbra{
\abra{
\sum_{i\in M_\ra}|c_i|^2
-
\frac{|M_\ra|}{D}
}
\leq\varepsilon_{6,2}
}
\geq
 1- \frac{1}{D\varepsilon_{6,2}^2}
\end{align}
holds for sufficiently large $N$.
Since $D$ exponentially increases in $N$,
for any $\varepsilon_{6,2} > 0$ and any $\tilde \varepsilon > 0$, 
we have
\begin{align}
\frac{1}{D\varepsilon_{6,2}^2}
< \tilde \varepsilon
\end{align}
for sufficiently large $N$.
By noting that $\|\hat{O}\|=1$,
for any $\varepsilon_{6,2}$ and for any $\tilde{\varepsilon}$,
we obtain
\begin{equation}
\begin{split}
\sum_{i\in M_\ra}
|c_i|^2
|O_i-\overline{O}|
&\leq
2
\frac{|M_\ra|}{D}
+
2
\abra{
\sum_{i\in M_\ra}
|c_i|^2-\frac{|M_\ra|}{D}
}
<
\varepsilon_{6,2}
\end{split}
\label{eq:rhodist_diagonal3}
\end{equation}
$\tilde{\varepsilon}$-typically for sufficiently large $N$.
Letting $\varepsilon_6:=\varepsilon_{6,1}+\varepsilon_{6,2}$,
we prove the corollary
from inequalities
(\ref{eq:rhodist_diagonal2}) and (\ref{eq:rhodist_diagonal3}).
\hspace{\fill}$\Box$

\

We now discuss a variant of Lemma \ref{lemma:Reference} in Sec.~\ref{sec:ref_dynamics}.
Let $\hat{\rho}(t):=e^{-i\hat{H}t}\hat{\rho}(0)e^{i\hat{H}t}$ with 
$\hat{\rho}(0):=\hat{\rho}_\rS(0)\otimes \hat{\rho}_\rB^\mathrm{diag}$.
We again introduce the reference dynamics 
$\hat{\rho}^\rR(t):=e^{-i\hat{H}t}\hat{\rho}^\rR(0)e^{i\hat{H}t}$
with 
$\hat{\rho}^\rR(0)$ defined in Eq.~(\ref{ref_initial}).
We define the reduced density operators on $\tilde{\rS}$ as
$\hat{\rho}_{\tilde{\rS}}(t):=\tr_{\tilde{\rB}}[\hat{\rho}(t)]$
and
$\hat{\rho}^\rR_{\tilde{\rS}}(t):=\tr_{\tilde{\rB}}[\hat{\rho}^\rR(t)]$.
From Proposition \ref{prop:EquiEns}, Corollary \ref{cor:DiagMC}, and Lemma \ref{lemma:TruncDynamics},
we can prove a similar lemma to Lemma \ref{lemma:Reference} for 
the initial diagonal ensemble (\ref{eq:initial_diag}),
by replacing $\tilde{\varepsilon}$-eigenstate-typically by $\tilde{\varepsilon}$-typically.

\begin{itembox}[l]
{{\bf Lemma \ref{lemma:Reference}$^\prime$}}
For any $\varepsilon_5>0$, any $\tilde{\varepsilon}>0$, and $t>0$, 
\begin{equation}
\|
\hat{\rho}_{\tilde{\mathrm{S}}}(t)
-
\hat{\rho}^\rR_{\tilde{\mathrm{S}}}(t)
\|_1
\leq
\varepsilon_5
\end{equation}
holds $\tilde{\varepsilon}$-typically for sufficiently large $N$.
\end{itembox}
From Lemma \ref{lemma:Reference}$^{\prime}$,
we can prove our main results (Theorems \ref{th:2nd} and \ref{th:FT}, and Corollary \ref{cor:IFT_pure}) also for typical pure states in the whole Hilbert space,
by again replacing $\tilde{\varepsilon}$-eigenstate-typically to $\tilde{\varepsilon}$-typically.
We note that 
the initial state is projected to the diagonal ensemble,
and therefore we should replace $\hat{\rho}_\rB$ by $\hat{\rho}_\rB^\mathrm{diag}$
in Definitions \ref{def:vNEntHeat} and \ref{def:CharaFunc}.

\section{Discussion}
\label{sec:Discussion}
We have proved the second law of thermodynamics and the fluctuation theorem with a pure-state bath.
To make the proof mathematically rigorous,
we made several technical assumptions.
For example, we assumed that the heat bath is on a hypercubic lattice with the periodic boundary condition and translation invariance.
Although these assumptions are technically necessary for our proof,
we can physically expect that the lattice structure is neither restricted to hypercubes nor to the periodic boundary condition, in order to obtain essentially the same results.

On the other hand, there is a subtle problem on the assumption of translation invariance.
We can again physically expect that local breaking of translation invariance (by for example a local defect) does not affect the essentials of our results.
However, if there is strong global disorder that leads to the Anderson localization or the many-body localization (MBL), it has been shown that the weak ETH is no longer valid~\cite{S_Pal2010,S_Imbrie2016}.
In this sense, the absence of localization is crucial to obtain our results.

We also emphasize that
the spatial locality of interaction,
represented by 
Assumptions \ref{assump:LRB}, \ref{assump:LocalInt}, and \ref{assump:LocalIntS},
is crucially important.
The spatial nonlocality leads to anomalous Lieb-Robinson time that is proportional to $\ln N$ with system size $N$~\cite{S_Lashkari2013}, as is the case for the Sachdev-Ye-Kitaev model~\cite{S_Maldacena2016}.

\

We next remark on the concept of typicality.
In the proof of our main results,
we used the concept of typicality 
in Corollary \ref{cor:EigenCan} for a typical energy eigenstates
or in Corollary  \ref{cor:DiagMC} for a typical pure state in the whole Hilbert space.
However, if we assume that inequality~(\ref{new_coro0}) in Corollary \ref{cor:EigenCan} holds for a given state, 
we can prove our main results without invoking the concept of typicality.
More explicitly, we can prove the following corollary.
\begin{itembox}[l]
{\Corollary[\label{cor:WithoutTypicality}] (Main results without typicality)}
Suppose that there exists a sequence of state vectors of bath B,
written as $\{\ket{\Psi_N}\}_{N\in\mathbb{N}}$,
which satisfies the following.  
Let $\rB_1$ be a hypercube in $\rB$, whose side length is $l=L^\alpha$ with $0\leq\alpha<1/2$.
Let $\hat{O}$ be any operator on $\rB_1$ with $\|\hat{O}\|=1$.
For any $\varepsilon >0$, 
\begin{equation}
\abra{
\tr_\rB\bbra{
\hat{O}\sbra{
\hat{\rho}^\mathrm{diag}[\ket{\Psi_N}]
-
\hat{\rho}_\rB^\can
}
}
}
\leq
\varepsilon
\end{equation}
holds for sufficiently large $N$,
where $\hat{\rho}^\mathrm{diag}[\ket{\Psi_N}]:=\sum_{i\in M_{U(\beta),\Delta}} \ket{E_i} \bra{E_i} |\langle\Psi_N|E_i\rangle|^2$.
Then, 
under Assumptions \ref{assump:LRB}, \ref{assump:LocalInt}, and \ref{assump:LocalIntS},
our main results (Theorems \ref{th:2nd} and \ref{th:FT}, and Corollary \ref{cor:IFT_pure}) hold true for $\{| \Psi_N \rangle\}$, by removing the phrases that ``for any $\tilde \varepsilon> 0$'' and that ``$\tilde \varepsilon$-eigenstate-typically'' or ``$\tilde \varepsilon$-typically''.
\end{itembox}

\section{Details of the numerical simulation}
\label{sec:Numerical}
In this section, we show the details of our numerical simulation
and the supplementary numerical results.

\subsection{The Hamiltonian}
We considered hard-core bosons with nearest-neighbor repulsions, 
whose Hamiltonian is given by:
\begin{align}
\hat{H}_\rS
&=
\omega\hat n_0
,
\quad
\hat{H}_\rI
=
-\gamma^\prime
\sum_{<0,j>}
(\hat{c}^\dag_0 \hat{c}_j+\hat{c}^\dag_j \hat{c}_0),\\
\hat{H}_\rB
&=
\omega
\sum_i \hat n_i
-
\gamma
\sum_{<i,j>}
(\hat{c}^\dag_i \hat{c}_j+\hat{c}^\dag_j \hat{c}_i)
+
g
\sum_{<i,j>}
\hat{n}_i \hat{n}_j,
\end{align}
where $\omega>0$ is the on-site potential,
$-\gamma$ is the hopping rate in bath B,
$-\gamma^\prime$ is the hopping rate between system S and bath B,
and $g>0$ is the repulsion energy.
The annihilation (creation) operator of bosons at site $i$ is written as $\hat{c}_i$ ($\hat{c}_i^\dag$),
which satisfies the commutation relations
$[\hat{c}_i,\hat{c}_j]=[\hat{c}^\dag_i,\hat{c}^\dag_j]=[\hat{c}_i,\hat{c}^\dag_j]=0$ for $i\neq j$, 
$\{\hat{c}_i,\hat{c}_i\}=\{\hat{c}^\dag_i,\hat{c}^\dag_i\}=0$, $\{\hat{c}_i,\hat{c}^\dag_i\}=1$.

Bath B is on the two-dimensional lattice with the open boundary condition.
We here employed the open boundary condition for our numerical simulation in order to make the Lieb-Robinson time larger,
while the periodic boundary condition is assumed in our rigorous proof.
We expect that this difference does not matter as discussed in Sec.~\ref{sec:Discussion}.
System S is on a single site that is attached to one of the sites in bath B.

The initial state is a product state of system S and bath B : 
\begin{equation}
\label{eq:initial_numerical}
\hat{\rho}(0)=\ket{\psi}\bra{\psi}\otimes\ket{\Psi}\bra{\Psi},
\end{equation}
where $\ket{\psi} :=\hat{c}_0^\dag\ket{0}$ and  $| \Psi \rangle$ is a pure state of bath B.
The time evolution operator 
$\hat{U}=\exp(-i\hat{H}t/\hbar)$
is calculated by the full exact diagonalization of  the total Hamiltonian $\hat H$.

\subsection{Temperature of bath B}
\begin{figure}[t]
\begin{center}
\includegraphics[width=0.5\linewidth]{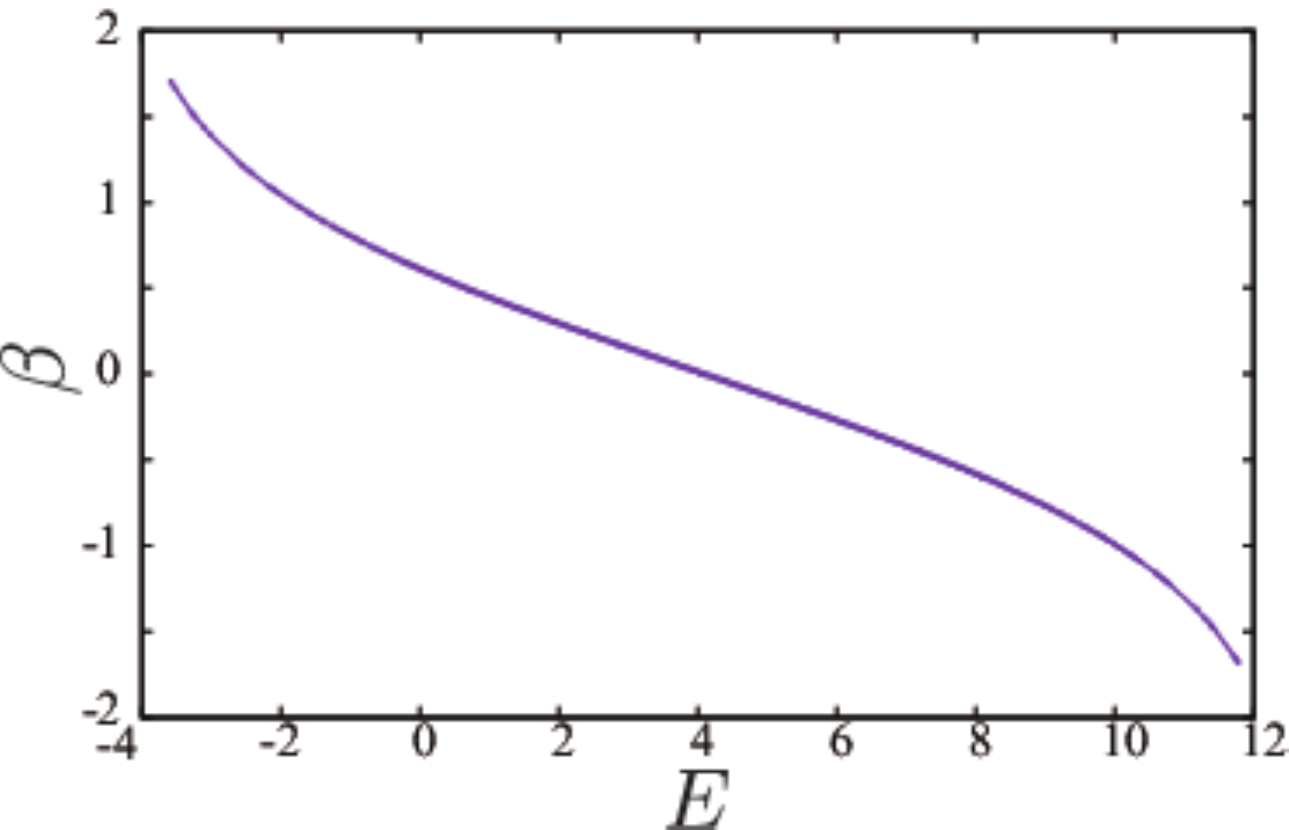}
\end{center}
\caption{
The inverse temperatures against the energy eigenvalues.
The bath is on $16=4\times4$ sites with 4 bosons.
The hopping rate within bath B is taken as $\gamma/\omega=1$,
and
the repulsion energy within bath B is taken as $g/\omega=0.1$.
}
\label{fig:fig4}
\end{figure}

In our rigorous theory,
temperature of bath B is defined by the corresponding energy shell in Definition \ref{def:EnergyShell}.
In practice, we numerically calculate the temperature of pure state $\ket{\Psi}$ as follows.
The average energy density of $| \Psi \rangle$ is given by $e (\Psi):= \langle \Psi | \hat{H}_\rB | \Psi \rangle / N$.
Then,
there exists  $\beta$ ($0<\beta < \infty$) such that
\begin{align}
\label{eq:def_temp_supppl}
e(\Psi )=u^\mathrm{can}(\beta),
\end{align}
where $u^\mathrm{can}(\beta)$ is given by Eq.~(\ref{u_beta}).
We refer to $\beta$ in Eq.~(\ref{eq:def_temp_supppl}) as the inverse temperature of $| \Psi \rangle$.
Figure~\ref{fig:fig4} shows the inverse temperatures of energy eigenstates, which are plotted against the energy eigenvalues.

\subsection{The Lieb-Robinson time}

We consider parameters  $\lambda_0$, $p_0$, $C$, $v$, and $\tau$ discussed in Sec.~\ref{sec:LRB} in terms of the Lieb-Robinson bound.
Because of the nearest neighbor interaction and the two-dimensional square lattice in our model,
the above parameters are given as follows:
$\lambda_0=\sqrt{\gamma^2+g^2}$,
$p_0=2$, $C=1/2$, 
$v=2\sqrt{\gamma^2+g^2}$,
and
$\tau=\dist(\tilde{S},\partial\rB_1) /( 2\sqrt{\gamma^2+g^2})$.
We choose $\mu=1$ to maximize $\tau$.
In addition, we set  $\rB_1$ such that $\dist(\tilde{S},\partial\rB_1)  =1$, and we have the Lieb-Robinson time $\tau=( 2\sqrt{\gamma^2+g^2})^{-1}$.

\subsection{Numerical verification of the detailed fluctuation theorem}

\begin{figure}[t]
\begin{center}
\includegraphics[width=\linewidth]{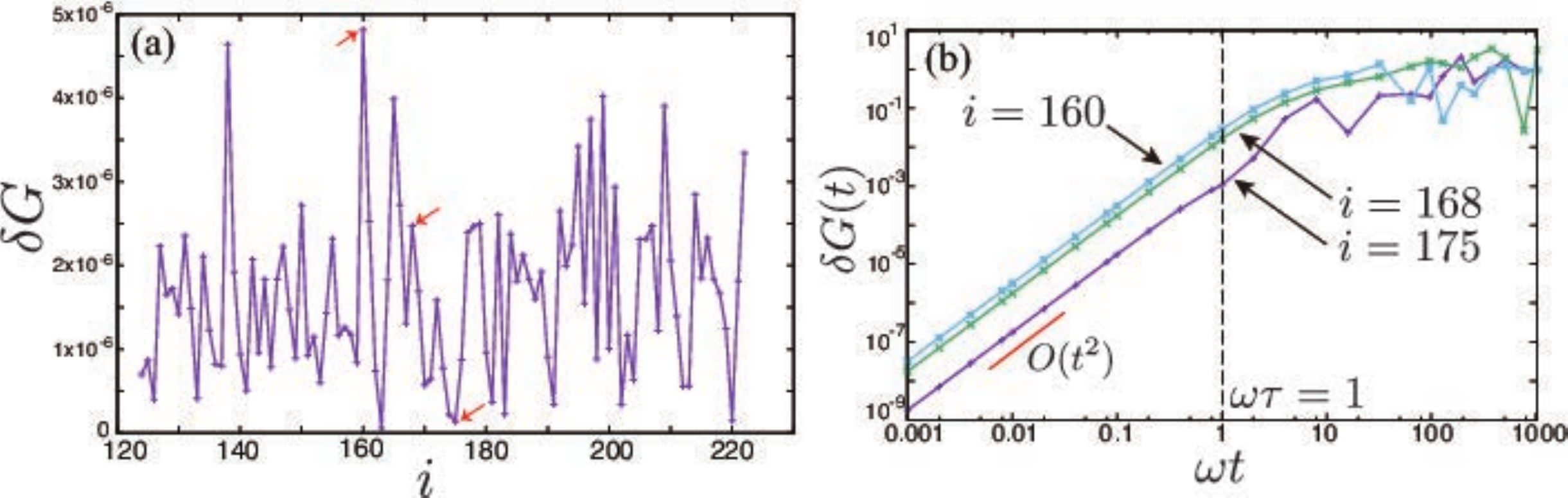}
\end{center}
\caption{
The eigenstate dependence and the time dependence of $\delta G(t)$.
The bath is  on $12=4\times3$ sites with 4 bosons, and the hopping rate within bath B is taken as $\gamma/\omega=1$, where the Lieb-Robinson time is $\tau \simeq 1/\omega$.
The repulsion energy within bath B is given by $g/\omega=0.1$,
and the system-bath coupling is given by $\gamma^\prime/\omega=0.1$.
(a)
The eigenstate dependence of $\delta G(t)$ with $\omega t=0.01$.
The eigenstates are chosen within $123\leq i\leq 222$.
The inverse temperatures of the initial eigenstates with $i=123$ and $i=222$
are $0.292$ and $0.052$, respectively.
The red arrows indicate $i=175, 168$ and $160$.
(b)
The time dependence of $\delta G(t)$.
The eigenstates are taken with $i=175, 168$ and 160.
In the short time regime,
the time dependence of the error is consistent with Observation \ref{obs:SizeDep}: $\delta G(t)\propto t^2$.
}
\label{fig:dft2}
\end{figure}

In the main text, we showed the numerical results on  the second law and the integral fluctuation theorem.
In this subsection,
we numerically study the detailed fluctuation theorem (Theorem \ref{th:FT}),
where the initial state is an energy eigenstate (i.e., $|\Psi \rangle = |E_i\rangle$ in Eq.~(\ref{eq:initial_numerical})).

We first discuss the eigenstate dependence of the difference between $G_\rF(u)$ and $G_\rR(-u+i)$.
We define an integrated error $\delta G(t)$ as
\begin{equation}
\delta G(t):=
\int_0^{2\pi}
du|G_\rF(u)-G_\rR(-u+i)|.
\end{equation}
Figure~\ref{fig:dft2} (a) shows  $\delta G(t)$ with $\varepsilon t=0.01$, which is plotted against index $i$ of the initial eigenstate $|E_i\rangle$ of bath B.
While $\delta G(t)$ is smaller than $3\times10^{-6}$ for most of the energy eigenstates,
there are atypical eigenstates with larger $\delta G(t)$, which is consistent to Theorem \ref{th:FT}.
Figure~\ref{fig:dft2} (b) shows time dependence of $\delta G(t)$,
which is proportional to $t^2$ in the short time regime.
This is consistent with the theoretical prediction based on the Lieb-Robinson bound (Observation  \ref{obs:SizeDep}). 

\begin{figure}[t]
\begin{center}
\includegraphics[width=\linewidth]{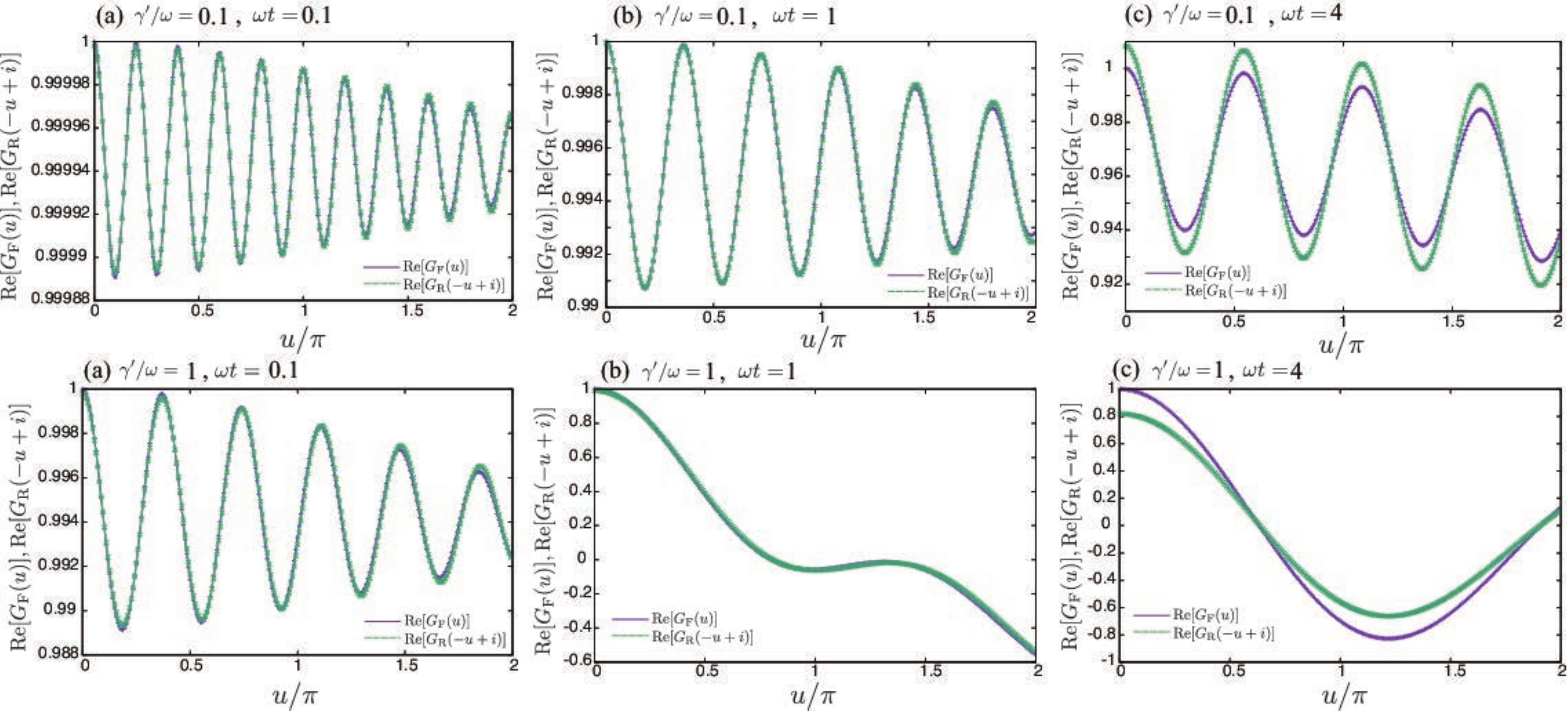}
\end{center}
\caption{
The real parts of $G_\rF(u)$ and $G_\rR(-u+i)$ for energy eigenstate $i=175$.
The bath is  on $12=4\times3$ sites with 4 bosons, and the hopping rate within bath B is taken as $\gamma/\omega=1$, where the Lieb-Robinson time is $\tau \simeq 1/\omega$.
The repulsion energy within bath B is given by $g/\omega=0.1$.
The upper panels (a)-(c) and the lower panels (d)-(f) show the weak and strong coupling cases, respectively.
In the short time regime (i.e., $\omega t\leq1$), 
the fluctuation theorem holds within small error.
As time increases, the fluctuation theorem tends to be violated as in (c) and (f).
}
\label{fig:dft3}
\end{figure}

We focus on the case that 
the initial state of bath B is taken as $\ket{E_i}$ with $i=175$,
whose $\delta G(t)$ is smallest in Fig.~\ref{fig:dft2} (b).
Figures~\ref{fig:dft3} (a)-(c) show the case of weak coupling between system S and bath B,
and 
Figs.~\ref{fig:dft3} (d)-(f) show the case of strong coupling.
In the short time regime,
the detailed fluctuation theorem holds within a small error for the both cases of the weak and the strong couplings.
As time increases, 
the fluctuation theorem tends to be violated as in Fig.~\ref{fig:dft3} (c) and (f).

\subsection{Numerical results on typical pure states in the Hilbert space}
In this subsection,
we show numerical results for the case that 
the initial state of bath B is the diagonal ensemble (\ref{eq:initial_diag}),
which is obtained by projecting $|\Psi\rangle$ in Eq.~(\ref{eq:initial_numerical}) that is sampled from the whole Hilbert space of the energy shell with respect to the Haar measure.
As discussed in Sec.~\ref{sec:typicality},
we can prove the second law and the fluctuation theorem for this setup.
Because the computational cost for mixed states is higher than that for pure states,
we consider a smaller bath ($9=3\times3$ sites with 3 bosons) in this subsection.

\begin{figure}[t]
\begin{center}
\includegraphics[width=\linewidth]{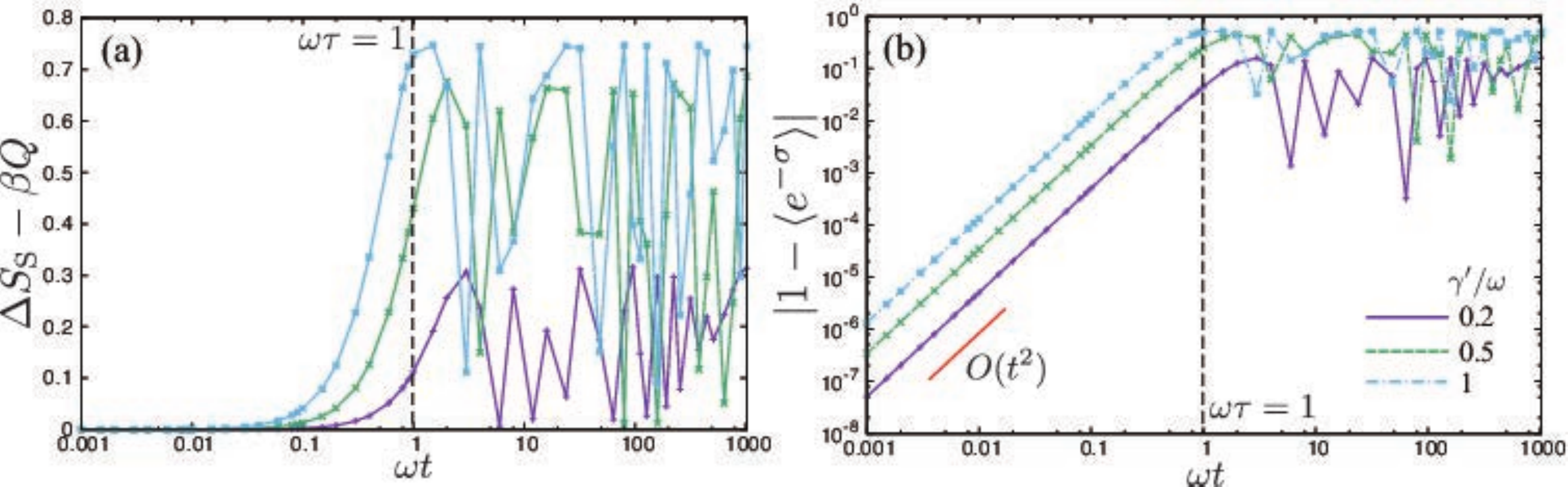}
\end{center}
\caption{
Numerical verification of the second law of thermodynamics and the fluctuation theorem for the diagonal ensemble, which is obtained from typical pure state $|\Psi\rangle$ that is sampled with respect to the Haar measure.
The bath is  on $9=3\times3$ sites with 3 bosons, and the hopping rate within bath B is taken as $\gamma/\omega=1$, where the Lieb-Robinson time is $\tau \simeq 1/\omega$.
The repulsion energy within bath B is given by $g/\omega=0.1$.
The initial inverse temperature of bath B is $\beta=0.1$.
(a)
Entropy production plotted against $\omega t$.
(b)
The deviation of $\langle e^{-\sigma}\rangle$ from unity plotted against $\omega t$.
The deviation is proportional to $t^2$ in the short time regime.
}
\label{fig:dft4}
\end{figure}

Figures~\ref{fig:dft4} (a) and (b) show the verifications of the second law and the integral fluctuation theorem, respectively.
Since bath B is smaller, 
the temporal fluctuations of $\langle \sigma\rangle$ and $\langle e^{-\sigma}\rangle$ are larger.

\end{document}